\documentclass[useAMS,a4paper,fleqn,usenatbib]{mnras}

\usepackage[T1]{fontenc}
\usepackage{ae,aecompl}

\usepackage{graphicx}	
\usepackage[hang,flushmargin]{footmisc}
\usepackage{txfonts}

\newcommand\HI{H\,{\small I}~}

\newcommand\blue{\color{blue}}

\title[Cosmic ray transport in galactic haloes]{Advective and diffusive cosmic
  ray transport in galactic haloes}

\author[V.~Heesen et al.]{Volker Heesen,$^{1}$\thanks{E-mail: {\blue V.Heesen@soton.ac.uk}} Ralf-J\"urgen Dettmar,$^{2}$
  Marita Krause,$^{3}$ Rainer Beck$^{3}$ and Yelena Stein$^{2}$\\
$^{1}$School of Physics and Astronomy, University of Southampton,
Southampton SO17 1BJ, UK\\
$^{2}$Astronomisches Institut der Ruhr-Universit\"at Bochum, Universit\"atsstr. 150, D-44780 Bochum, Germany\\
$^{3}$Max-Planck-Institut f\"ur Radioastronomie, Auf dem H\"ugel 69, D-53121 Bonn, Germany}

\date{Accepted 2016 February 12. Received 2016 February 12; in original form
  2015 March 27.}

\pubyear {2016}

\begin{document}
\label{firstpage}
\pagerange{\pageref{firstpage}--\pageref{lastpage}}

\maketitle

\begin{abstract}
We present 1D cosmic ray transport models, numerically solving equations of pure
advection and diffusion for the electrons and calculating synchrotron emission spectra. We find that for exponential halo magnetic field
distributions advection leads to
approximately exponential radio continuum intensity profiles, whereas
diffusion leads to profiles
that can be better approximated by a Gaussian function. Accordingly, the vertical radio spectral
profiles for advection are approximately linear, whereas for diffusion they are of
`parabolic' shape. We compare our models with deep ATCA observations of two edge-on galaxies, NGC~7090 and 7462, at
$\lambda\lambda$ 22 and 6~cm.  Our result is that the cosmic ray transport in NGC~7090 is
advection dominated with $V=150^{+80}_{-30}~\rm km\,s^{-1}$, and that the one in
NGC~7462 is diffusion dominated with
$D=3.0\pm 1.0 \times 10^{28}E_{\rm GeV}^{0.5}~\rm cm^2\,s^{-1}$. NGC~7090 has both
a thin and thick radio disc with respective magnetic field
scale heights of $h_{\rm B1}=0.8\pm 0.1$~kpc and $h_{\rm B2}=4.7\pm
1.0$~kpc. NGC~7462 has only a thick radio disc with $h_{\rm B2}=3.8\pm
1.0$~kpc. In both galaxies, the magnetic field scale heights are significantly smaller than
what estimates from energy equipartition would suggest. A non-negligible
fraction of cosmic ray electrons can escape from NGC~7090, so that this galaxy is not an electron calorimeter.
\end{abstract}


\section{Introduction}
\label{sec:intro}
Following the first detection of a radio halo in an external galaxy by
\citet{ekers_77a}, it has become clear that radio haloes are ubiquitous among late-type, star forming spiral galaxies
\citep[e.g.][]{hummel_88a,dahlem_97a,irwin_99a,dahlem_01a,krause_06a,oosterloo_07a,heesen_09a,soida_11a,wiegert_15a}
and starburst irregular dwarf galaxies \citep{kepley_10a, adebahr_13a}. Such radio
haloes indicate the presence of extra-planar cosmic rays and magnetic fields, because the radio continuum emission in
galactic haloes originates predominantly from synchrotron emission of cosmic ray
electrons (CRe) spiraling along magnetic field lines. The CRe cooling
time-scales are a few $10^7$~yr, so that we can infer that they come from the star forming
(SF) disc, accelerated and injected by supernovae (SNe). Notably, the radio continuum emission
extends in the vertical direction, visible as haloes, but only above star
formation sites in the disc plane \citep*{dahlem_06a}, whereas they hardly
extend radially beyond the optical disc \citep{mulcahy_14a}. This has reinforced the view that radio haloes are connected with
galactic outflows such as predicted by the `Galactic
fountain' model \citep{shapiro_76a, bregman_80a}, where the hot, X-ray emitting gas
in superbubbles, heated by multiple SNe, is vented into the halo. 

This so-called `disc--halo connection' was first established by the finding
that many star-forming galaxies possess a layer of extra-planar
diffuse ionized gas \citep[eDIG;][]{dettmar_92a,rossa_03a,rossa_03b}. The picture was soon
extended to include also the thermal X-ray emission \citep{tuellmann_06a},
a tracer for supernova heated gas. It now has become clear that galaxies also
possess extra-planar dust, first seen as sub-mm emission \citep{neininger_99a}
and as absorbing filaments in the optical light \citep{rossa_04a}, meanwhile observed as thermal (far)-infrared emission
\citep{hughes_14a} or as reflected UV-emission \citep{hodges-kluck_14a}. The
thick dusty discs of spiral galaxies are in part produced by the
expulsion of dust grains from the thin disc, again showing the existence of a
disc--halo connection. It has been noted early on that there is a
spatial correlation between the eDIG and the radio continuum emission
\citep{dettmar_92a}. The magnetic field is to very good approximation an ideal
MHD plasma and `frozen-in' the ionized gas, although a small amount of
resistivity remains, which is in fact essential to facilitate magnetic field
amplification by the galactic dynamo. Hence, to first order, in an outflow of ionized
gas the magnetic field and cosmic rays are advectively transported together. Vertical gas motions in the disc--halo
interface may also open up magnetic field lines, allowing cosmic
rays to diffuse along these field lines into the halo. Indeed, magnetic fields
in radio haloes
have in many cases a significant vertical component. Often, the field lines assume a
distinct `X'-shaped pattern
\citep[e.g.][]{dahlem_97a,tuellmann_00a,krause_06a,heesen_09b,krause_09a,soida_11a,mora_13a},
which may be caused by a superposition of a plane-parallel and vertical
component. Alternatively, the field lines may deviate from the flow lines as
expected for a galactic dynamo; although flow lines in
hydrodynamical simulations also have a radial component, so that they are
neither purely vertical \citep[e.g.][]{dalla_vecchia_08a}.
\begin{table}
\caption{Summary of galaxy parameters.\label{tab:par}}
\begin{tabular}{lllc}
\hline\hline
Parameter & NGC~7090 & NGC~7462 & Ref.\\
\hline
$D$ (Mpc) & $10.6$ & $13.6$ & 1\\
Redshift & $0.002825$ & $0.003549$ & 1\\
Scale (pc\,$\rm arcsec^{-1})$ & 51 & 65 & 1\\
Morphological type & Scd & Scd & 2\\
$M_{\rm tot}$ $(10^{10}~\rm M_{\sun})$ & $0.6$ & $0.3$ & 3\\
 $i$ (from face-on) & $89\degr$ & $90\degr$ & 2\\
$M_{\rm atom}$ ($10^9~\rm M_{\sun}$) & $1.73$ & $2.37$ & 2\\
$V_{\rm rot}$ ($\rm km\,s^{-1}$) & 124 & 112 & 3\\
PA & $128\degr$ & $73\degr$ &  4\\
$S_{\rm 1.4\,GHz}$ (mJy)  & $55\pm 3$ & $25.6\pm 1.3$ & This paper\\
$S_{\rm th, 1.4\,GHz}$ (mJy) & $4.5\pm 0.9$ & $1.8\pm 0.9$ & $\cdots$\\
$\rm SFR_{\rm hyb}$ $(\rm M_{\sun}\,yr^{-1})^*$ & $0.34\pm 0.02$ & $0.26\pm 0.05$ & $\cdots$\\
$\rm SFR_{\rm rad}$ $(\rm M_{\sun}\,yr^{-1})$ & $0.58\pm 0.03$ & $0.43\pm 0.02$ & $\cdots$\\
$\Sigma_{\rm SFR}$ $({\rm M_{\sun}yr^{-1}kpc^{-2}})^*$ & $1.0\times
10^{-3}$ & $7.7\times 10^{-4}$ & $\cdots$\\
$B$ ($\umu\rm G$) & $8.7\pm 1.1$ & $8.4\pm 1.1$ & $\cdots$\\
$\alpha$ & $-1.06\pm 0.09$ & $-1.12\pm 0.09$ & $\cdots$\\
$\alpha_{\rm nt}$ & $-1.24\pm 0.11$ & $-1.29\pm 0.18$ & $\cdots$\\
$U_{\rm IRF}/U_{\rm B}$ & $0.31$ & $0.18$ & This paper\\
\hline
\end{tabular}
\medskip\\References -- 1: NED Virgo Infall only ($H_0=73~\rm
km\,s^{-1}\,Mpc^{-1}$), 2: \citet{karachentsev_13a}, 3: \citet{dahlem_06a}, 4: for
NGC~7090 determined from $R$-band image, for NGC~7462 from
\citet*{dahlem_01a}. $^*$ based on hybrid (24~$\umu$m $+$ FUV) SFR
\end{table}

Cosmic rays do not only
have the potential to trace outflows, it has been theorized early on that they can be pivotal in
driving them, being not subjected to
strong radiative losses as the hot, X-ray emitting gas
\citep{ipavich_75a}. A series of papers by D.~Breitschwerdt and co-workers
\citep*[e.g.][]{breitschwerdt_91a,breitschwerdt_93a,dorfi_12a} have since then laid the
theoretical foundation. They found that a
cosmic ray driven galactic wind is likely to form. Furthermore,
the cosmic rays are able to transfer part of their momentum and
energy to the gas thus leading to mass-loaded winds thanks to the coupling between
the cosmic rays and the ionized gas, which is usually referred to as the `streaming instability' \citep{kulsrud_69a}. Cosmic ray driven winds
remove significant
amounts of mass, energy and angular momentum during the evolution of a
galaxy. Application to soft X-ray observations of the Milky Way by
\citet{everett_08a} has underlined the importance of cosmic rays in launching
galactic winds. In their `best-fit' model, the initial wind speed
is below the sound speed of the combined thermal
and cosmic ray gas (Mach number $M<1)$, accelerates in the halo, where it goes through the
critical point ($M$=1) at a distance of 2--3~kpc away from the disc and
accelerates further to reach an asymptotic velocity. A consequence of the
buoyancy of the relativistic cosmic ray gas is the cosmic ray driven galactic dynamo
\citep{hanasz_09a}, where magnetic field lines are stretched and twisted by a
combination of cosmic ray buoyancy force, Coriolis force and differential
rotation. A galactic wind has profound consequences for the directional symmetry of the magnetic field structure in
disc and halo. \citet{moss_10a} could show that mixed parity solutions, e.g.\
even disc and odd halo parity, can be obtained only if a galactic wind is
included in the modelling.

In this paper, we present a radio continuum polarimetry study of two
edge-on spiral galaxies, NGC~7090 and NGC~7462, at $\lambda\lambda$ 22 and
6~cm, observed at sub-kpc spatial resolution. We combine our radio observations with Balmer H$\alpha$
emission maps to correct for the contribution of thermal radio continuum
emission and use a combination of \emph{GALEX}
FUV and \emph{Spitzer} and \emph{WISE} mid-infrared emission to measure
spatially resolved star formation rate surface densities ($\Sigma_{\rm SFR}$).
Both NGC~7090 and 7462 have been scrutinized in earlier studies. \citet*{dahlem_01a} presented an ATCA radio continuum study at
$\lambda\lambda$ 22 and 13~cm, showing that both of them possess radio
haloes. In a follow-up study, extending this research to H\,{\small I} observations,
\citet{dahlem_05a} showed that both galaxies have thick discs of atomic
hydrogen, with irregularities detected both in the H\,{\small I} distribution and
velocity field. Furthermore, \citet{rossa_03a,rossa_03b} found extra-planar
Balmer H$\alpha$ emission, revealing the existence of ionized $H^+$ gas in their haloes. Here, we present the re-reduced combined data at $\lambda 22$~cm of
\citet{dahlem_01a,dahlem_05a} with 230~h on-source time and newly acquired observations at $\lambda
6$~cm with 160~h on-source time; we note that we omitted a re-analysis of the
$\lambda$13~cm data of \citet*{dahlem_01a}, because the observations are less
sensitive than either at $\lambda\lambda$ 22 or 6~cm and no further data were
acquired. We present also for the first time maps of the linearly polarized
radio continuum emission of both galaxies. Some fundamental galaxy parameters are summarized in Table~\ref{tab:par}.

This paper is organized as follows. In Section~\ref{sec:obs} we describe our 
observations and data reduction, followed by
Sect.~\ref{sec:meth}, where we describe our analysis of the vertical
distribution of the
non-thermal radio continuum emission. Our work probes the
application of the diffusion--loss equation to describe the transport of cosmic
rays as explained in Sect.~\ref{mod:CR}, results of which are presented in
Sect.~\ref{res:CRt}.  In Sect.~\ref{res:mag} we make use of the linear
polarization to study the halo magnetic
field structure. We discuss implications in Sect.~\ref{dis} and finish off with a summary of our
conclusions in Sect.~\ref{conc}. Throughout the paper, the radio spectral index
$\alpha$ is defined in the sense $S_{\nu}\propto\nu^{\alpha}$.
\begin{table}
\caption{ATCA observing journal. \label{tab:ATCA}}
\centering
\begin{tabular}{lccccr}
\hline\hline
Array & Project ID & Obs Dates & Freq & BWidth & Time\\
& & & (MHz) & (MHz) & (h)\\
\hline
\multicolumn{6}{c}{---NGC~7090---}\\\hline
750B & C655 & 1997 Aug 6 & $1384.0$ & $128.0$ & $16.6$\\
750C & C655 & 1997 Oct 18 & $1384.0$ & $128.0$& $15.8$\\
750A & C655 & 1998 May 1 & $1384.0$ & $128.0$& $14.2$\\
$1.5$D & C655 & 1998 Oct 16 & $1384.0$ & $128.0$& $14.0$\\
$1.5$D & C655 & 1998 Oct 19 & $1384.0$ & $128.0$& $6.0$\\
$1.5$B & C655 & 1999 Apr 2 & $1384.0$ & $128.0$& $14.1$\\
EW352 & C1005 & 2001 Oct 14 & $1384.0$ & $128.0$& $8.3$\\
$1.5$D & C1005 & 2001 Nov 16 & $1384.0$ & $128.0$& $8.0$\\
750D & C1005 & 2003 Feb 23 & $1384.0$ & $128.0$& $1.4$\\
750D & C1005 & 2003 Feb 24 & $1384.0$ & $128.0$& $8.0$\\
$1.5$B & C1005 & 2003 Jan 13 & $1384.0$ & $128.0$& $8.5$\\
6D & C1005 & 2003 Jul 15 & $1384.0$ & $128.0$ & $5.6$\\
6A & C1005 & 2003 Dec 13 & $1384.0$ & $128.0$ & $8.6$\\
$1.5$C & C1487 & 2005 Nov 29 & $4672.0$ & $256.0$ & $15.8$\\
6A & C1487 & 2005 Dec 8 & $4672.0$ & $256.0$ & $15.5$\\
6A & C1487 & 2005 Dec 9 & $4672.0$ & $256.0$ & $15.7$\\
EW352 & C1487 & 2006 Jan 17 & $4672.0$ & $256.0$ & $14.2$\\
750A & C1487 & 2006 Jan 22 & $4672.0$ & $256.0$ & $16.4$\\\hline
\multicolumn{6}{c}{---NGC~7462---}\\\hline
750B & C655 & 1997 Aug 8 & $1384.0$ & $128.0$ & $13.9$\\
750C & C655 & 1997 Oct 19 & $1384.0$ & $128.0$ & $13.8$\\
750A & C655 & 1998 May 3 & $1384.0$ & $128.0$ & $14.6$\\
$1.5$D & C655 & 1998 Oct 18 & $1384.0$ & $128.0$ & $14.8$\\
$1.5$D & C655 & 1998 Oct 20 & $1384.0$ & $128.0$ & $1.6$\\ 
750D & C1005 & 2001 Sep 27 & $1384.0$ & $128.0$ & $7.4$\\
EW352 & C1005 & 2001 Oct 16 & $1384.0$ & $128.0$ & $8.3$\\
$1.5$D & C1005 & 2001 Nov 20 & $1384.0$ & $128.0$ & $6.6$\\
$1.5$B & C1005 & 2003 Jan 14 & $1384.0$ & $128.0$ & $8.2$\\
6D  & C1005 & 2003 Jul 16 & $1384.0$ & $128.0$ & $4.8$\\
6A & C1005 & 2003 Dec 13 & $1384.0$ & $128.0$ & $8.7$\\
$1.5$C & C1487 & 2005 Nov 28 & $4672.0$ & $256.0$ & $15.2$\\
6A & C1487 & 2005 Dec 10 & $4672.0$ & $256.0$ & $15.1$\\
6A & C1487 & 2005 Dec 11 & $4672.0$ & $256.0$ & $17.7$\\
EW352 & C1487 & 2005 Jan 15 & $4672.0$ & $256.0$ & $14.2$\\
750A & C1487 & 2006 Jan 23 & $4672.0$ & $256.0$ & $16.4$\\\hline
\end{tabular}
\medskip\\ \flushleft Notes -- Observing times are on-source, corrected for 20
per cent calib-\\ ration overheads.
\end{table}

\section{Observations and data reduction}
\label{sec:obs}
\subsection{Radio continuum polarimetry}

\subsubsection{Observational setup, calibration and imaging}

Observations were taken with the Australia Telescope Compact
Array (ATCA), prior to the correlator upgrade with its then radio
continuum mode.\footnote{The Australia Telescope is funded by the
  Commonwealth of Australia for operation as a National Facility managed by
  CSIRO.} We observed at $\lambda 6$~cm with 2 $\times$ 128~MHz bandwidth
distributed in two `IFs' (intermediate frequencies) centered at $4.800$ and
$4.544$~GHz, respectively. At $\lambda$22~cm we used a single IF with a
bandwidth of $128$~MHz centered at $1.384$~GHz, where the other IF was used
either for a simultaneous observing of continuum emission at $\lambda$13~cm
(project code C655) or for observing of the H\,{\small I} line of atomic hydrogen
(C1005). Observing runs lasted typically 10~h, or longer, to ensure good
(u,v)-coverage, and started
with a 10--15~min long scan of J1938$-$634 as flux (primary) calibrator,
which is also used to measure the absolute polarization angle. Observations of
our source were interleaved every 30~min with a 5~min observation of a phase
(secondary) calibrator, which were J2106$-$413 and J2117$-$642 for NGC~7090 and
7462, respectively. Each observing run had enough parallactic angle
coverage ($\goa 60\degr$), so that we were able to calibrate for the instrumental polarization
of all antennae. The ATCA is an East--West interferometer, consisting of six
22-m antennae, five of which (CA01--CA05) are movable along a railway track. We used a variety of configurations ranging from the compact
EW352 (350-m maximum baseline) to the extended 6A ($2.9$-km maximum baseline)
configuration in order to fill in the (u,v)-plane with visibilities, so that we
are sensitive to extended emission.\footnote{Maximum baselines lengths do not include the fixed
  antenna CA06, which provides baseline lengths of 4--6~km in all configurations.} A journal of the observations is presented
in Table~\ref{tab:ATCA}.

We followed standard data
reduction procedures using {\small MIRIAD} \citep*{sault_95a}, setting our flux
densities according to the \citet{baars_77a} flux scale.  For imaging, we
used the Multi-Scale Multi-Frequency Synthesis (MS--MFS) cleaning algorithm
\citep{rau_11a} of the Common Astronomy
  Software Applications \citep[{\small CASA};][]{mcmullin_07a}. MS--MFS cleaning removes any residual flux densities
very efficiently \citep{hunter_12a}, which is particularly important for
extended sources that are large in comparison to the synthesized beam. We inverted the (u,v)-data where we used
a  `Briggs' robust weighting of $0.5$ in 
total power radio continuum (Stokes $I$) and an outer (u,v)-taper of $12\times 24~\rm
k\lambda$ ($\rm PA=90\degr$) in order to boost the signal-to-noise ratio
(S/N). Antenna CA06 is at a fixed location, approximately 3~km away from
the nearest of the other antennae. Including it results
in having baselines up to 6~km length, whereas without it the baseline lengths
are smaller than 3~km. Hence, we included antenna CA06 for the $\lambda
22$~cm maps, but not in the $\lambda 6$~cm maps, resulting in a similar
(u,v)-coverage and angular resolution. We used cleaning scales of up to 300~arcsec
angular size, similar to the angular extent of our galaxies, and we cleaned all maps until we
reached components with amplitudes of 2 $\times$ the rms noise level. For the maps in polarization (Stokes $Q$ and $U$) we used natural weighting
and left out antenna CA06 both at $\lambda\lambda$ 22 and 6~cm, in order
to boost the sensitivity of weak extended emission. Because our synthesized
beam in polarization was relatively large (27--34~arcsec) and the emission in Stokes $Q$
and $U$ is relatively compact, we chose to clean the
maps with the MFS implementation of clean in {\small MIRIAD}, without the use of
multiple angular cleaning scales.

\subsubsection{Specific data reduction issues}
We encountered several specific issues with our data, which required a special
treatment, which we outline in the following. Firstly, the wide field-of-view
(FoV) of the $\lambda 22$~cm maps meant that a number of
bright sources, mostly unresolved, were in the outskirts of the map close to
the half-power point or beyond. Because of the uncertainty of the primary beam
model in this region, and the fact that small positional changes (phase
errors) will result in large changes of the antenna amplitude gains, sources
can have significant residual
sidelobes after cleaning, increasing the rms noise level of the
maps. Hence, we attempted to
`peel' the sources away, by tailoring antenna gain solutions to them and
subsequently subtracting them from the (u,v)-data.  We applied the following
procedure: firstly, we subtracted all sources in the field except
the source to be peeled from the (u,v)-data using clean components. Then a self-calibration in
amplitude and phase ({\small GAINCAL} in {\small CASA} with a 600~s solution
interval) was performed using a model of the to be peeled source
only, resulting in amplitude and gain corrections to the antennas. The model
of the source was inserted into the model column of the measurement set with
{\small FT} in {\small CASA} and the gain corrections were applied to this model column using
the {\small CASA} toolkit function {\small CB.CORRUPT}. The source
was subtracted with {\small UVSUB}. The map was cleaned
again in order to create an updated model of the field and the procedure was
repeated with the next source to be peeled. We subtracted eight
sources from each map in this way, where we grouped several unresolved sources
together if they were nearby to each other. As the last step, we performed two
rounds of ordinary self-calibration in phase only on the entire field, using a
solution interval of 200~s. In this way we were able to remove
the sidelobes of the bright sources in a sufficient way, leading to an
improved rms noise level, 25~per cent lower than in the un-peeled maps.
We checked that the flux densities of our galaxies changed by only 3~per cent at most.
\begin{table}
\caption{Total power (TP) and polarized intensity (PI) maps.\label{tab:maps}}
\centering
\begin{tabular}{lccccc}
\hline\hline
Galaxy & $\lambda$ & \multicolumn{2}{c}{Resolution (arcsec)} & \multicolumn{2}{c}{Noise ($\rm \umu Jy\,beam^{-1}$)} \\
& (cm) & TP & PI & $\sigma_{\rm TP}$ & $\sigma_{\rm PI}$ \\\hline
NGC~7090 & 22 & $13.8$ & $30.4$ & 28 & 27\\
NGC~7090 & 6  & $13.8$ & $27.5$ & 13 & 12\\
NGC~7462 & 22 & $13.9$ & $37.4$ & 25 & 26\\
NGC~7462 & 6 & $13.9$ & $27.0$ & 9 & 12\\\hline
\end{tabular}
\end{table}

\begin{figure*}
\includegraphics[width=0.9\hsize]{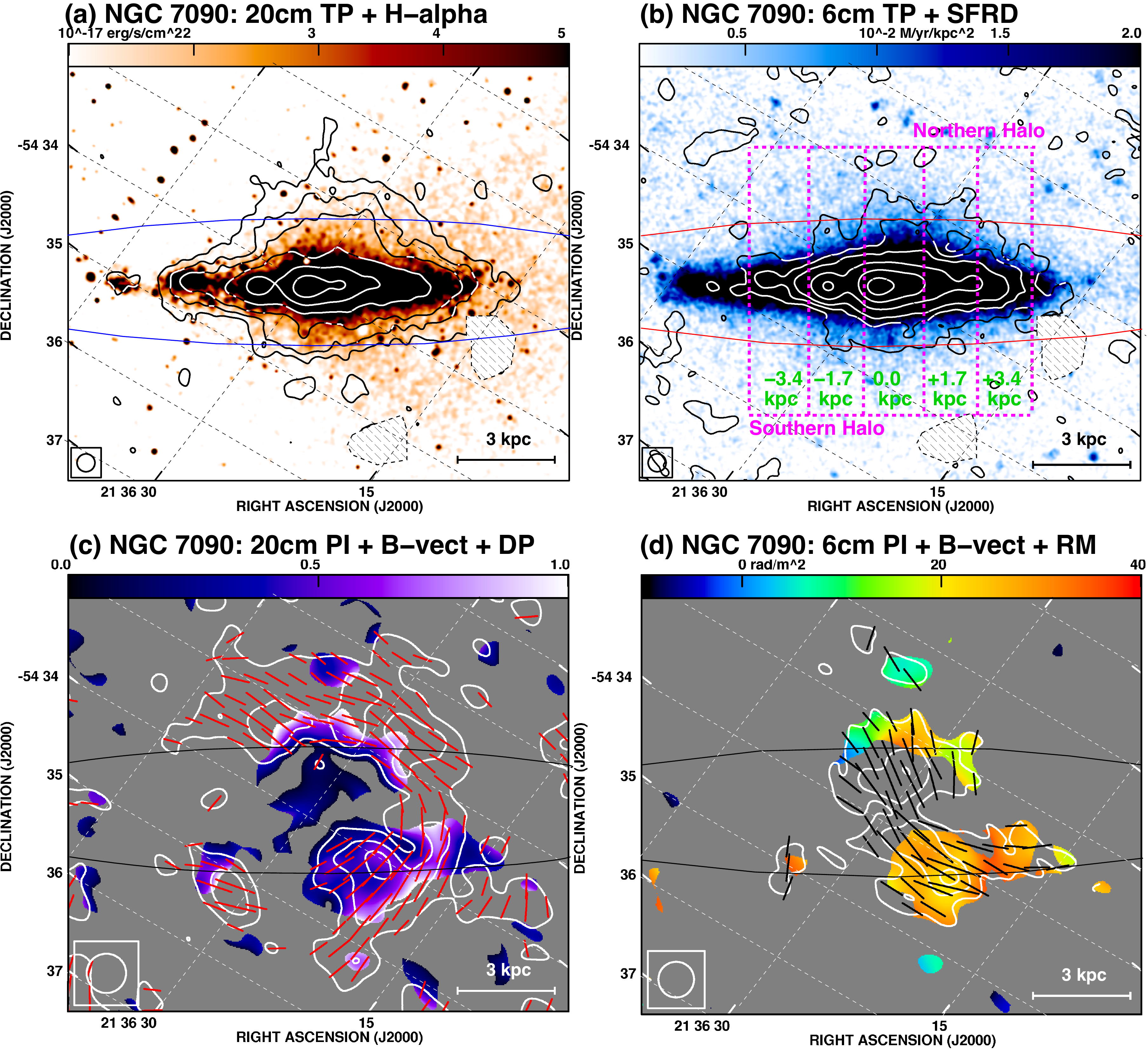}
\caption{ATCA observations of NGC~7090. \emph{Top left:} $\lambda$22~cm total power
  (TP) radio
  continuum emission at
  $13.8$~arcsec resolution overlaid on a Balmer H$\alpha$ emission map, convolved with
  a 3~arcsec Gaussian kernel to highlight weak, diffuse emission. Contours are at 3, 6, 10, 20, 40 and 80
  $\times$ 28~$\rm\umu Jy\,beam^{-1}$. Hatched areas show masked map
  regions. \emph{Top right:} $\lambda$6~cm total power (TP) radio
  continuum emission at
  $13.8$~arcsec resolution, overlaid on a $\Sigma_{\rm SFR}$ map ($\rm FWHM=6.7~arcsec$) from a linear
  combination of \emph{GALEX} FUV and \emph{Spitzer} 24~$\umu$m
  emission. Contours are at $-3$, 3, 6, 10, 20,
  40, and 80 $\times$ 13~$\rm \umu Jy\,beam^{-1}$. Hatched areas show masked
  map regions. The stripes used for measuring the non-thermal
  radio continuum scale heights are shown as well. \emph{Bottom left:} $\lambda$22~cm linearly polarized
  intensity (PI) and $B$-vectors at $30.4$~arcsec resolution, overlaid on
  the degree of depolarization (DP) between $\lambda\lambda$ 22 and 6~cm (0:
  entirely depolarized, 1: no depolarization, $\sigma_{\rm DP}=0.3$), where a constant non-thermal
  radio spectral index of $-1.26$ was assumed. Contours are at
  2, 4, and 6 $\times$ 27~$\rm \umu Jy\,beam^{-1}$. \emph{Bottom right:} $\lambda$6~cm
  linearly polarized intensity (PI) and $B$-vectors
   at $27.0$~arcsec resolution, overlaid on the
  rotation measure (RM) between $\lambda\lambda$ 22 and 6~cm ($\sigma_{\rm RM}=10~\rm rad\,m^{-2})$. Contours are at
  2, 4, and 6 $\times$ 12~$\rm \umu Jy\,beam^{-1}$. $B$-vectors are of
  polarized emission: the position angle of the polarized electric field
  rotated by $90\degr$, not corrected for Faraday rotation, with the length
  proportional to the polarized intensity and only plotted where $\rm PI \geq
  2\sigma_{PI}$. Ellipses show the optical extent of the disc
  ($7.4\times 1.3$~arcmin, source: NED). The maps were rotated, so that the
  major axis is aligned in East--West orientation.}
\label{fig:n7090}
\end{figure*}

\begin{figure*}
\includegraphics[width=0.9\hsize]{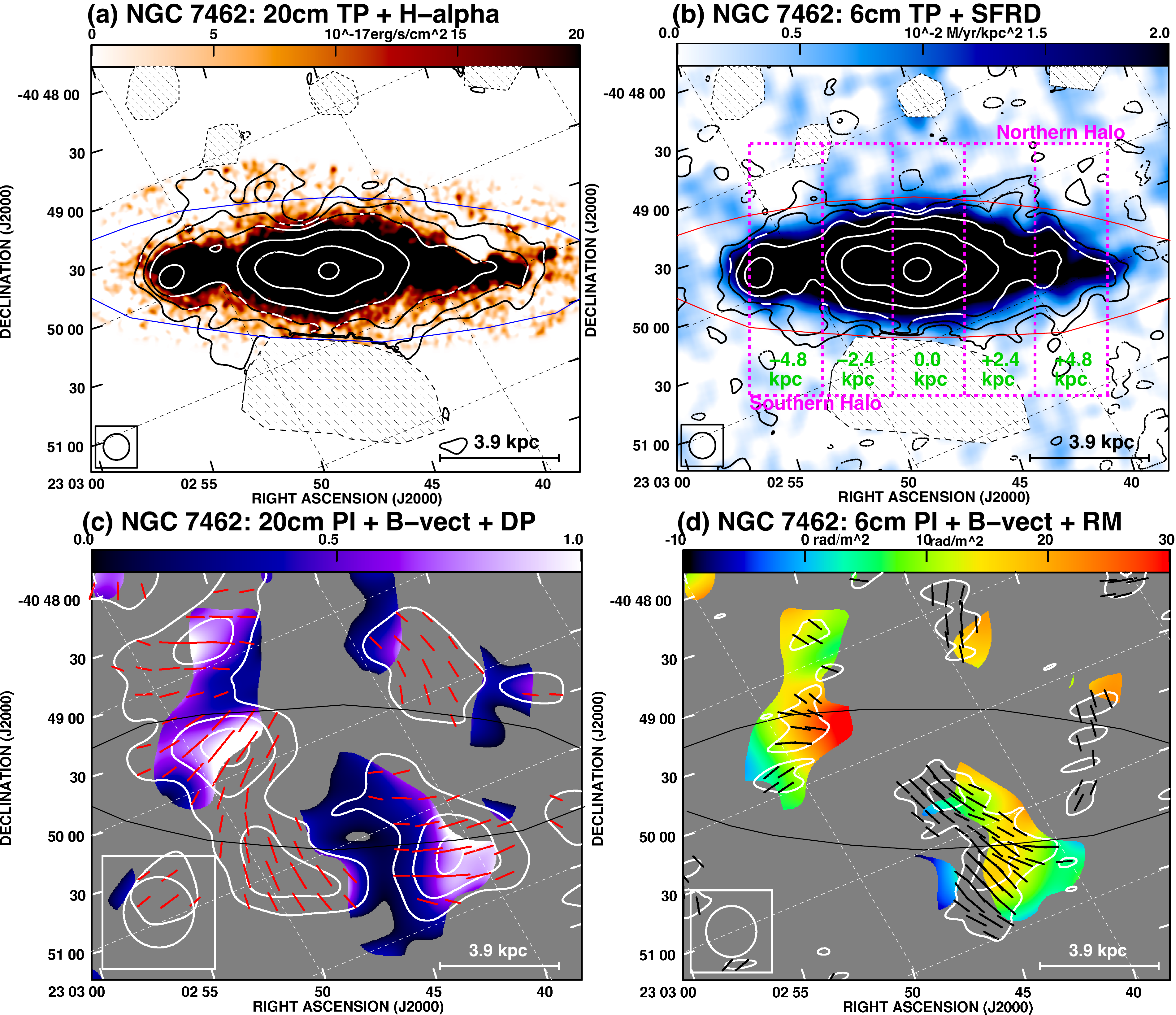}
\caption{ATCA observations of NGC~7462. \emph{Top left:} $\lambda$22~cm total power
  (TP) radio
  continuum emission at
  $13.9$~arcsec resolution overlaid on a Balmer H$\alpha$ emission map, convolved with
  a 3~arcsec Gaussian kernel to highlight weak, diffuse H$\alpha$ emission. Contours are at $-3$, 3, 6, 10, 20, 40 and 80
  $\times$ 25~$\rm\umu Jy\,beam^{-1}$. Hatched areas show masked map
  regions. \emph{Top right:} $\lambda$6~cm total power (TP) radio continuum emission at
  $13.9$~arcsec resolution, overlaid on a $\Sigma_{\rm SFR}$ map ($\rm FWHM=12~arcsec$) from a linear
  combination of \emph{GALEX} FUV and \emph{WISE} 22~$\umu$m emission. Contours
  are at $-2$, 2, 4, 8, 16, 32, and 64 $\times$ 9~$\rm \umu
  Jy\,beam^{-1}$. Hatched areas show masked map regions. The stripes used for measuring the non-thermal
  radio continuum scale heights are shown as well. \emph{Bottom left:} $\lambda$22~cm
  linearly polarized intensity (PI) and $B$-vectors at $37.4$~arcsec resolution, overlaid on the
  degree of depolarization (DP) between $\lambda\lambda$ 22 and 6~cm (0: entirely
  depolarized, 1: no depolarization, $\sigma_{\rm DP}=0.3$),  where a constant non-thermal
  radio spectral index of $-1.26$ was assumed. Contours are at
  2, 4, and 6 $\times$ 26~$\rm \umu Jy\,beam^{-1}$. \emph{Bottom right:} $\lambda$6~cm linearly polarized
  intensity (PI) and $B$-vectors at $27.0$~arcsec resolution, overlaid on the
  rotation measure (RM) between $\lambda\lambda$ 22 and 6~cm ($\sigma_{\rm RM}=10~\rm rad\,m^{-2})$. Contours are at
  2 and 4 $\times$ 12~$\rm \umu Jy\,beam^{-1}$. $B$-vectors are of
  polarized emission: the position angle of the polarized electric field
  rotated by $90\degr$, not corrected for Faraday rotation, with the length
  proportional to the polarized intensity and only plotted where $\rm PI \geq
  2\sigma_{PI}$. Ellipses show the optical extent of the disc
  ($4.6\times 1.2$~arcmin, source: DSS blue). The maps were rotated, so that the
  major axis is aligned in East--West orientation.}
\label{fig:n7462}
\end{figure*}

Another issue we encountered was that we were not able to use the first five channels
(of the 13 channels in total)
at $\lambda$22~cm in polarization, because ring-like structures appeared in
the maps that could not be identified and flagged in the (u,v)-data. It
should be noted that channel number three was flagged by the online flagging
system, and the adjacent channels (two and four) showed the strongest ring-like
structures, so that the problem was obviously in the data recording and not
some transient radio frequency interference (RFI) affecting the data, which we
could have flagged. Lastly, we note that the largest angular structure the
ATCA can image well at $\lambda$6~cm in the EW352 array is 460~arcsec,
assuming an observing time of 1~d. Both our galaxies fall comfortably within
this limit, with the largest angular scale of 300~arcsec, as estimated from
the star formation rate surface density maps (Figs~\ref{fig:n7090}(b) and \ref{fig:n7462}(b)).

All maps were primary beam corrected using {\small LINMOS} (part of {\small MIRIAD}),
before we exported them into {\small FITS} format and took them for further
analysis into {\small AIPS}. \footnote{{\scriptsize AIPS}, the Astronomical Image Processing Software, is free software
 available from NRAO.} Details of
the maps presented in this paper can be found in Table~\ref{tab:maps},
where angular resolutions are referred to as the full-width at half-mean (FWHM).
\subsection{Thermal radio continuum emission}
\label{subsec:Ha}
We created maps of the thermal (free--free) radio continuum emission using Balmer
H$\alpha$ emission line maps following standard conversion
\citep*[e.g.][equation~3, electron temperature $T=10^4$~K]{deeg_97a}. NGC~7090 is
part of the Local Volume Legacy (LVL) survey \citep{kennicutt_08a}, where
calibrated H$\alpha$ and continuum subtracted H$\alpha$ maps are readily
available. NGC~7090 and 7462 are part of the H$\alpha$ survey by
\citet{rossa_03a,rossa_03b}, who supplied us with un-calibrated, but
continuum subtracted H$\alpha$ and an $R$-band optical maps. In order to
bootstrap the H$\alpha$ flux of NGC~7462, we assumed that the ratio of the H$\alpha$ to the
$R$-band flux is identical for both NGC~7090 and 7462 in instrumental
units. We obtained $R$-band fluxes
from \citet{doyle_05a} and scaled the H$\alpha$ map of NGC~7462
accordingly. We estimate the accuracy of this method to be accurate within
50~per cent. This is, however, for the purpose of this work accurate enough. At a
reference height of 2~kpc, the thermal radio continuum fraction is only 5~per cent at
$\lambda 6$~cm in NGC~7462. Hence, even a 50~per cent uncertainty will hardly
change the results of our scale height analysis.

Foreground stars in the H$\alpha$ maps were masked, and pixels replaced with values of surrounding pixels. NGC~7462 was
surrounded by a bowl of negative emission, probably due to a too conservative
continuum subtraction. We corrected this
using a constant offset, where all emission outside of the bowl was
blanked. All data were corrected for foreground absorption, using
$A_{V}=2.5\times E(B-V)~\rm mag$, where we used the $E(B-V)$ magnitudes
from \citet{schlafly_11a}. We did not correct for internal absorption due to
dust in the galaxies, which means that we may underestimate the amount of
thermal radio continuum emission, particularly in the disc plane. We will come
to this limitation in the following analysis and discuss its implications
there (Sects~\ref{res:n7090} and \ref{res:n7462}).

\begin{figure*}
 \includegraphics[width=0.99\hsize]{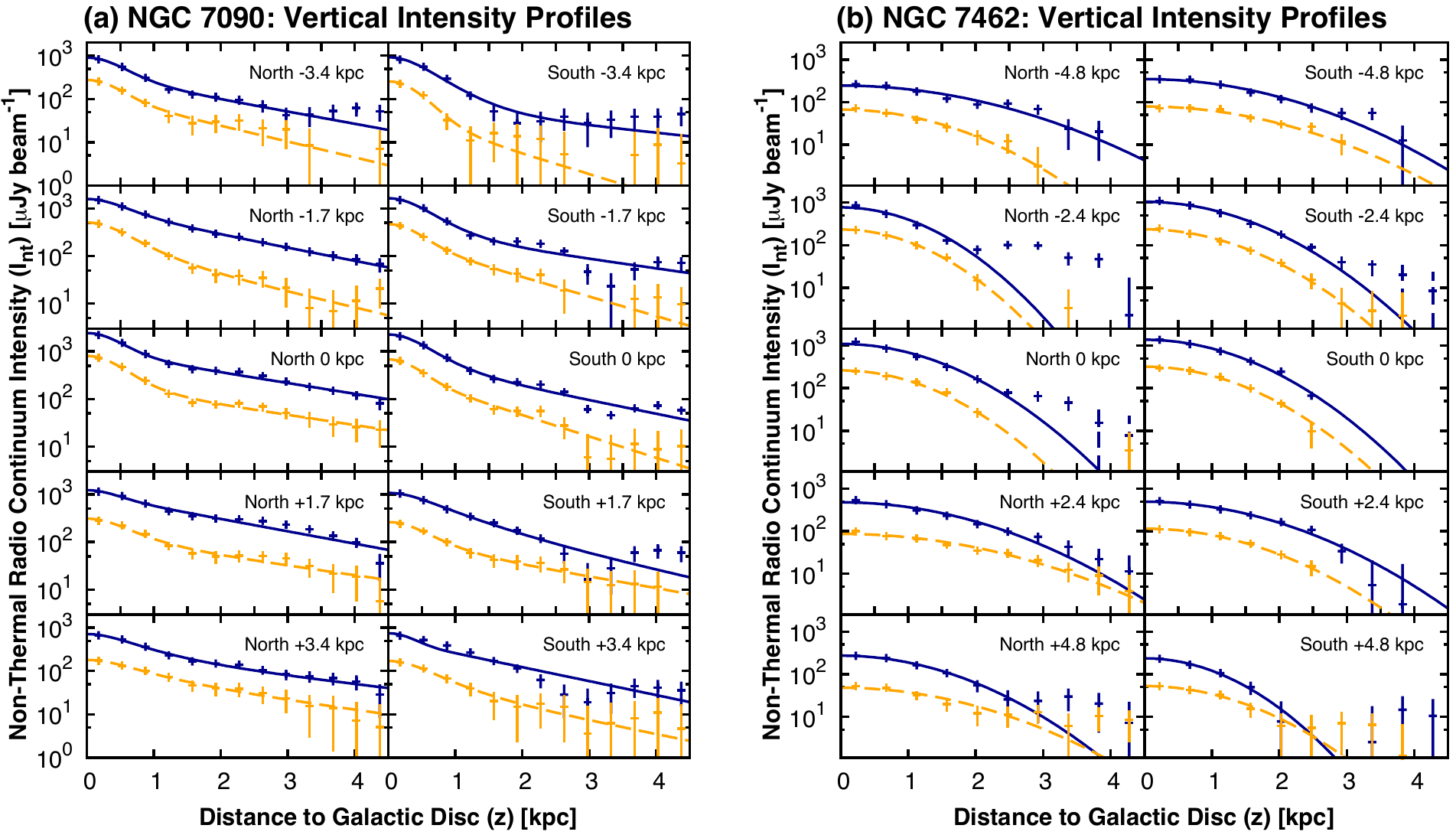}
\caption{Vertical profiles of the non-thermal radio continuum intensity as function of the distance to the galactic midplane. The offset of each stripe centre is indicated in
  each panel where `$+$' indicates an offset to the East and `$-$' to the West
  (see Figs~\ref{fig:n7090}(b) and \ref{fig:n7462}(b)). Data
  points denote the average radio continuum intensity at $\lambda\lambda$ 22
  and 6~cm. Lines indicate least-squares fits to the data points, where
  solid, dark-blue lines are fits to the $\lambda$22~cm data points and 
  dashed, orange lines to
  those at $\lambda$6~cm. \emph{Left:}
  NGC~7090, where the fits are two-component exponential functions. \emph{Right:}
  NGC~7462, where the fits are one-component Gaussian functions. Three stripes
  at $\lambda$22~cm ($-2.4$~kpc North and South, 0~kpc North) show halo
  emission as a second, more extended, component (see text).}
\label{fig:sh}
\end{figure*}

\section{Analysis}
\label{sec:meth}
We present the radio continuum maps of NGC~7090 and 7462 in
Figs~\ref{fig:n7090} and \ref{fig:n7462}, respectively. Both galaxies clearly
display extra-planar radio continuum emission, extending beyond the optical
disc, which we will analyse in this
section. Before we proceed, we subtract the thermal radio continuum (free--free) emission using
Balmer H$\alpha$ as a tracer, so that we are left over with the non-thermal
radio continuum emission only. Furthermore, we have masked unrelated background
sources, which are mostly point-like, i.e.\ unresolved. The exception is an
area in the southern halo of NGC~7462, centred on $\rm J2000.0$ RA $23^{\rm h}02^{\rm
  m}48\fs 3$ Dec.\
$-40\degr 51\arcmin 12\arcsec$, which contains diffuse emission. It also
harbours one double-peaked ($65~\rm\umu Jy$ each)
background source at RA $23^{\rm h}02^{\rm m}48\fs 7$ Dec.\
$-40\degr 50\arcmin 48\arcsec$, with 3~arcsec separation ($\rm PA\approx 240\degr$),
which we could split with the full resolution (6-km baseline) $\lambda$6~cm
observations. Because the emission in this area is
morphologically very different from the remaining radio continuum emission,
and clearly separated, we masked it as unrelated background emission. It may be
speculated that the masked spurious emission is related to an active galactic
nucleus (AGN); but we did not find any prominent unresolved source
($>$$70~\rm\umu Jy$) in the centre of NGC~7462, which would hint at any past or current AGN activity.

\subsection{Non-thermal radio continuum emission}
\label{an:tp}

\subsubsection{Scale heights}
\label{sh}
Following \citet{dumke_95a}, we used analytical
functions that allow the fitting of vertical radio continuum emission profiles with one or two
component Gaussian or exponential functions, where the instrumental point
spread function (PSF) is
assumed to be of Gaussian shape. In addition to the PSF, a correction for
projected disc emission can be added, for instance by fitting a Gaussian and
adding the FWHM in quadrature. Assuming that both of our galaxies are in an almost exact edge-on position ($i\geq 89\degr$), no such
correction is required. Hence, we use an effective beam size $\sigma = 0.5 \cdot {\rm FWHM} / \sqrt
{2\ln(2)}\approx 0.425\cdot {\rm FWHM}$, where FWHM is the angular (spatial)
resolution of our maps, resulting in $\sigma=0.30$ and $0.38$~kpc for NGC~7090
and 7462, respectively. As we shall see, in our case a one-component Gaussian function is appropriate
for one of our galaxies, for which the analytical function is:
\begin{equation}
  W_{\rm Gauss}(z) = \frac{w_o z_0}{\sqrt{2\sigma^2+z_0^2}} \exp\left (-\frac{z^2}{2\sigma^2+z_0^2} \right ),
\end{equation}
with $\sigma$ being the effective beam size, $w_0$ the maximum of the
distribution and $z_0$ the (Gaussian) scale height. For the second galaxy in
our study, a two-component
function is appropriate, of which one component takes the following form:
\begin{eqnarray} 
  W_{\rm exp}(z) & = & \frac{w_0}{2}\exp\left (-\frac{z^2}{2\sigma^2}\right
  )\nonumber\\
& & \times\left [ \exp\left ( \frac{\sigma^2-zz_0}{\sqrt{2}\sigma z_0} \right
  )^2
{\rm erfc} \left (\frac{\sigma^2-zz_0}{\sqrt{2}\sigma z_0} \right )\right .\nonumber\\
& & \left . + \exp\left ( \frac{\sigma^2+zz_0}{\sqrt{2}\sigma z_0} \right  
  )^2
{\rm erfc} \left (\frac{\sigma^2+zz_0}{\sqrt{2}\sigma z_0} \right )\right ],
\end{eqnarray}
with the complementary error function:
\begin{equation}
{\rm erfc}(x) = 1 - {\rm erf}(x) = \frac{2}{\sqrt{\upi}}\int_0^\infty
\exp(-r^2){\rm d}r.
\end{equation}
Again, $\sigma$ is the effective beam size, $w_0$ the maximum of the
distribution and $z_0$ the (exponential) scale height. We averaged the maps in stripes in order to raise the S/N of the weak
emission in the halo. A stripe width around 2~kpc is ideal to do
this, because this is the maximum typical diffusion length of cosmic rays in
the disc \citep{murphy_08a, tabatabaei_13b}, so that we do not smear out a
change of the local scale height. Hence, we used widths
of $1.7$~kpc (NGC~7090) and $2.4$~kpc (NGC~7462), sampled with a vertical
spacing of 7~arcsec, which is half of our angular resolution, equivalent to 350 and
460~pc in NGC~7090 and 7462, respectively. The stripes for the extraction of
the vertical profiles are shown in Figs~\ref{fig:n7090}(b) and
\ref{fig:n7462}(b); these were fitted using a least-squares routine, which
provides a formal error of the fit parameters. The profiles along with the fits are presented in
Fig.~\ref{fig:sh} and averaged scale heights are tabulated in
Table~\ref{tab:sh} (see Appendix~\ref{data_tables} for the scale heights in
the individual stripes).

We find that in NGC~7090 the vertical profiles can be equally well
described by a two-component exponential ($\chi^2=1.3$) or a two-component Gaussian fit ($\chi^2=1.2$). Presented in Fig.~\ref{fig:sh}(a) are the exponential fits, because, as we shall
see in Sect.~\ref{res:CRt}, the advective transport of cosmic rays in NGC~7090
favours them over Gaussian profiles. The thin disc in NGC~7090 has non-thermal scale
heights of $0.35\pm 0.06$ and $0.37\pm 0.08$~kpc ($\lambda\lambda$ 22 and 6~cm)
in the northern and $0.40\pm 0.06$ and $0.29\pm 0.10$~kpc in the
southern thin disc. In
NGC~7462, the profiles shown in Fig.~\ref{fig:sh}(b) can be better fitted with a one-component Gaussian fit
($\chi^2=0.6$) than by a one-component exponential fit ($\chi^2=1.5$), except at $\lambda$22~cm in the northern halo, where the
one-component exponential fit is better ($\chi^2=1.2$) than a
one-component Gaussian fit ($\chi^2=2.7$). This is largely because of a
second, more extended component, which can be traced in the northern
stripes at offsets of $-2.4$ and 0~kpc; a similar, albeit less prominent, component can
be found in the southern stripe at an offset of $-2.4$~kpc. Because we do not detect this
emission in any of the other stripes, nor at $\lambda$6~cm, we omit it in the
following analysis.

Notably, NGC~7462 does
not have a thin disc. The only other galaxy known to have a Gaussian vertical
profile is NGC~4594 (M104, the Sombrero Galaxy), an Sa galaxy, which is dominated by its
large central bulge \citep*{krause_06a}. However, we have to caution that this
result can be changed when also fitting for the baselevel. A baselevel of
$\approx$$-1\sigma_{\rm TP}$ allows us to fit the profiles with an
exponential function with a similar reduced $\chi^2$ as for the Gaussian
fit. We will come back to the significance of this finding in the further
analysis of NGC~7462 (Sect.~\ref{res:n7462}). If we were to fit the profiles with a one-component
exponential function, the scale height would be only $0.9\pm 0.1$~kpc, both at
$\lambda\lambda$ 22 and 6~cm, significantly less than in NGC~7090 and observed
in other galaxies \citep{krause_11a}, underlining that the radio halo of
NGC~7462 is different from that of most other galaxies.

It should be noted that the inclination angles we
used were derived from the ratio of the optical minor to major axis, so that
they are not very reliable. But we found that for an inclination angle of
$i=85\degr$ the scale height of the thick disc for both a one-component
Gaussian or a two-component
exponential fit decreases
by two per cent only. Hence, we conclude that our scale height measurements are
robust with respect to the uncertainty of the inclination angle.
\begin{table}
\caption{Non-thermal radio continuum scale heights $z_0$. \label{tab:sh}}
\begin{tabular}{lrlll}
\hline\hline
Galaxy & $\lambda$ & North & South & Weighted mean\\
& (cm) &  (kpc) & (kpc) & (kpc) \\\hline
NGC~7090 & 22 & $1.73\pm 0.17$ & $1.53\pm 0.39$ & $1.70\pm 0.22$\\
NGC~7090  & 6 & $1.90\pm 0.69$ & $1.14\pm 0.44$ & $1.35\pm 0.51$\\
NGC~7462  & 22 & $1.57\pm 0.10$ & $1.44\pm 0.05$ & $1.47\pm 0.06$\\
NGC~7462  & 6 & $1.31\pm 0.07$ & $1.39\pm 0.04$ & $1.36\pm 0.05$\\
\hline
\end{tabular}
\medskip\\Notes -- Values are weighted arithmetic means of the
scale heights measured in the stripes. Scale heights in NGC~7090 are referring
to the
thick disc from the
two-component exponential fitting functions; in NGC~7462 they refer to the
disc from the one-component Gaussian fitting functions.
\end{table}

\subsubsection{Radio spectral index}
\label{alpha}
The galaxy-wide integrated radio spectral indices for both galaxies are $\approx$$-1.1$ and
$\approx$$-1.3$, for the total and non-thermal radio continuum emission, respectively. The steep
non-thermal spectral index indicates that in both galaxies
CRe radiation losses, synchrotron and inverse Compton radiation, are
important and the CRe escape fraction is small \citep{lisenfeld_00a}. Hence, as a second characterization of the vertical distribution of the non-thermal
radio continuum emission, we measure profiles
of the non-thermal radio spectral index. The radio continuum emission in the
halo is weak, adversely affecting the accuracy of spectral index
measurements. A higher S/N value reduces the uncertainty, so that we averaged
only in one stripe between the same offsets along the major axis as used for
measuring the scale heights ($\pm 4.25$~kpc for NGC~7090, $\pm 6.0$~kpc for
NGC~7462). We present the vertical
profiles of the non-thermal radio spectral index in
Fig.~\ref{fig:alpha} (these values can be found in Appendix~\ref{data_tables}). 

In NGC~7090, the non-thermal radio
spectral index is $-1.0$ in the galactic midplane and steepens rapidly to
$-1.4\leq \alpha_{\rm nt}\leq -1.2$ within heights of 2~kpc. From then on, the radio spectral index
steepens less rapidly reaching values of $\approx$$-1.5$ at a height of 4~kpc,
the limit to which we can measure its value with a sufficient degree of certainty. In contrast, the radio spectral index in NGC~7462 shows only
little steepening within heights of 2~kpc, where values of
$-1.3\leq \alpha_{\rm nt}\leq -1.2$ are found. 
At heights larger than 2~kpc, however, the spectral index steepens rapidly,
particularly in the northern halo, reaching there a value of $-2.2\pm 0.4$. In
Sect.~\ref{mod:CR} we will present idealized models for the transport of cosmic
rays, in order to calculate synthetic spectral index profiles, which we can
compare with the observations.

\begin{figure*}
  \includegraphics[width=0.9\hsize]{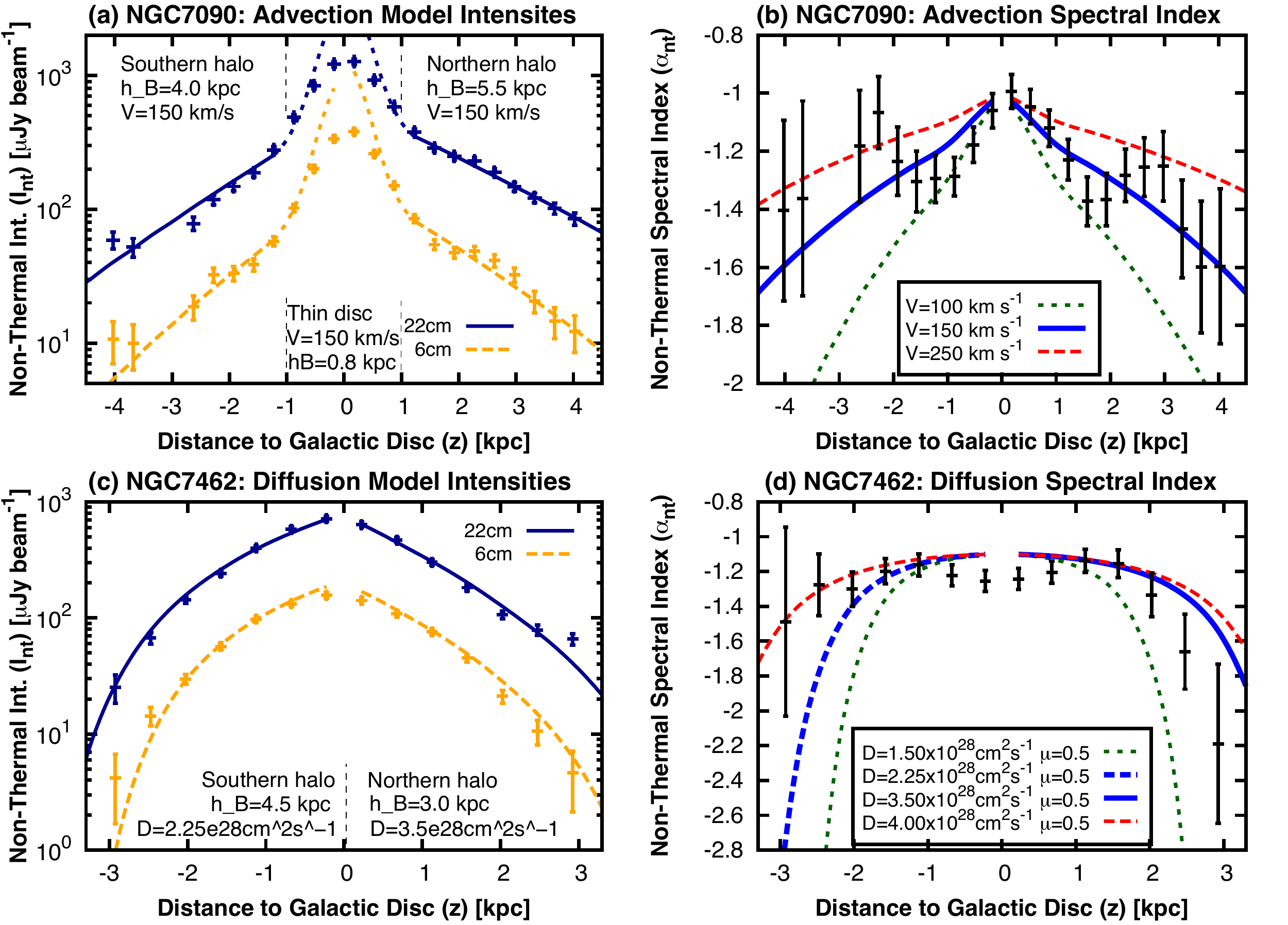}
\caption{Vertical profiles of the non-thermal radio continuum emission and the non-thermal
  spectral index in NGC~7090 (averaged between offsets of $\pm 4.25$~kpc,
  top row) and NGC~7462 (averaged between offsets of $\pm 6.0$~kpc, bottom row). Negative values of $z$ are for the southern halo, and
  positive ones are for the northern halo. \emph{Left panels:} modelled profiles from
  the solution of the 1D cosmic ray transport equations for advection
  (NGC~7090) and diffusion (NGC~7462) shown as solid, dark-blue ($\lambda$22~cm) and
  as dashed, orange ($\lambda$6~cm) lines. The short-dashed lines show the thin disc
  intensities in NGC~7090, not convolved to the resolution, so that they
  appear not to fit to the data. \emph{Right panels:} lines show our 1D cosmic ray transport models, with the best-fitting model shown as
  thick (solid, except in the S halo of NGC~7462) blue line and the error
  interval indicated by the dashed lines.}
\label{fig:alpha}
\end{figure*}

\subsection{Line-of-sight averaged quantities}
The geometry of an edge-on galaxy restricts measurements to averages
along the line-of-sight (LoS). The lengths of the lines-of-sight range from several
kpc to at least $\approx$10~kpc and, consequently, quantities are averaged over a
large variation of galactocentric radii.

\subsubsection{Magnetic field strengths}
\label{an:mag}
We determine (total) magnetic field strengths $B$ from the revised equipartition formula, using the non-thermal radio continuum intensities at $\lambda
22$~cm as input for the program {\small BFIELD} of \citet{beck_05a}. Intensities are converted into surface brightnesses appropriate for a
`face-on' view ($i=0\degr$) as function
of the offset along the major axis. The non-thermal radio spectral index is
another input for {\small BFIELD} and is assumed to be $-1.0$. Our measurement
between $\lambda
\lambda$ 22 and 6~cm resulted into a steeper non-thermal spectral index of
$\approx$$-1.25$ (Sect.~\ref{alpha}), but is this probably not representative for the
entire spectrum and with most of the energy contained in low-energy cosmic
rays, we revert to a more conservative choice. We assume a polarization degree
of 10~per cent, where we note that the resulting estimate of the
total magnetic
field strength depends only weakly on this particular choice (choosing 5
per cent instead makes a difference of less than 1 per cent). Furthermore,
we assumed an integration length of 1~kpc perpendicular to the galactic
plane, which is appropriate for the radio continuum emission in the thin disc
(Sect.~\ref{sh}) and in accordance with usually used assumptions in the
literature \citep[e.g.][]{niklas_97a, heesen_14a}. Lastly, we use $\mathcal
K=100$ as the number density ratio of the cosmic ray protons to the CRe.

In order to calculate globally averaged magnetic field strengths, we convert
the integrated radio continuum flux density into an averaged surface
brightness and repeat above analysis. We find
that the globally averaged magnetic field strength in NGC~7090 is
$8.7~\umu\rm G$ and NGC~7462 it is $8.4~\umu\rm G$, which is close to
$8.4~\umu\rm G$, the
mean value of the WSRT SINGS sample \citep{heesen_14a}, meaning that our
galaxies have very average magnetic field strengths. The average magnetic
field strength along the LoS varies between 5 and $13~\rm \umu G$, the lower
value found in the outskirts of the galaxies and the higher one in the
centre. For the area in which we measure scale heights and non-thermal spectral
indices, the magnetic field strength is mostly within a range of 8--12~$\umu\rm
G$, so that we assume an average magnetic field strength of 10~$\umu\rm
G$ in what follows. From the linearly polarized intensity, we can measure the magnetic field strengths of the
ordered component in the halo, results of which we will present in Sect.~\ref{res:mag}.

\subsubsection{Photon energy densities}
\label{urad}
Inverse Compton (IC) radiation losses depend on the energy density of the
interstellar radiation field (IRF). It can be described as the
superposition of the energy densities of the Cosmic Microwave Background
(CMB) with $U_{\rm CMB}=4.2\times 10^{-13}~\rm erg\, cm^{-3}$ at redshift
zero, the infrared (IR) radiation field, and the stellar (optical) radiation
field. Usually the stellar radiation field is scaled to
the IR radiation field with $U_{\rm star}=1.73\times U_{\rm IR}$, as derived for the solar neighbourhood \citep{draine_11a,tabatabaei_13a}. The infrared radiation field
can be measured from the total IR luminosity. For NGC~7090 we use the
prescription by \citet{dale_02a}, using \emph{Spitzer} 24, 70 and $160~\rm\umu
m$ luminosities from \citet{dale_09a}. For NGC~7462, we use the prescription
by \citet{condon_92a}, using the \emph{IRAS} flux densities at 60 and $100~\rm\umu m$. A global IR radiation energy density can then be calculated as
$ U_{\rm IR} = L_{\rm
 FIR}/(2\upi r_{\rm int}^2 c)$. Here,
$r_{\rm int}$ is the galactocentric radius of the galaxy
 to which we integrated the radio
flux densities, and $c$ is the speed of light. We can calculate the global
energy density of the IRF as $U_{\rm rad}=U_{\rm IRF}+U_{\rm CMB}$, where
$U_{\rm IRF}=2.73\,U_{\rm IR}$.

In order to obtain a local (kpc scale) IRF energy density, we assume that it
scales with $\Sigma_{\rm SFR}$, because the bulk of the dust heating
UV-radiation comes from young, massive stars. We find that the
average ratio of the photon energy density to that of the magnetic field is
$0.31$ (NGC~7090) and $0.18$ (NGC~7462). These values mean that the CRe energy
loss rate is dominated by synchrotron over IC radiation losses.

\subsubsection{CRe lifetime}
\label{sec:life}

In the interstellar medium (ISM), CRe are losing their energy mainly due to
synchrotron and IC radiation, so that GeV-electrons have lifetimes of a few $10^7~\rm
yr$. The ionization and bremsstrahlung losses for typical ISM
densities of $n=0.05~\rm cm^{-3}$ result in lifetimes of the order of $10^9~\rm yr$
and can hence be neglected \citep{heesen_09a}, where we
assume that cosmic rays are tracing the average density of the ISM within a factor of
a few \citep{boettcher_13a}. The combined synchrotron and IC loss rate for CRe is given by \citep[e.g.][]{longair_11a}:
\begin{equation}
-\left (\frac{{\rm d}E}{{\rm d}t}\right )=b(E)=\frac{4}{3} \sigma_{\rm T} c \left (\frac{E}{m_{\rm
      e}c^2} \right )^2 (U_{\rm rad}+U_{\rm B}),
\label{eq:be}
\end{equation}
where $U_{\rm rad}$ is the radiation energy density, $U_{\rm B}=B^2/8\upi$ is the magnetic
field energy density, $\sigma_{\rm T}=6.65\times 10^{-25}~\rm cm^2$ is the
Thomson cross section and $m_{\rm e}=511~\rm keV\,c^{-2}$ is the electron
rest mass. The time dependence of the energy is $E(t)=E_0(1+t/t_{\rm rad})^{-1}$,
so that at $t=t_{\rm rad}$ the CRe energy has dropped to half of its initial
energy $E_0$. The CRe lifetime, as determined by synchrotron and IC radiation losses, can be expressed by:
\begin{equation}
t_{\rm rad} = 34.2 \left (\frac{\nu}{\rm 1~GHz}\right )^{-0.5}
\left (\frac{B}{\rm 10~\umu G}\right )^{-1.5} \left
  (1+\frac{U_{\rm rad}}{U_{\rm B}}\right )^{-1}~{\rm Myr}.
\label{eq:t_rad}
\end{equation}
In NGC~7090 and 7462, we find CRe lifetimes, averaged along the lines-of-sight, of 16--60~Myr, with the shorter
lifetimes in the centre of the galaxies, where the magnetic field strengths
and photon energy densities are highest.

\subsubsection{Uncertainties}
The equipartition estimate of the magnetic field strength depends on the input
values with a power of only $1/(3-\alpha_{\rm nt})=0.25$, so that even
large input errors hardly affect the results. If we assume a combined
input uncertainty of 50~per cent, arising from
the number
density ratio, integration length and subtraction of thermal radio continuum emission, our magnetic field
uncertainty is $12.5$~per cent. We checked that an uncertainty of $\pm 0.1$ in the
non-thermal spectral index has only 3~per cent influence on the magnetic field
strength. Consequently, the magnetic field energy density has an uncertainty of
$25$~per cent, which is then also the uncertainty of the CRe loss term $b(E)$. In what follows, we do not
take this systematic error into account and only derive statistical
least-squares fitting errors. But for deriving absolute values of the
advection speed and diffusion coefficient, one would have to add it to the
error budget.

\begin{figure*}
 \includegraphics[width=0.8\hsize]{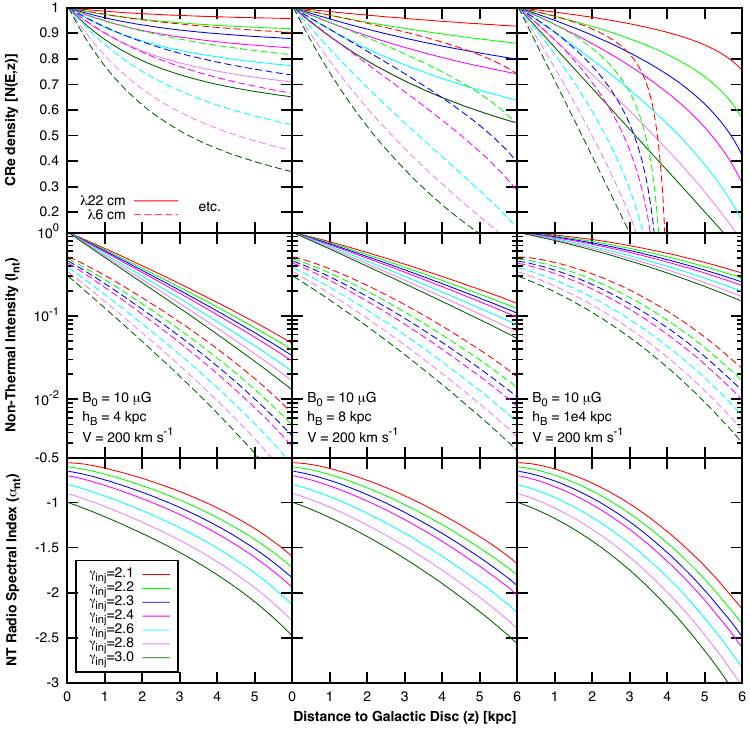}
\caption{Family of one-zone {\scriptsize SPINNAKER} advection models with various CRe
  energy injection indices ($\gamma_{\rm inj}=2.1\ldots 3.0$, note that the
  order of the lines in the plots and in the legend are identical). The advection speed is $V=200~\rm
  km\,s^{-1}$, the magnetic field strength in the galactic disc is
  $B_0=10~\umu\rm G$ and the ratio of the energy densities of the interstellar
  radiation field to that of the magnetic field is $U_{\rm IRF}/U_{\rm B}=0.3$
  (constant ratio everywhere). The magnetic
field scale height was varied with $h_{\rm B}=4$~kpc (\emph{left panels}), $h_{\rm
  B}=8$~kpc (\emph{middle panels}) and $h_{\rm B}=10^4$~kpc (i.e.\ constant magnetic
field, \emph{right panels}). Vertical profiles of
the CRe number density (\emph{top panels}) and of the non-thermal radio continuum intensity
(\emph{middle panels}) are for $\lambda$22~cm (solid lines) and for
$\lambda$6~cm (dashed line), respectively. The non-thermal radio spectral index between
$\lambda\lambda$ 22 and 6~cm is shown in the \emph{bottom panels}.}
\label{fig:adv}
\end{figure*}

\begin{figure*}
  \includegraphics[width=0.8\hsize]{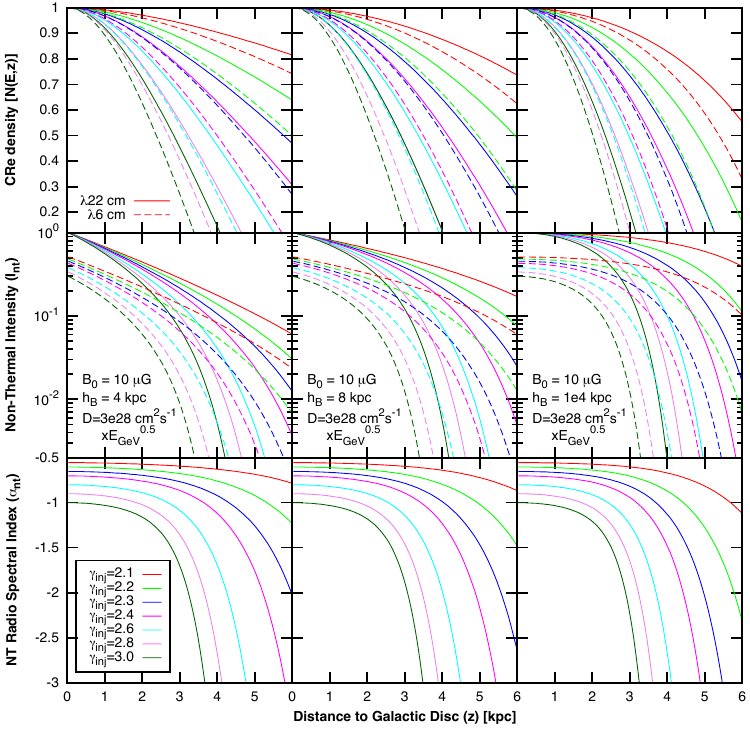}
\caption{Family of one-zone {\scriptsize SPINNAKER} diffusion models with various CRe
  energy injection indices ($\gamma_{\rm inj}=2.1\ldots 3.0$, note that the
  order of the lines in the plots and in the legend are identical) The diffusion coefficient is $D=3\times 10^{28}E_{\rm
    GeV}^{0.5}~\rm
  cm^2\,s^{-1}$. Otherwise same as Fig.~\ref{fig:adv}.}
\label{fig:diff}
\end{figure*}

\section{Cosmic ray transport models}
\label{mod:CR}
\subsection{Non-calorimetric numerical models}
The vertical profiles of the non-thermal radio continuum emission can be
modelled by solving
the diffusion--loss equation for the CRe number density $N(E)$ and
convolving it with the synchrotron emission spectrum of a single CRe \citep[e.g.][]{longair_11a}:
\begin{equation}
  \frac{{\rm d} N(E)}{{\rm d}t} = D \nabla^2 N(E) +
  \frac{\upartial}{\upartial E}\left[ b(E) N(E)\right ] + Q(E,t),
\label{eq:diffloss}
\end{equation}
where $b(E)=-{\rm d}E/{\rm d}t$ for a single CRe as given by Equation~(\ref{eq:be}). Massive
spiral galaxies have rather constant SF histories, so that the CRe injection
rate is constant and the source term $Q(E,t)$ has no explicit time dependence. We assume that all sources of CRe a located in the disc plane, neglecting a possible in-situ CRe re-acceleration in the halo, so
that for the source term we have $Q(E,t)=0$ for
$|z|>0$. Equation~(\ref{eq:diffloss}) can be numerically integrated in
time until a stationary solution is found.

We take here a slightly different
approach, first by restricting ourselves to a one-dimensional (1D) problem,
and second by imposing a fixed inner
boundary condition of $N(E,0)=N_0E^{-\gamma_{\rm inj}}$. In the stationary case, the change of
the CRe number density $\upartial N/\upartial t$ is solely determined by the
energy loss term (second term on the right hand side of
Equation~(\ref{eq:diffloss})). Noticing that for advection we
have $\upartial N / \upartial t = V\upartial N / \upartial z $, we can
re-write Equation~(\ref{eq:diffloss}) for the case of pure advection to:
\begin{equation}
  \frac{\upartial N(E,z)} {\upartial z} = \frac{1}{V}\left \lbrace
    \frac{\upartial}{\upartial E}\left [ b(E)
    N(E,z)\right ]\right \rbrace\qquad ({\rm Advection}),
\label{eq:conv}
\end{equation}
where $V$ is the advection speed, assumed here to be constant. Similarly, for diffusion we have $\upartial N/\upartial t
= D \upartial N^2/\upartial z^2$ (Fick's second law of diffusion\footnote{See
  any physics textbook on heat conduction.}), so that we can re-write
Equation~(\ref{eq:diffloss}) for the case of pure diffusion to:
\begin{equation}
  \frac{\upartial^2N(E,z)}{\upartial z^2} = \frac{1}{D}\left \lbrace\frac{\upartial}{\upartial
    E}\left [ b(E) N(E,z)\right ]\right\rbrace\qquad ({\rm Diffusion}),
\label{eq:diff}
\end{equation}
where we parametrize the diffusion coefficient as function of the CRe energy
as $D=D_0E_{\rm GeV}^{\umu}$. Usually, values for $\umu$ are thought to be between
$0.3$ and $0.6$ \citep*{strong_07a}. We assume an exponential magnetic field
distribution:
\begin{equation}
  B(z) = \left \{ \begin{array}{ll}B_0\cdot \exp ( -|z|/ h_{\rm B1})
      \qquad\qquad\qquad\quad (|z|\leq z_1)\\[5pt]
  B_0\cdot \exp (-z_1/h_{\rm B1}) \cdot \exp(-|z|/h_{\rm B2}) \quad (|z|>z_1).\end{array}
\right .
\label{eq:b_distribution}
\end{equation}
Here, $h_{\rm B1}$ and $h_{\rm B2}$ are the magnetic field scale heights in
the thin and thick disc, respectively, and $z_1$ is the transition height
between the two.
Equations~(\ref{eq:conv}) and (\ref{eq:diff}) can be
integrated numerically from the inner boundary. The CRe energy injection
spectral index can be obtained from the non-thermal radio spectral index via $\alpha_{\rm
  nt}=(1-\gamma_{\rm inj})/2$. Further details about the numerical
solution of the CRe number density and the calculation of the model synchrotron spectra are provided in
Appendix~A.

\subsection{Fitting procedure}
\begin{figure*}
  \includegraphics[width=0.9\hsize]{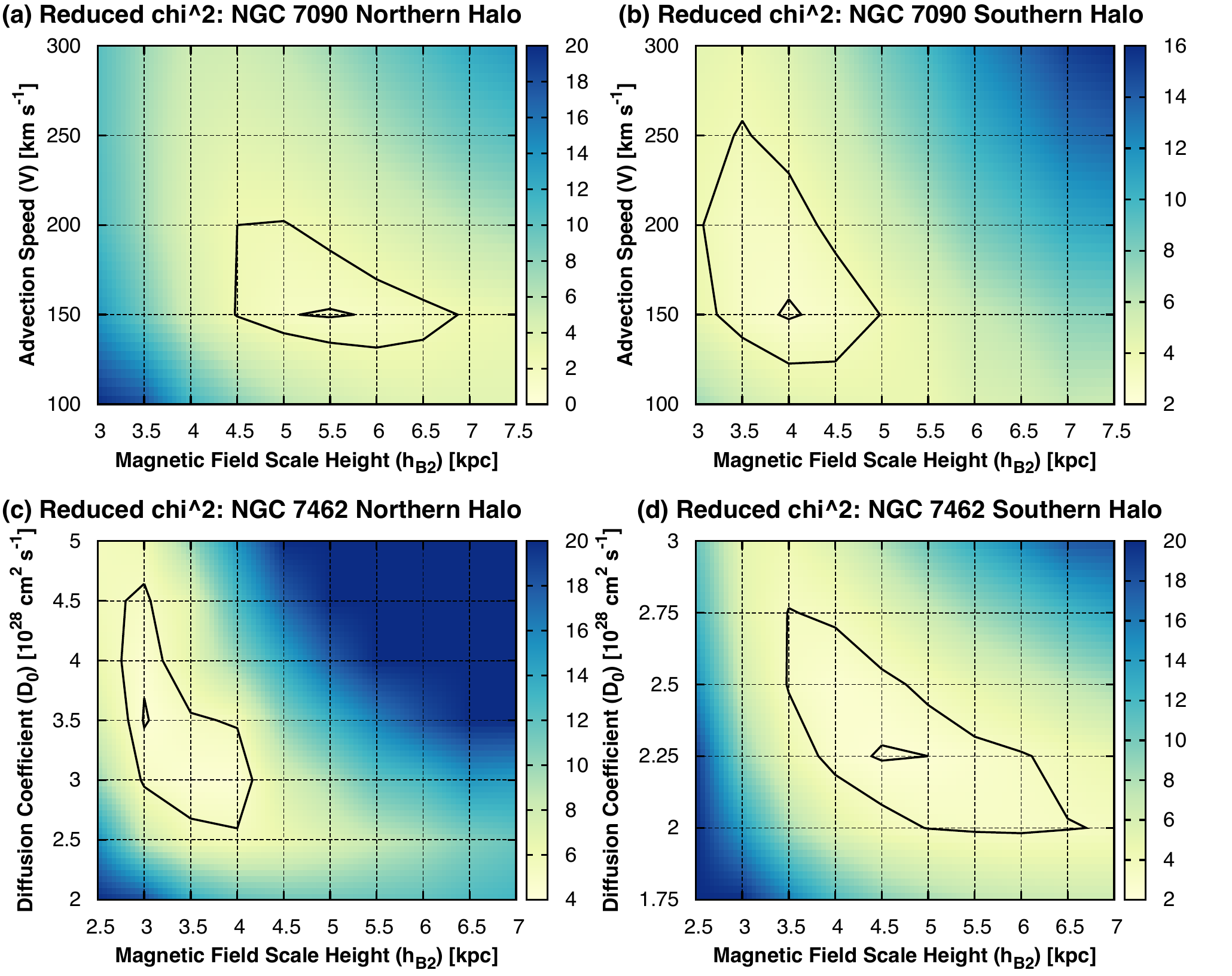}
    \caption{Reduced $\chi^2$ for the advection model in NGC~7090 (\emph{top panels})
      and the diffusion model in NGC~7462 (\emph{bottom panels}). Contours
      indicate the best-fitting reduced $\chi_{\rm min}^2$ (inner contours) and $\chi_{\rm
        min}^2+1$ (outer contours). Values in NGC~7462 were clipped at
      $\chi^2=20$.}
    \label{fig:chi}
\end{figure*}

We use the following procedure to fit cosmic ray transport models of either pure diffusion or pure advection to our data: first, we fix the electron energy spectral index in the
disc plane, to match the observed non-thermal radio spectral index. Also, we use a
representative magnetic field strength in the disc of $B_0=10~\umu\rm G$ for both galaxies,
as averaged in the area, where we have measured non-thermal spectral index
profiles (Sect.~\ref{alpha}). IC
radiation losses are calculated based on the IRF photon energy density
(Sect.~\ref{urad}). Second, we
compute profiles of the CRe number density at $\lambda\lambda$ 22
and 6~cm, to obtain model profiles of both the non-thermal radio spectral
index and the non-thermal radio
continuum emission. We then fit
the radio intensity profiles by varying both the magnetic field scale
height and the advection speed (or diffusion coefficient), resulting in the reduced $\chi^2$:
\begin{equation}
  \chi^2 = \frac{1}{N-n} \sum \left ( \frac{I_i -
      M_i}{\sigma_i} \right ).
\end{equation}
Here, $I_i$ is the $i$th non-thermal intensity measurement, $M_i$ the
corresponding model value, $\sigma_i$ the error of the measured intensity, $N$
the number of data points and $n$ the number of fitted parameters. We use
$n=3$, because we fit also for the intensity normalization. The reduced
$\chi^2$ is shown in
Fig.~\ref{fig:chi} with our best-fitting models
presented in Fig.~\ref{fig:alpha} (see Table~\ref{tab:model} for the best-fitting
parameters).

\subsection{Calorimetric analytical models}
We can compare our results with the `calorimetric' advection and diffusion
models employed by \citet{heesen_09a}, who assumed:
\begin{equation}
  h_{\rm e} = \left \{ \begin{array}{ll}V \cdot t_{\rm rad} \qquad ({\rm
        Advection}),\\[5pt]
  \sqrt{D\cdot t_{\rm rad}} \quad ({\rm Diffusion}).\end{array}
\right .
\label{eq:n253}
\end{equation}
This are meaningful only if the electron radiation losses are dominating the
vertical decrease in the non-thermal radio continuum intensity, an assumption
that appeared to be justified in the starburst galaxy NGC~253, where above relations
provided good fits to the data \citep{heesen_09a}. This is the case the magnetic field
strength is constant in the halo and the galaxy constitutes and electron
calorimeter, hence we refer to these models as the calorimetric ones. It is
also noted that \citet{heesen_09a} calculated $h_{\rm e}$ assuming energy equipartition,
which may not hold in the halo, where the CRe either escape almost
freely or are losing the bulk of their energy via synchrotron and IC
radiation. Furthermore, the relation for diffusion neglects any possible
energy dependence of the diffusion coefficient. It is one of the scopes of this work, to test
whether the calorimetric models can be applied to normal star forming galaxies, where the
outflow speeds may be sufficiently high, so that CRe radiation losses do not
dominate in the halo.

\section{Cosmic ray transport}
\label{res:CRt}
In this section, we apply our 1D cosmic ray transport models created
  with {\small SPINNAKER} to the radio
continuum observations of NGC~7090 and 7462 (Sects~\ref{res:n7090} and
\ref{res:n7462}).\footnote{{\scriptsize
  SPINNAKER},  SPectral INdex
Numerical Analysis of K(c)osmic-ray Electron Radio-emission, is a computer
program created by us and will be made publicly available at a later date.} This is preceded
by Sect.~\ref{res:mod}, where we present some general results of our
modelling that can help to decide whether diffusion or advection is dominating. In Sect.~\ref{res:ht} we investigate the scale height--CRe lifetime
relation, an analysis equivalent to that of \citet{heesen_09a}.

\subsection{The distinction between advection and diffusion}
\label{res:mod}
In Figs~\ref{fig:adv} and \ref{fig:diff} we present families of one-zone advection and
diffusion models created with our numerical cosmic ray transport models (Sect.~\ref{mod:CR}). We use a typical magnetic field strength of $B_0=10~\umu\rm G$
and a ratio of the interstellar radiation field energy density to magnetic
field energy density of $U_{\rm IRF}/U_{\rm B}=0.3$, assumed to be constant
everywhere. We varied the magnetic field scale height, with one set of models
using $h_{\rm B}=4$~kpc, the second one $h_{\rm B}=8$~kpc and the third one
$h_{\rm B}=10^4$~kpc (constant magnetic field). We varied the CRe injection
spectral index between $\gamma_{\rm inj}=2.1\ldots 3.0$. The CRe number
density was calculated for $\lambda\lambda$ 22 and 6~cm as well as the resulting
non-thermal radio continuum intensities and corresponding non-thermal radio
spectral indices. For the advection models we used a constant advection speed
of $V=200~\rm km\,s^{-1}$, whereas for the diffusion models we used $D=3\times
10^{28}E_{\rm GeV}^{0.5}~\rm cm^2\,s^{-1}$.
\begin{table}
\centering
\caption{Model parameters in the northern (N) and
  southern (S) haloes.\label{tab:crtrans}}
\begin{tabular}{lcc}
\hline\hline
Parameter & NGC~7090 (N/S) & NGC~7462 (N/S)\\\hline
$B_0$ (fixed) [$\umu$G] & 10 &  10\\
$z_1$ (fixed) [kpc] & 1 & $-$\\
$h_{\rm B1}$ (fixed) [kpc] & $0.8$ & $-$\\
$\gamma_{\rm inj}$ (fixed) & $3.0$ & $3.2$\\
$U_{\rm IRF}/U_{\rm B}$ (fixed) & 0.31 & 0.18\\
$V$ (var.) [$\rm km\,s^{-1}$] & $150^{+50}_{-20}/150^{+100}_{-30}$ & $-$\\
$D_0$ (var.) [$\rm cm^2\,s^{-1}$] & $-$ & $(3.5^{+1.2}_{-0.8}/2.25^{+0.5}_{-0.2}) 10^{28}E_{\rm GeV}^{0.5}$ \\
$h_{\rm B2}$ (varied) [kpc] & $5.5^{+1.3}_{-1.0}/ 4.0^{+1.0}_{-0.9}$ & $3.0^{+1.2}_{-0.25}/4.5^{+2.2}_{-1.0}$ \\
$\chi^2$ (minimum) & $1.5$/$2.6$ & $4.3$/$2.4$\\
\hline
\end{tabular}
\label{tab:model}
\end{table}

\emph{Cosmic ray electron number density}. It is instructive to first
  analyse how the vertical profile the CRe number density evolves for either a
  pure advection or diffusion model. If the advection speed is
sufficiently high (or the magnetic field scale height sufficiently small), the CRe number density decreases monotonically, but with a decreasing
slope, so that it almost approaches an
asymptotic value (Fig.~\ref{fig:adv}, top panels). In contrast,
diffusion will always lead to a sharp cut-off in the CRe number density
(Fig.~\ref{fig:diff}, top panels).

\emph{Non-thermal radio continuum emission profiles.} We found that advection results in approximately exponential profiles (Fig.~\ref{fig:adv},
middle panels), whereas diffusion results in profiles, which are reminiscent of Gaussian
functions (Fig.~\ref{fig:diff}, middle panels). The exact shape of
the profile depends obviously on the combination of the magnetic field scale
height and either diffusion coefficient or advection speed. If CRe radiation losses
are unimportant, we expect exponential profiles caused by the decrease of the
magnetic field strength only. If CRe radiation losses are important, both
diffusion and advection will have cut-offs in their profiles, where the
intensity decreases rapidly. If radiation losses are important, but the
decrease of the magnetic field plays a role as well, we expect to see a mix of
the profiles of the CRe number density and the magnetic field. In general it can be said, that advection
will result in profiles that are to a good approximation exponential, although
at large $z$ it is steeper than an exponential profile. Diffusion leads to profiles that are exponential at
small $z$, where the decrease of the magnetic field dominates. At large $z$, a
strong steepening is observed due to large CRe radiation losses. Hence, a
diffusion profile is steeper both at small and large $z$ than a Gaussian
profile would predict.

\emph{Non-thermal radio spectral index profiles.} For
advection, the non-thermal radio spectral index steepens rather gradually with increasing
distance from the disc (Fig.~\ref{fig:adv}, bottom panels). The profile can be
approximated by a linear function, with the curvature only becoming more
prominent if the magnetic field is constant. For diffusion, the profile of the
non-thermal radio spectral index has a characteristic shape, where the slope is very flat
in the inner part and progressively steepens dramatically in the outer parts
(Fig.~\ref{fig:diff}, bottom panels). We have chosen $\umu=0.5$; a higher energy dependence 
of the diffusion coefficient ($0.5< \umu\leq 0.7$), leads to less steepening, contrary, a
lower energy dependence ($0\leq \umu<0.5$) steepens the spectral index.

\emph{Non-thermal radio continuum scale heights.} One important result of our
modelling is that it is always possible to find a good fit to the profile of the 
non-thermal spectral index, choosing an appropriate combination of
advection speed (or diffusion coefficient) and magnetic field scale height. A
large magnetic field scale height can be balanced by either a large advection speed
or diffusion coefficient. But this has an effect on the non-thermal radio
continuum scale heights, leading to higher values as well. The combination of spectral index profiles and
radio continuum scale heights constrains both
parameters fairly well.

\subsection{Cosmic ray transport in NGC~7090}
\label{res:n7090}
\subsubsection{Advection or diffusion?}
NGC~7090 has a thin and a thick radio disc (Sect.~\ref{sh}), so that we modelled the magnetic
field accordingly with a two-component exponential profile (with the thin disc
at $|z|\leq 1$~kpc). As pointed out in
Sect.~\ref{sh}, we can not distinguish between an approximately exponential
or Gaussian profile, so that  we modelled the thin/thick
disc with an advection/advection, advection/diffusion, diffusion/advection and
diffusion/diffusion model in order to cover all possible combinations. The vertical
profiles of the radio spectral index shown in Fig.~\ref{fig:alpha}(b) display
a characteristic steepening in the inner parts at $|z|\leq 2$~kpc and a possible flattening
at larger heights, which has to be explained by the modelling. We can rule out
a diffusion/diffusion model ($\chi^2=8$, both haloes), where the steep spectral
index in the thin disc at $|z|\leq 1$~kpc
requires a small diffusion coefficient of  $D\approx 0.5\times
10^{28}E_{\rm GeV}^{0.5}~\rm cm^2\,s^{-1}$, which results in no significant emission in the
thick disc regardless of the diffusion coefficient chosen there. This can be
understood in such a way that the diffusion coefficient
prescribes the curvature rather than the slope of $N(E,z)$. Hence, even
increasing the diffusion coefficient in the thick disc to large values
(a few $10^{29}E_{\rm GeV}^{0.5}~\rm cm^2\,s^{-1}$), will
still result in too low emission levels. Notably an
advection/diffusion model works, but it is unphysical, because we would expect a
galactic wind or outflow to become more efficient further away from the
disc. 

Hence, advection has to
dominate in the thick disc, which leaves the advection/advection and diffusion/advection models that
describe the data equally well ($\chi^2=1.3$--$2.3$). The thin disc can be fitted by $h_{\rm
  B1}=0.8\pm 0.1$~kpc with either $D=0.5^{+0.2}_{-0.1}\times 10^{28}E_{\rm GeV}^{0.5}~\rm cm^2\,s^{-1}$ or $V=150\pm 50~\rm km\,s^{-1}$. In the thick disc, the minimum advection speed
is $100~\rm km\,s^{-1}$ with the upper limit unconstrained, since the radio
spectral index does not steepen in the halo (Fig.~\ref{fig:alpha}(b)). This shows that the advection speed is sufficiently
high, so that radiation losses are not dominating in the halo.

\subsubsection{Best-fitting advection model}
In order to better constrain the value of the advection speed, we fitted an advection/advection
model with equal advection speeds in the thin and thick disc. The upper limit of the
advection speed is then largely determined by
the thin disc and may be potentially higher in the thick disc. 
In Fig.~\ref{fig:chi}(a) we show the reduced
$\chi^2$ as function of advection speed and magnetic field scale height in the
thick disc. The best-fitting
parameters are an advection speed of $V=150^{+50}_{-20}~\rm km\,s^{-1}$ and a magnetic field
scale height of $h_{\rm B2}=5.5^{+1.3}_{-1.0}$~kpc in the northern halo and
$V=150^{+100}_{-30}~\rm km\,s^{-1}$ and $h_{\rm B2}=4.0^{+1.0}_{-0.9}$~kpc in the
southern halo. The parameters used in the fitting procedure are summarized in
Table~\ref{tab:crtrans}. It is instructive to compare the magnetic field scale
heights with the values that we would obtain from equipartition. In the
northern halo we would expect a magnetic field scale height of $6.8$~kpc
(assuming a non-thermal spectral index of $\alpha_{\rm nt}=-1$) and in the
southern halo of $6.1$~kpc (from the $\lambda$22~cm scale heights). Hence, the
scale heights as inferred from our modelling are 20--35 per cent lower than
the equipartition values. This is suggests that the assumption of
energy equipartition breaks down in the halo.

A complication we have not yet discussed is the fact that the thin disc
consists of a mix of young and old CRe, so that the radio spectral index may
not be representative for the CRe that actually enter the outflow. It is
conceivable that the spectral index steepening within the first 2~kpc distance
from the galactic disc is largely due to the decreasing influence of the thin
disc, containing some young, freshly injected CRe. Such a scenario is
consistent with a diffusion dominated transport in the thin disc,
where the CRe population at any given position is a mix of
young and old electrons. As pointed out above, if we fit a diffusion/advection
model, the advection speed can be potentially significantly higher in the
halo. An
advection of $V=250~\rm km\,s^{-1}$, at the upper end of the error interval,
would fit well (compare Fig.~\ref{fig:alpha}(a)). The same effect can be created
by underestimating the thermal radio continuum emission due to internal
absorption of Balmer H$\alpha$ line emission by dust (Sect.~\ref{subsec:Ha}). Hence, our advection speeds
are conservative lower limits.

%
\begin{figure*}
  \includegraphics[width=0.8\hsize]{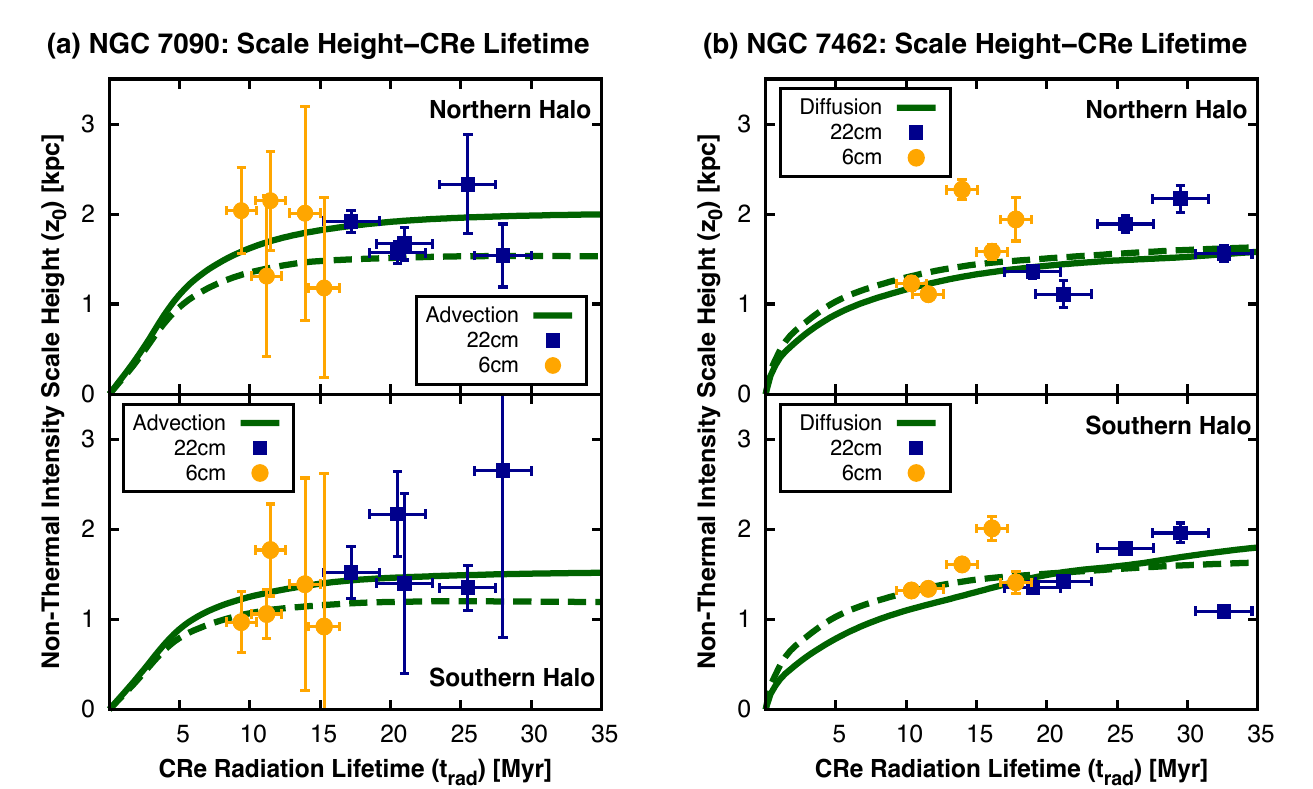}
\caption{Non-thermal radio continuum scale heights as function of the CRe
  lifetime at $\lambda\lambda$ 22~cm (dark-blue) and 6~cm (orange). \emph{ Left:} NGC~7090, where scale heights of the thick exponential
  disc are plotted. \emph{Right:} NGC~7462, where scale heights of
  the Gaussian disc are plotted. The green, solid (dashed) dark-green line shows the predicted
  scale heights from the best-fitting transport model at $\lambda$22~cm
  ($\lambda6~\rm cm)$.}
\label{fig:t_syn}
\end{figure*}

\subsection{Diffusive cosmic ray transport in NGC~7462} 
\label{res:n7462}
In NGC~7462, we did not find a thin disc (Sect.~\ref{sh}), so that we modelled the magnetic
field with a one-component exponential function describing the thick disc
only. The
Gaussian
vertical radio continuum profiles suggest diffusion as the dominating
transport mode (Fig.~\ref{fig:alpha}(c)). The vertical spectral
index profile corroborates shown in Fig.~\ref{fig:alpha}(d) this scenario, with a flat spectral index within
$|z|\leq  2$~kpc and a significant steepening for heights $|z|>2$~kpc; the spectral
index profile has a parabolic shape. In the southern halo, diffusion ($\chi^2=2.4$) describes
the data significantly better than advection ($\chi^2=12$). In the
northern halo, diffusion ($\chi^2=4.3$) describes the data 
marginally worse than advection ($\chi^2=3.9$); this is largely due
to the extended emission that can be seen at $\lambda$22~cm at $z>2$~kpc
(Fig.~\ref{fig:sh}(b) at offsets $-2.4$ and 0~kpc). We believe that diffusion
is the more likely transport mode, because a `one-sided' galactic wind appears
to be an unlikely scenario. Also, advection does not reproduce the parabolic
spectral index shape, but leads to a linear steepening of the spectral index,
in contrast to what is seen in the observations. For completeness, we present
here the best-fitting advection 
parameters in the northern halo as $V=200^{+200}_{-30}~\rm km\,s^{-1}$ and $h_{\rm B2}=3.0^{+0.25}_{-1.2}$~kpc.

In
Fig.~\ref{fig:chi}(b), the reduced $\chi^2$ of the spectral index
fitting as function of the diffusion coefficient and magnetic field
scale height is presented. Our best-fitting
parameters are a diffusion coefficient of $3.5^{+1.2}_{-0.8}$ and $2.25^{+0.5}_{-0.2}\times 10^{28}E_{\rm GeV}^{0.5}~{\rm
cm^2\,s^{-1}}$ and a magnetic field scale height of $3.0^{+1.2}_{-0.25}$ and
$4.5^{+2.2}_{-1.0}$~kpc in the northern and southern halo, respectively. We note that the average CRe energy within our wavelength range is
$4.2$~GeV, so that the actual diffusion coefficient is a factor of $\approx$2
higher than $D_0$. We have varied the energy dependence of the diffusion coefficient
$\umu$ and found that for $\umu=0.3\ldots 0.7$, the diffusion coefficient
(normalized to $E=4.2$~GeV)  changes by less than 10 per cent and the magnetic
field scale height changes by less than 17 per cent.\footnote{The best-fitting
  parameters (N/S) for $\umu=0.3$ are $h_{\rm B}=3.5/4.0$~kpc, $D=4.25/3.25\times
  10^{28}E_{\rm GeV}^{0.3}~\rm cm^2\,s^{-1}$ and for $\umu=0.7$ are $h_{\rm B}=3.0/4.0$~kpc, $D=2.5/1.75\times
  10^{28}E_{\rm GeV}^{0.7}~\rm cm^2\,s^{-1}$.}
The parameters used in the fitting procedure are summarized in
Table~\ref{tab:crtrans}. The diffusion coefficients we find are in good
agreement what is reported in the literature for diffusion \emph{within} the
disc plane
\citep[e.g.][]{murphy_08a,fletcher_11a,heesen_11a,buffie_13a,berkhuijsen_13a,tabatabaei_13b,mulcahy_14a}. This
means that NGC~7462 is an example for a galaxy, where cosmic rays are just
diffusing from the disc into the halo and advection plays no role.

Finally, as noted earlier (Sect.~\ref{sh}), it may be possible that the
radio continuum profiles are exponential, if the baselevel of our map is
negative by $\approx$$-1\sigma_{\rm TP}$. However, the profiles of the
non-thermal radio spectral index are still in agreement with diffusion
only. Our model has to explain why the spectral index remains constant
  within $|z|\leq 1.1$~kpc, where $S/N>10$, so that the baselevel uncertainty
  does not play a role. Advection only leads to parabolic spectral index profiles, if the
magnetic field scale height is large ($\gtrsim$10~kpc). In this case, the
resulting radio continuum scale height is, however, much larger than the measured
$0.9$~kpc for an exponential profile. The only way out is that our non-thermal
radio spectral index profiles are biased to too steep spectral indices within
$|z|<2$~kpc, for instance by subtracting a too high thermal fraction of radio
continuum emission due to our uncertain calibration procedure
(Sect.~\ref{subsec:Ha}); this is corroborated by the local minimum of the
non-thermal spectral index in the disc plane. On the other hand, we expect the
Balmer H$\alpha$ emission to underestimate the thermal radio continuum,
because we did not take absorption due to dust into account. Thus, we conclude
that the measurement of a steep spectral index in the disc plane is robust and,
consequently, cosmic ray diffusion is the dominating transport mode in NGC~7462.

\subsection{Scale height--lifetime relation}
\label{res:ht}
Our cosmic ray transport models (Sect.~\ref{mod:CR}) are fitted to the data averaged over
a wide range of galactocentric radii, so that we are not sensitive to variations of
the advection speed or diffusion coefficient as function of position. Ideally,
we would like to apply our models locally (kpc scale), but the low S/N of the
halo emission prevents us from
doing so. What we can do instead, is to analyse the dependence of the local scale heights derived in
the stripes of Sect.~\ref{sh}) ($\approx$2~kpc width), as function of
the CRe lifetime; so we can check whether they are consistent with our cosmic
ray transport models.

The scale heights of both galaxies, presented in
Fig.~\ref{fig:t_syn}, do not show any clear
dependence of the CRe lifetime. The average Pearson's rank order correlation coefficient is with $r_{\rm s}=0.23$ so low
($r_{\rm s}=0\ldots 1$, for a positive correlation),
that we conclude that no linear trend is observed. Our models (shown as green
lines) reproduce this (non-existing trend) well: for CRe lifetimes in excess
of 15~Myr the scale heights are almost constant, since the decrease of the
magnetic field dominates the decrease of the non-thermal intensity
(non-calorimetric halo). Only for lifetimes smaller than 10~Myr we would
expect a decrease of the scale heights, but since we are not probing this
regime we can not verify this particular prediction. The model reproduces the scale heights of
NGC~7090 well within the uncertainties, although it fails to explain the
variation of the local scale heights in NGC~7462. In the latter case, the
error bars are small because the
single Gaussian fits are well constrained by data points within $|z|<3$~kpc,
which have a high S/N (Sect.~\ref{sh}).

As a further consistency check, we fit our calorimetric transport models from Equation~(\ref{eq:n253}),
where we utilise the usual assumption of
energy equipartition for which the CRe scale height is:
\begin{equation}
  h_{\rm e} = \frac{3-\alpha_{\rm nt}}{2}z_0,
\label{electron_scaleheight}
\end{equation}
where we assume $\alpha_{\rm nt}\approx -1$ and where $z_0$ is the non-thermal
radio scale height of the thick disc (Sect.~\ref{sh}). In
NGC~7090, we
find advection speeds of $186$ and $144~\rm km\,s^{-1}$ in the northern
and southern halo, respectively. This is in good agreement with our
models. However, we point out that the magnetic field strength in the halo ($|z|>1$~kpc)
is lower, $B\leq 3~\rm\umu$G, than what we used for our calorimetric
approximation ($B=10~\umu$G, $B$=const.). Therefore,
the actual CRe lifetime is a factor of five longer
(Equation~(\ref{eq:t_rad})), so that CRe scale height must be a factor of
five higher as well. The agreement between calorimetric and non-calorimetric
model is in the case of NGC~7090 thus coincidental. Similarly, in NGC~7462 we find a calorimetric diffusion coefficient of
$1.4\times 10^{29}~\rm cm^2\,s^{-1}$ both in the northern and southern halo of
NGC~7462 (no energy dependence, $\umu=0$). This is a factor of two higher than
the value derived by our models. The reason is again that we have
overestimated $B$; at 2~kpc height, the average height where we
determined the scale height, the magnetic field strength is only 60 per cent of the one in the disc, so
that the CRe lifetime increases by a factor of two. 

In summary, we find that the scale heights measured in stripes
are broadly consistent with a constant diffusion coefficient and advection
speed across the width of the galactic disc, as assumed for our modelling. But the numerical values obtained from the
calorimetric approximation can be used only if (i) the CRe scale heights
estimated from equipartition are accurate and (ii) the average halo magnetic
field strength can be estimated.

\begin{figure*}
  \includegraphics[width=0.8\hsize]{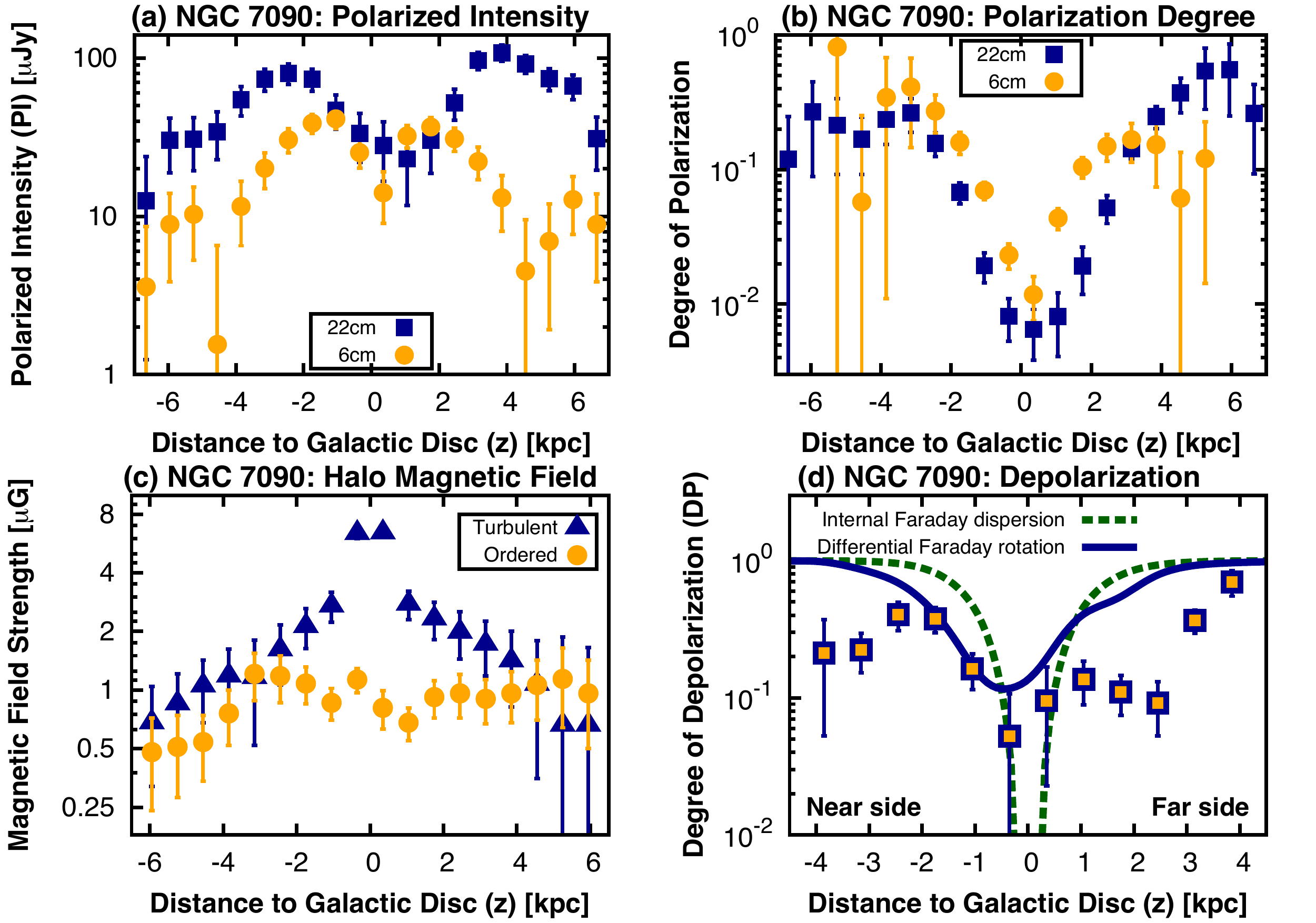}
    \caption{Vertical profiles used for the analysis of the linearly polarized
      intensity in NGC~7090, averaged between
offsets of $\pm 4.25$~kpc (panels a,b,c) and $\pm 0.75$~kpc (d) along the major axis. Negative values of $z$ are for the southern halo, and
  positive ones are for the northern halo. \emph{Top left:} polarized intensity at $\lambda\lambda$ 22 (dark-blue) and 6~cm
      (orange). \emph{Top right:} Degree of
    polarization at $\lambda\lambda$ 22 (dark-blue) and 6~cm (orange). \emph{Bottom
      left:} magnetic field strength of the turbulent and ordered magnetic field in the halo. \emph{Bottom right:} Degree of depolarization between $\lambda\lambda$ 22 and 6~cm. The dark-green line shows a model profile
of internal Faraday dispersion by turbulence in the magneto-ionized ISM and
the dark-blue line shows a model for Faraday dispersion by the regular halo magnetic
field (see text for details).}
    \label{fig:n7090_pi}
\end{figure*}

\section{Halo magnetic field}
\label{res:mag}
\subsection{Morphology}
In Figs~\ref{fig:n7090} and \ref{fig:n7462} (panels (c) and (d)) we present the distribution of the
linearly polarized intensity in NGC~7090 and 7462 at $\lambda\lambda$ 22 and
6~cm, respectively. In a magnetized plasma a polarized wave vector is rotated by Faraday
rotation. For a uniform magnetic field and (thermal) electron density, we define the rotation measure (RM) via:
\begin{eqnarray}
  \frac{\Delta \phi} {\rm rad} & = &0.808 \times \left ( \frac{\lambda}{\rm
      m}\right )^2 {\int}_0^{\frac{L}{\rm pc}}\left ( \frac{B_{\rm
          reg\parallel}}{\umu \rm G} \right ) \left ( \frac{n_{\rm e}}{\rm
        cm^{-3}}\right ) {\rm d}\left ( \frac{l}{\rm pc}\right )\\\nonumber
& = & 0.808\lambda^2 {\rm RM},
  \label{eq:rm}
\end{eqnarray}
where $n_{\rm e}$ is the thermal electron density, $B_{\rm reg\parallel}$ is the
LoS component of the regular magnetic field and $L$ is the
integration length through the magneto-ionized medium. The maximum RM value we
measure between $\lambda\lambda$ 22 and 6~cm is $\pm 40~\rm
rad\,m^{-2}$ (Figs \ref{fig:n7090}(d) and \ref{fig:n7462}(d)), which is not enough to correct for Faraday rotation in an
edge-on galaxy; the galaxies are partially depolarized at $\lambda$22~cm, so
that one can not measure the RM reliably. Thus, the magnetic
field orientations presented in this work are not corrected for Faraday
rotation. A maximum RM amplitude of $40~\rm
rad\,m^{-2}$  requires differential Faraday
rotation \citep{sokoloff_98a}, either by an ordered disc or halo magnetic
field, with a coherent direction over kpc-scales, also referred to as
`regular' magnetic field.
 
We find that
NGC~7090 features a prominent polarized radio halo, particularly at
$\lambda$22~cm (Fig.~\ref{fig:n7090}(c)),
where we can detect polarized emission up to heights of 6~kpc in the
northern halo and 4~kpc in the southern halo. The polarized emission at
$\lambda$22~cm is asymmetric with respect to the minor axis, where the bulk of
polarized emission is located on the approaching side of the galaxy
\citep[see][for an H\,{\small I} rotation curve of the
galaxy]{dahlem_05a}. This is in agreement with findings of \citet*{braun_10a}, who postulated that this
kind of
asymmetry can be explained by depolarization due to differential Faraday rotation caused by a
magnetic field structure, which consists of a disc magnetic field and a
quadrupolar halo magnetic field. But their model is not applicable in our
case, since it works only for galaxies at moderate inclination angles from
20--$70\degr$. For an alternative explanation we note that the western side of
the galaxy has a higher star formation rate, as
indicated by the higher radio continuum intensity (Figs~\ref{fig:n7090}(a) and
(b)). The asymmetry of polarized emission with respect to the minor axis may
thus be a consequence of the asymmetric S/N. In Figs
\ref{fig:n7090}(c) and (d) we can see evidence of depolarization by the
combined effects of the turbulent magnetic field in the disc and by
differential Faraday rotation due to the ordered (more precisely regular)
magnetic field in the disc and halo, both at $\lambda\lambda$ 22 and 6~cm; but
no direct evidence of an ordered disc magnetic field itself is seen, probably due to
depolarization as well, in part also caused by depolarization due to unresolved vertical
$E$-vectors within the beam area.

In NGC~7462, the bulk of the polarized emission at $\lambda$22~cm
(Fig.~\ref{fig:n7462}(c)) is again found
on the approaching side of the galaxy, concentrated in a filamentary region of
10~kpc length and $2.6$~kpc width. It is probably related to the very extended
extra-planar component only detected at $\lambda$22~cm (Sect.~\ref{sh}); this
would explain why almost none of the polarized emission is seen at
$\lambda$6~cm (Fig.~\ref{fig:n7462}(d)). The integrated \HI map of \citet{dahlem_05a} shows that the \HI
distribution in the eastern half of the galaxy is thicker than in the western
half. Hence, we may speculate that we are seeing the remnants of an outflow,
which stirred up the \HI disc and where the CRe are still visible as
aged radio continuum emission with a steep spectral index. Another prominent feature is a `spur' of
polarized emission centred on RA $23^{\rm h}02^{\rm m}47^{\rm s}$ Dec.\
$-40\degr 50\arcmin 00\arcsec$. It may be connected to the extended emission
which we masked (Sect.~\ref{sec:meth}), although it does not align exactly
with it and is more reminiscent of a jet-driven lobe, emerging from the
galaxy's centre. As mentioned earlier, we did not find any unresolved source
in the centre of the galaxy, hinting at past or current AGN activity. In any
case, the emission is probably not related to any galactic outflow, which
NGC~7462 probably does currently not possess anyway, being diffusion
dominated. Hence, we continue our analysis with NGC~7090 only.

\subsection{Magnetic field strengths and depolarization}
We can now measure the strength of the ordered halo magnetic
field. To this end, we investigate vertical profiles of the linearly
  polarized emission and degree of polarization as presented in
  Fig.~\ref{fig:n7090_pi}(a) and (b) (see Appendix~\ref{data_tables} for the
  tabulated values). The degree of
polarization $p$ relates to the ratio $q=B_{\rm turb}/B_{\rm ord,\perp}$ of the isotropic turbulent magnetic field
$B_{\rm turb}$ to
the ordered magnetic field in the plane of the sky $B_{\rm ord,\perp}$ via:
\begin{equation}
  \frac{p}{p_0} =  \frac{1}{1+q^2}, \qquad p_0 =
  \frac{3-3\alpha_{\rm nt}}{5-3\alpha_{\rm nt}},
\end{equation}
where $p_0$ is the theoretically maximum polarization degree for a purely
uniform magnetic field. In above relation we used that the polarized (total) synchrotron intensity scales with $PI\propto B_{\rm ord,
  \perp}^2$ ($I_{\rm nt}\propto B^2$) for $\alpha_{\rm nt}=-1$, so that
$p/p_0=B_{\rm ord, \perp}^2/B^2$. We now use the assumption that the ordered
magnetic field is in the plane of the sky, $B_{\rm ord}=B_{\rm ord,\perp}$,
so that we can calculate the ordered field strength as $B_{\rm ord}=B
(1+q^2)^{-1/2}$ and the turbulent magnetic field strength as $B_{\rm turb}=(B^2-B_{\rm ord}^2)^{1/2}$. We use the
$\lambda$22~cm degree of polarization at $|z|>3.5$~kpc and the $\lambda$6~cm
degree of polarization elsewhere. This is because the $\lambda$22~cm
observations are more severely affected by depolarization, but extend away
further from the galactic midplane. The vertical profiles of the turbulent
and ordered magnetic field strength are shown in Fig.~\ref{fig:n7090_pi}(c). The ordered magnetic field strength ranges largely between
$0.5$ and
$1~\rm\umu G$, whereas the turbulent magnetic field strength decreases
monotonically from $10~\umu\rm G$ at $z=0$ to $1~\umu\rm G$ at $z=5$~kpc.

For further analysis, we introduce 
the degree of depolarization (DP), defined as the ratio of the polarization degrees
at two wavelengths, which is usually calculated using the ratio of the
(linearly) polarized intensities and the non-thermal radio spectral index:
\begin{equation}
  {\rm DP} \equiv \frac{PI_{22}}{PI_{6}}\times \left ( \frac{\nu_6}{\nu_{22}}\right  )^{\alpha_{\rm nt}},
\end{equation}
where $PI_{22}$ and $PI_6$ are the polarized intensities at
$\lambda\lambda$ 22 and 6~cm, respectively, and $\nu_{22}=1.4$ and
$\nu_6=4.7$~GHz are the corresponding frequencies. We found a pronounced minimum of the degree of depolarization surrounding the
galactic midplane of NGC~7090, as shown in Fig.~\ref{fig:n7090_pi}(d). There is a steep rise in the
degree of depolarization in the southern halo ($-3\leq z\leq 0~\rm
kpc$). For
illustration, we compare it with a depolarization model for Faraday dispersion by turbulence in the
magneto-ionized ISM \citep{burn_66a,sokoloff_98a}:
\begin{equation}
  \mathcal{P}_{\rm disp}  = p_{\rm i} \frac{1-\exp(-2\sigma_{\rm RM}^2\lambda^4Ld^{-1})} {2\sigma_{\rm RM}^2\lambda^4Ld^{-1}}.
\end{equation}
Here, $\sigma_{\rm RM}(z)=0.808\times n_{\rm e}(z)\times d\times
B_{\rm turb,\parallel}(z)$, where $d$ is
the size of the turbulent eddies in pc, $B_{\rm
  turb,\parallel}$, the LoS component of the turbulent magnetic
field in $\umu\rm G$, $L$ the length of the LoS in pc and $n_{\rm e}$
the thermal electron density in $\rm cm^{-3}$. The intrinsic degree of
polarization $p_{\rm i}$ can be eliminated by using two wavelengths and
forming the ratio to measure the degree of depolarization. We have assumed an electron
density of $0.02~\rm cm^{-3}$ and a
scale height of $1.5$~kpc, in loose agreement with values for the Milky Way
\citep{ferriere_01a}. We used 100~pc as the size of the
turbulent eddy in the disc plane \citep[see][for a measurement in an
external galaxy]{fletcher_11a}. We find that Faraday dispersion explains
our data partially, although the model over-predicts the depolarization in the
disc and under-predicts it in the halo. One would have to increase the
electron scale height to values of 4~kpc in order to get agreement with the
data in the southern halo. Another source of depolarization is differential Faraday
rotation due to a regular disc magnetic field with a coherent direction, for which the model is:
\begin{equation}
  \mathcal{P}_{\rm diff} = p_i \frac{\sin(\mathcal{R}\lambda^2)}{\mathcal{R}\lambda^2},
\end{equation}
where $\mathcal{R}=0.808\times n_{\rm e}(z)\times d\times B_{\rm
  reg,\parallel}(z)$ and $B_{\rm
  reg,\parallel}(z)$ is the regular magnetic field component along the
LoS (which we assumed to be equal to the ordered field strength). We find that this model can explain the depolarization in the
southern halo ($-2\leq z \leq -1$~kpc) better. Where our depolarization
models all fail is north of the
galactic midplane ($0\leq z\leq 2$~kpc), where the degree of
depolarization is low and rises steeply
further away from the disc. A possible explanation is that the turbulent magnetic field in the disc depolarizes the northern
halo, which hence is on the \emph{far
  side}, whereas the southern halo has to be on the \emph{near side} (we have overlaid in Fig.~\ref{fig:n7090}(c)  the optical extent of the
disc for comparison). However, it is difficult to come to any firm
  conclusion without a more comprehensive modelling of the magnetic field
  structure, such as the three-dimensional magnetic field models employed by
  \citet{heesen_09b, heesen_11a}.
 
The magnetic field in NGC~7090 appears to be dominated by the halo
  component (Fig.~\ref{fig:n7090}(d)), although shorter wavelengths would be required to exclude
  depolarization of a disc magnetic field. This suggests a relation to a galactic wind or outflow. A galactic wind facilitates an
efficient halo dynamo \citep{moss_10a}; an outflow stretches magnetic field lines into Parker-type loops \citep{mao_15a}. The former process results in regular magnetic fields, the
latter in anisotropic ones. Observations at shorter wavelengths are
needed to measure Faraday rotation and distinguish
between the two scenarios.



\section{Discussion}
\label{dis}
The outflow of cosmic rays has important ramifications for the observed relation between the
star formation rate (SFR) and the radio continuum (RC) luminosity, in the following RC--SFR
relation \citep{condon_92a, heesen_14a}. The widely used semi-empirical `Condon relation' assumes that galaxies are electron calorimeters. This would
imply that the CRe lose all their energy exclusively within the galaxy, or,
more precisely, we are observing always the \emph{same fraction} of possible
non-thermal synchrotron emission. This was already postulated by
\citet{voelk_89a}, who predicted that the radio continuum emission of a galaxy
should, in case of a CRe calorimeter, only be a function of the ratio of
IC to synchrotron losses. But observational tests are at odds with the
calorimetric theory. The non-thermal radio spectral index would have to be
quite steep, $\approx$$-1.2$ \citep{lisenfeld_00a}, which does not agree with
the observed spectral index, where studies find spectral indices of $-0.83$
\citep*{niklas_97b}, $-0.8$ \citep*{marvil_15a}. Furthermore, \citet{heesen_14a} found that in spiral
galaxies the average ratio of IC to synchrotron losses is only 30~per cent,
which means that we cannot easily test the prediction by
\citet{voelk_89a}. The alternative are non-calorimetric models, which assume
(i) energy equipartition between the cosmic rays and the magnetic field,  and
(ii), a relation between either the magnetic field strength and gas density
\citep{niklas_97a}, or a relation between the magnetic field strength and star
formation rate \citep{heesen_14a}. 

Non-electron calorimetry can be caused, if the advection speed is
sufficiently high, so that the CRe lifetime is longer than the CRe escape
time: $t_{\rm rad}\gg t_{\rm esc}$. The escape time is here defined as the $t_{\rm esc}=h_{\rm B}/(2V)$,
where $h_{\rm B}/2$ is the scale height of the magnetic field energy
density, assuming that the non-thermal intensity scales with the
magnetic field as $I_{\rm nt}\propto B^2$ (exact for $\alpha_{\rm nt}=-1$). We
now define the equivalent advection speed, where the CRe
lifetime (Equation~(\ref{eq:t_rad})) is equal to the CRe escape time:  

\begin{equation}
  V_{\rm equiv}\equiv \frac{h_{\rm B}}{2\,t_{\rm rad}} = 200~{\rm km\,s^{-1}}~
  \left ( \frac{h_{\rm B}}{\rm 4~kpc}\right ) \left ( \frac{t_{\rm rad}}{\rm
      10~Myr}\right )^{-1}.
\label{eq:v_esc}
\end{equation}
For example, in the northern halo of NGC~7090 ($B=3.3~\umu\rm G$ at $z=1~\rm kpc$,
the base of the halo), the CRe lifetimes are
$110$ and $60$~Myr at $\lambda\lambda$ 22 and 6~cm, so that the equivalent
advection speeds are 25 and 50~km\,s$^{-1}$, respectively. At higher wind speeds, the CRe are transported faster along the gradient of the
magnetic field and stellar photon energy density than they lose their energy via synchrotron and IC radiation. They are able to retain a
non-vanishing fraction of their initial energy and the galaxy is not an
electron calorimeter. 
\begin{figure}
 \includegraphics[width=0.99\hsize]{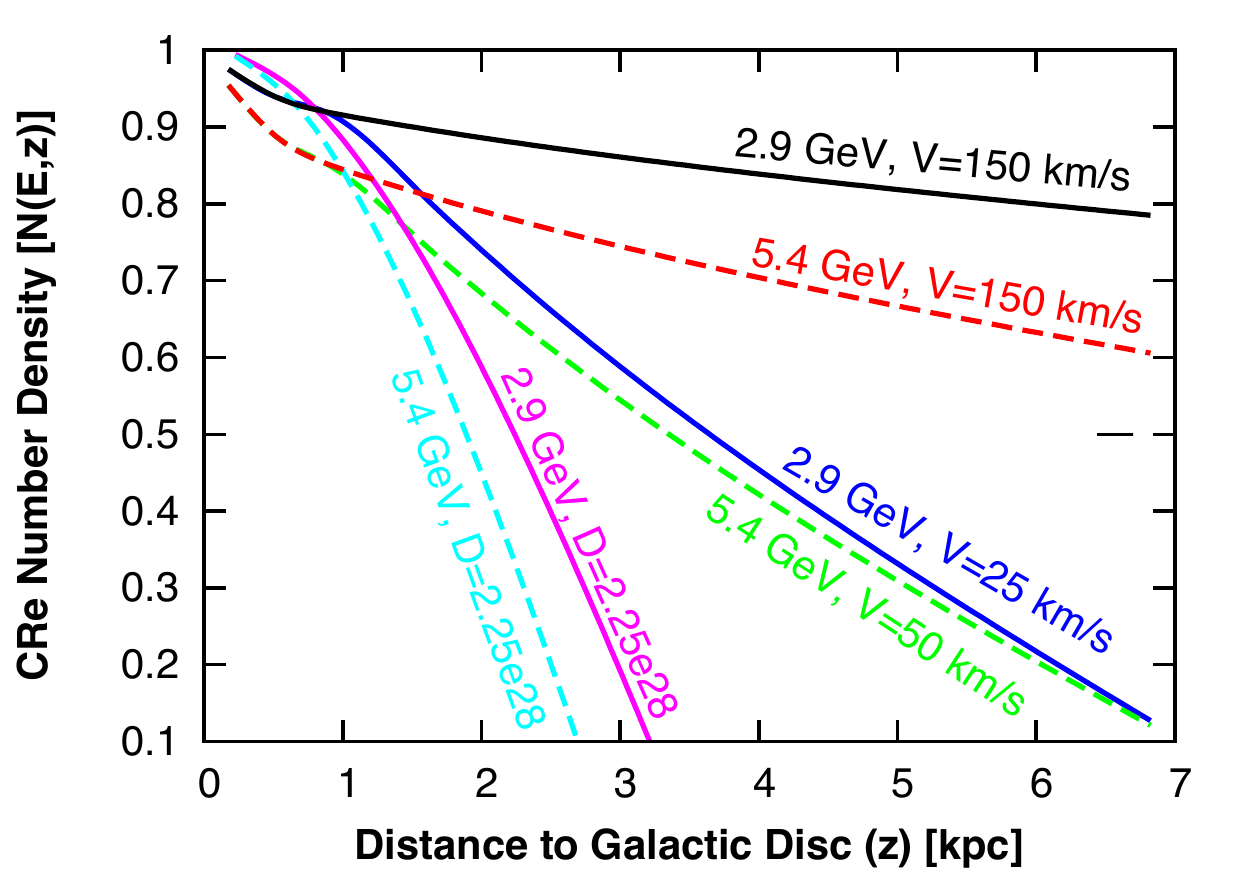}
    \caption{Profiles of the CRe number density at
      $\lambda\lambda$ 22 and 6~cm (solid and dashed lines), equivalent to CRe energies of $2.9$ and
      $5.4$~GeV, respectively, for the northern halo of NGC~7090 (advection)
      and for the southern halo of NGC~7462 (diffusion).}
    \label{fig:calor}
\end{figure}

Profiles of the CRe electron energy
distributions for the northern NGC~7090 are shown in
Fig.~\ref{fig:calor}.\footnote{The CRe energies quoted here are computed
  according to their critical frequency (Equation~(\ref{eq:crit})) using $B_0$, the magnetic field
  strength in the disc. For a fixed observing frequency the CRe in the halo
  have a higher energy, because the critical frequency decreases.} For
our best-fitting model,
at the detection limit of the halo, $z=7$~kpc, the CRe number density has
dropped by less than 40 per cent, so that the CRe retain at least 60 per cent of their energy at 7~kpc
height. In contrast, if the advection speed is dropped to the critical speed
(at $z>1$~kpc), the CRe number density drops by 90 per cent. At $\lambda$22~cm, equivalent to a CRe
energy of $2.9$~GeV, the CRe number density is at least 70 per cent at 7~kpc,
the limit to which we can observe the halo. This means that we potentially
lose up to 70 per cent of the potential radio luminosity, but we should keep
in mind that the CRe number density refers to those electrons that are able to
leave the thin radio disc. The thin radio disc has a non-thermal radio
spectral index of $-1.0$, so that the electron have already lost some of their
energy, before they are able to enter the outflow. In the
thin radio disc, one probably has a mix of young, freshly injected CRe, and
older ones, which are enclosed in the SNe heated superbubbles, before they are able to
break out from the disc \citep{heesen_15a}. In contrast, diffusion leads
always to calorimetric haloes (for usual values of diffusion coefficient and
magnetic field scale height) as the CRe number density profiles of NGC~7462
show.

Do we trace a galactic wind in NGC~7090? The minimum advective transport
speed of $150~\rm km\,s^{-1}$ is similar to the escape velocity of $V_{\rm
  esc}=\sqrt{2}V_{\rm rot}$ of $180~\rm km\,s^{-1}$. Also, we have to keep in mind
that the advection speed is the sum of the wind speed and the Alfv\'en speed,
as cosmic rays can stream along the vertical magnetic field lines. If the
magnetic field is immersed in an outflow of hot X-ray emitting gas, typical
electron densities are 4--$6\times 10^{-3}~\rm cm^{-3}$ with scale heights of
3--7~kpc \citep{hodges-kluck_13a}. With an ordered magnetic field of
$1~\umu\rm G$ in the halo, the Alfv\'en speed $V_{\rm A}=B/\sqrt{4\upi \rho}$
is 30~km\,s$^{-1}$. It remains thus difficult to exactly ascertain the actual
wind speed, with $V_{\rm
  w}=120~\rm km\,s^{-1}$ our best-fitting value (for the better fitted
northern halo), but with a large error
interval of 120--170~km\,s$^{-1}$, stemming from the uncertainty of the
advection speed. Also, as pointed out earlier, the advection
speeds may be significantly higher in the halo, where our data provides us
only with a lower limit (Sect.~\ref{res:n7090}).

\section{Conclusions}
\label{conc}
We have used ATCA radio continuum polarimetry observations at $\lambda\lambda$
22 and 6~cm, in order to search for non-thermal radio haloes with
relativistic CRe and magnetic fields in two late-type edge-on spiral galaxies. We have measured non-thermal radio
continuum scale heights and profiles of the vertical non-thermal radio spectral
index distribution. We modelled our measurements by solving equations for
stationary 1D transport of cosmic rays, estimating the CRe synchrotron and IC
radiation losses from equipartition magnetic field strengths and a combination of
IR, optical and CMB radiation energy densities. This allowed us to measure
advection speeds, diffusion coefficients and magnetic field scale heights
without having to use the assumption of \emph{local}  energy equipartition in the halo. These are our conclusions:

\begin{enumerate}
\item NGC~7090 has a prominent radio halo with polarized emission detected to heights
  of up to 6~kpc. Vertical radio continuum profiles can be described by 
  two-component exponential functions, with scale heights at $\lambda 22$~cm of $0.38\pm 0.06$~kpc and
  $1.7\pm 0.2$~kpc of the thin and thick disc, respectively. NGC~7462 has a radio halo, very different from most other known
  cases. The vertical radio continuum profiles can be best approximated by 
  one-component Gaussian functions, hence this galaxy lacks a thin disc
  component. The Gaussian profiles mean that the emission decreases rapidly
  with height, so that it is debatable, whether this galaxy possesses a
  `bona-fide' radio
  halo. We find a scale height
  of $1.47\pm 0.06$~kpc at $\lambda$22~cm.
\item Advection dominated cosmic ray transport leads to approximately exponential vertical radio
  continuum profiles, while diffusion dominated transport leads to profiles, which can
  be better approximated by a Gaussian function. The non-thermal radio
  spectral profiles are very different as well: advective transport leads to
  a gradual steepening of the radio spectral index, whereas in case of
  diffusion the spectral index steepens hardly within one scale height and very
  rapidly at larger heights, so that profiles are of `parabolic' shape.
\item The cosmic ray transport in NGC~7090 can be approximated by a two-zone
  model, with the thin disc at $|z|\leq 1$~kpc and the thick disc at
  $|z|>1$~kpc. The thin disc can be modelled equally well with either a pure
  diffusion or advection model with $h_{\rm B}=0.8\pm 0.1$~kpc and either
  $D=0.5^{+0.2}_{-0.1}\times 10^{28}E^{0.5}_{\rm GeV}~\rm cm^2\,s^{-1}$ or
  $V=150\pm 50~\rm km\,s^{-1}$. The thick disc can only be described with
  advection, with a minimum advection speed of $100~\rm km\,s^{-1}$. Assuming a pure
  advection model in both the thin and thick disc, we find a best-fitting advection speed of $150^{+50}_{-20}/150^{+100}_{-30}~\rm
  km\,s^{-1}$ and a magnetic field scale height of
  $5.5^{+1.3}_{-1.0}/4.0^{1.0}_{-0.9}$~kpc in the northern and southern halo, respectively.
\item The cosmic ray transport in NGC~7462 can be
  approximated by a pure diffusion model, with a diffusion coefficient of
  $3.5^{+1.2}_{-0.8}/2.25^{+0.5}_{-0.2}~\times 10^{28}\,E_{\rm GeV}^{0.5}~{\rm cm\,s^{-2}}$ and a magnetic field
  scale height of $3.0^{+1.2}_{-0.25}/4.5^{+2.2}_{-1.0}$~kpc in the northern
  and southern halo, respectively. 
\item The magnetic field scale heights both in NGC~7090 and 7462 are 25--35
  per cent lower than the value suggested by energy equipartition. This
  suggest that energy equipartition breaks down in the halo.
\item If the advection speed is sufficiently
  high, the CRe radiation loss time can become longer than the CRe escape
  time. For the
  specific case of NGC~7090, CRe at $\lambda$22~cm ($2.9$~GeV)
  retain 70~per cent of their energy (after they left the thin disc) at $|z|=7$~kpc, the height to which we can
  detect the halo. NGC~7090 is thus no electron calorimeter.
\end{enumerate}

Our findings can explain why globally averaged scale heights of galaxies are
surprisingly constant, with a mean of $1.8$~kpc at $\lambda 6$~cm, independent of the SFR and 
magnetic field strength \citep{krause_11a}. Galaxies may have magnetic field scale heights of
4--6~kpc, with only little variation. Increasing the outflow speed as
expected for higher star formation rates would then only slightly change the radio continuum
scale heights. However, the sample of galaxies with well studied
radio haloes is still small (approximately ten). The constant global scale heights
can alternatively be explained by increased transport speeds at higher SFRs,
for instance by higher wind speeds \citep{arribas_14a}, or due to faster streaming
of cosmic rays along the magnetic field lines. Hence, it would be desirable to
apply our cosmic ray transport modelling to a larger sample of
radio haloes, such as
the CHANG-ES survey \citep{irwin_12a,wiegert_15a}, or archive data from
the various radio interferometers (ATCA, VLA, WSRT). Another important test would be the extension to
low-frequency observations, as now made possible by LOFAR
\citep{vanHaarlem_13a}. This would allow us to measure stringent upper limits for the magnetic field scale heights as we are
able to probe the oldest CRe, so far undetected at GHz-frequencies.

\section*{Acknowledegements}
We would like to thank Michael Dahlem for initializing this
project and for his warm welcome at the ATCA in Narrabri. VH and RJD also wish to thank
ATCA staff for their hospitality during their visit. Urvashi Rau and George
Moellenbrock from the NRAO are thanked for their assistance to the data
reduction. We thank Sui Ann Mao from the MPIfR for carefully reading the
manuscript. We are grateful to an anonymous referee for an insightful report. VH
acknowledges support from the Science and Technology Facilities Council (STFC)
under grant ST/J001600/1. RB and RJD
are supported by the Deutsche Forschungsgemeinschaft (DFG) through Research
Unit FOR~1254. This research has made use of the NASA/IPAC
Extragalactic Database (NED) which is operated by the Jet Propulsion
Laboratory, California Institute of Technology, under contract with the
National Aeronautics and Space Administration. The Digitized Sky Surveys were produced at the Space Telescope Science Institute under U.S.\ Government grant NAG W-2166.

\bibliographystyle{mnras}
{\small \bibliography{sh}}

\begin{thebibliography}{}
\makeatletter
\relax
\def\mn@urlcharsother{\let\do\@makeother \do\$\do\&\do\#\do\^\do\_\do\%\do\~}
\def\mn@doi{\begingroup\mn@urlcharsother \@ifnextchar [ {\mn@doi@}
  {\mn@doi@[]}}
\def\mn@doi@[#1]#2{\def\@tempa{#1}\ifx\@tempa\@empty \href
  {http://dx.doi.org/#2} {doi:#2}\else \href {http://dx.doi.org/#2} {#1}\fi
  \endgroup}
\def\mn@eprint#1#2{\mn@eprint@#1:#2::\@nil}
\def\mn@eprint@arXiv#1{\href {http://arxiv.org/abs/#1} {{\tt arXiv:#1}}}
\def\mn@eprint@dblp#1{\href {http://dblp.uni-trier.de/rec/bibtex/#1.xml}
  {dblp:#1}}
\def\mn@eprint@#1:#2:#3:#4\@nil{\def\@tempa {#1}\def\@tempb {#2}\def\@tempc
  {#3}\ifx \@tempc \@empty \let \@tempc \@tempb \let \@tempb \@tempa \fi \ifx
  \@tempb \@empty \def\@tempb {arXiv}\fi \@ifundefined
  {mn@eprint@\@tempb}{\@tempb:\@tempc}{\expandafter \expandafter \csname
  mn@eprint@\@tempb\endcsname \expandafter{\@tempc}}}

\bibitem[\protect\citeauthoryear{{Adebahr}, {Krause}, {Klein}, {We{\.z}gowiec},
  {Bomans}  \& {Dettmar}}{{Adebahr} et~al.}{2013}]{adebahr_13a}
{Adebahr} B.,  {Krause} M.,  {Klein} U.,  {We{\.z}gowiec} M.,  {Bomans} D.~J.,
   {Dettmar} R.-J.,  2013, \mn@doi [\aap] {10.1051/0004-6361/201220226}, \href
  {http://esoads.eso.org/abs/2013A%26A...555A..23A} {555, A23}

\bibitem[\protect\citeauthoryear{{Arribas}, {Colina}, {Bellocchi}, {Maiolino}
  \& {Villar-Mart{\'{\i}}n}}{{Arribas} et~al.}{2014}]{arribas_14a}
{Arribas} S.,  {Colina} L.,  {Bellocchi} E.,  {Maiolino} R.,
  {Villar-Mart{\'{\i}}n} M.,  2014, \mn@doi [\aap]
  {10.1051/0004-6361/201323324}, \href
  {http://esoads.eso.org/abs/2014A%26A...568A..14A} {568, A14}

\bibitem[\protect\citeauthoryear{{Baars}, {Genzel}, {Pauliny-Toth}  \&
  {Witzel}}{{Baars} et~al.}{1977}]{baars_77a}
{Baars} J.~W.~M.,  {Genzel} R.,  {Pauliny-Toth} I.~I.~K.,   {Witzel} A.,  1977,
  \aap, \href {http://adsabs.harvard.edu/abs/1977A%26A....61...99B} {61, 99}

\bibitem[\protect\citeauthoryear{{Beck} \& {Krause}}{{Beck} \&
  {Krause}}{2005}]{beck_05a}
{Beck} R.,  {Krause} M.,  2005, \mn@doi [Astron.\ Nachr.]
  {10.1002/asna.200510366}, \href
  {http://adsabs.harvard.edu/abs/2005AN....326..414B} {326, 414}

\bibitem[\protect\citeauthoryear{{Berkhuijsen}, {Beck}  \&
  {Tabatabaei}}{{Berkhuijsen} et~al.}{2013}]{berkhuijsen_13a}
{Berkhuijsen} E.~M.,  {Beck} R.,   {Tabatabaei} F.~S.,  2013, \mn@doi [\mnras]
  {10.1093/mnras/stt1400}, \href
  {http://esoads.eso.org/abs/2013MNRAS.435.1598B} {435, 1598}

\bibitem[\protect\citeauthoryear{{Boettcher}, {Zweibel}, {Yoast-Hull}  \&
  {Gallagher}}{{Boettcher} et~al.}{2013}]{boettcher_13a}
{Boettcher} E.,  {Zweibel} E.~G.,  {Yoast-Hull} T.~M.,   {Gallagher} III J.~S.,
   2013, \mn@doi [\apj] {10.1088/0004-637X/779/1/12}, \href
  {http://esoads.eso.org/abs/2013ApJ...779...12B} {779, 12}

\bibitem[\protect\citeauthoryear{{Braun}, {Heald}  \& {Beck}}{{Braun}
  et~al.}{2010}]{braun_10a}
{Braun} R.,  {Heald} G.,   {Beck} R.,  2010, \mn@doi [\aap]
  {10.1051/0004-6361/200913375}, \href
  {http://adsabs.harvard.edu/abs/2010A%26A...514A..42B} {514, A42}

\bibitem[\protect\citeauthoryear{{Bregman}}{{Bregman}}{1980}]{bregman_80a}
{Bregman} J.~N.,  1980, \mn@doi [\apj] {10.1086/157776}, \href
  {http://esoads.eso.org/abs/1980ApJ...236..577B} {236, 577}

\bibitem[\protect\citeauthoryear{{Breitschwerdt}, {McKenzie}  \&
  {V{\"o}lk}}{{Breitschwerdt} et~al.}{1991}]{breitschwerdt_91a}
{Breitschwerdt} D.,  {McKenzie} J.~F.,   {V{\"o}lk} H.~J.,  1991, \aap, \href
  {http://adsabs.harvard.edu/abs/1991A%26A...245...79B} {245, 79}

\bibitem[\protect\citeauthoryear{{Breitschwerdt}, {McKenzie}  \&
  {V{\"o}lk}}{{Breitschwerdt} et~al.}{1993}]{breitschwerdt_93a}
{Breitschwerdt} D.,  {McKenzie} J.~F.,   {V{\"o}lk} H.~J.,  1993, \aap, \href
  {http://adsabs.harvard.edu/abs/1993A%26A...269...54B} {269, 54}

\bibitem[\protect\citeauthoryear{{Buffie}, {Heesen}  \& {Shalchi}}{{Buffie}
  et~al.}{2013}]{buffie_13a}
{Buffie} K.,  {Heesen} V.,   {Shalchi} A.,  2013, \mn@doi [\apj]
  {10.1088/0004-637X/764/1/37}, \href
  {http://esoads.eso.org/abs/2013ApJ...764...37B} {764, 37}

\bibitem[\protect\citeauthoryear{{Burn}}{{Burn}}{1966}]{burn_66a}
{Burn} B.~J.,  1966, \mnras, \href
  {http://adsabs.harvard.edu/abs/1966MNRAS.133...67B} {133, 67}

\bibitem[\protect\citeauthoryear{{Condon}}{{Condon}}{1992}]{condon_92a}
{Condon} J.~J.,  1992, \mn@doi [\araa] {10.1146/annurev.aa.30.090192.003043},
  \href {http://adsabs.harvard.edu/abs/1992ARA%26A..30..575C} {30, 575}

\bibitem[\protect\citeauthoryear{{Dahlem}, {Petr}, {Lehnert}, {Heckman}  \&
  {Ehle}}{{Dahlem} et~al.}{1997}]{dahlem_97a}
{Dahlem} M.,  {Petr} M.~G.,  {Lehnert} M.~D.,  {Heckman} T.~M.,   {Ehle} M.,
  1997, \aap, \href {http://esoads.eso.org/abs/1997A%26A...320..731D} {320,
  731}

\bibitem[\protect\citeauthoryear{{Dahlem}, {Ehle}  \& {Ryder}}{{Dahlem}
  et~al.}{2001}]{dahlem_01a}
{Dahlem} M.,  {Ehle} M.,   {Ryder} S.~D.,  2001, \mn@doi [\aap]
  {10.1051/0004-6361:20010614}, \href
  {http://adsabs.harvard.edu/abs/2001A%26A...373..485D} {373, 485}

\bibitem[\protect\citeauthoryear{{Dahlem}, {Ehle}, {Ryder}, {Vlaji{\'c}}  \&
  {Haynes}}{{Dahlem} et~al.}{2005}]{dahlem_05a}
{Dahlem} M.,  {Ehle} M.,  {Ryder} S.~D.,  {Vlaji{\'c}} M.,   {Haynes} R.~F.,
  2005, \mn@doi [\aap] {10.1051/0004-6361:20041671}, \href
  {http://adsabs.harvard.edu/abs/2005A%26A...432..475D} {432, 475}

\bibitem[\protect\citeauthoryear{{Dahlem}, {Lisenfeld}  \& {Rossa}}{{Dahlem}
  et~al.}{2006}]{dahlem_06a}
{Dahlem} M.,  {Lisenfeld} U.,   {Rossa} J.,  2006, \mn@doi [\aap]
  {10.1051/0004-6361:20054787}, \href
  {http://adsabs.harvard.edu/abs/2006A%26A...457..121D} {457, 121}

\bibitem[\protect\citeauthoryear{{Dale} \& {Helou}}{{Dale} \&
  {Helou}}{2002}]{dale_02a}
{Dale} D.~A.,  {Helou} G.,  2002, \mn@doi [\apj] {10.1086/341632}, \href
  {http://esoads.eso.org/abs/2002ApJ...576..159D} {576, 159}

\bibitem[\protect\citeauthoryear{{Dale} et~al.,}{{Dale}
  et~al.}{2009}]{dale_09a}
{Dale} D.~A.,  et~al., 2009, \mn@doi [\apj] {10.1088/0004-637X/703/1/517},
  \href {http://adsabs.harvard.edu/abs/2009ApJ...703..517D} {703, 517}

\bibitem[\protect\citeauthoryear{{Dalla Vecchia} \& {Schaye}}{{Dalla Vecchia}
  \& {Schaye}}{2008}]{dalla_vecchia_08a}
{Dalla Vecchia} C.,  {Schaye} J.,  2008, \mn@doi [\mnras]
  {10.1111/j.1365-2966.2008.13322.x}, \href
  {http://esoads.eso.org/abs/2008MNRAS.387.1431D} {387, 1431}

\bibitem[\protect\citeauthoryear{{Deeg}, {Duric}  \& {Brinks}}{{Deeg}
  et~al.}{1997}]{deeg_97a}
{Deeg} H.-J.,  {Duric} N.,   {Brinks} E.,  1997, \aap, \href
  {http://esoads.eso.org/abs/1997A%26A...323..323D} {323, 323}

\bibitem[\protect\citeauthoryear{{Dettmar}}{{Dettmar}}{1992}]{dettmar_92a}
{Dettmar} R.-J.,  1992, Fundamentals Cosmic Phys., \href
  {http://adsabs.harvard.edu/abs/1992FCPh...15..143D} {15, 143}

\bibitem[\protect\citeauthoryear{{Dorfi} \& {Breitschwerdt}}{{Dorfi} \&
  {Breitschwerdt}}{2012}]{dorfi_12a}
{Dorfi} E.~A.,  {Breitschwerdt} D.,  2012, \mn@doi [\aap]
  {10.1051/0004-6361/201118082}, \href
  {http://esoads.eso.org/abs/2012A%26A...540A..77D} {540, A77}

\bibitem[\protect\citeauthoryear{{Doyle} et~al.,}{{Doyle}
  et~al.}{2005}]{doyle_05a}
{Doyle} M.~T.,  et~al., 2005, \mn@doi [\mnras]
  {10.1111/j.1365-2966.2005.09159.x}, \href
  {http://esoads.eso.org/abs/2005MNRAS.361...34D} {361, 34}

\bibitem[\protect\citeauthoryear{{Draine}}{{Draine}}{2011}]{draine_11a}
{Draine} B.~T.,  2011, {Physics of the Interstellar and Intergalactic Medium}.
Princeton University Press, Princeton, NJ

\bibitem[\protect\citeauthoryear{{Dumke}, {Krause}, {Wielebinski}  \&
  {Klein}}{{Dumke} et~al.}{1995}]{dumke_95a}
{Dumke} M.,  {Krause} M.,  {Wielebinski} R.,   {Klein} U.,  1995, \aap, \href
  {http://adsabs.harvard.edu/abs/1995A%26A...302..691D} {302, 691}

\bibitem[\protect\citeauthoryear{{Ekers} \& {Sancisi}}{{Ekers} \&
  {Sancisi}}{1977}]{ekers_77a}
{Ekers} R.~D.,  {Sancisi} R.,  1977, \aap, \href
  {http://esoads.eso.org/abs/1977A%26A....54..973E} {54, 973}

\bibitem[\protect\citeauthoryear{{Everett}, {Zweibel}, {Benjamin}, {McCammon},
  {Rocks}  \& {Gallagher}}{{Everett} et~al.}{2008}]{everett_08a}
{Everett} J.~E.,  {Zweibel} E.~G.,  {Benjamin} R.~A.,  {McCammon} D.,  {Rocks}
  L.,   {Gallagher} III J.~S.,  2008, \mn@doi [\apj] {10.1086/524766}, \href
  {http://adsabs.harvard.edu/abs/2008ApJ...674..258E} {674, 258}

\bibitem[\protect\citeauthoryear{{Ferri{\`e}re}}{{Ferri{\`e}re}}{2001}]{ferriere_01a}
{Ferri{\`e}re} K.~M.,  2001, \mn@doi [Reviews of Modern Physics]
  {10.1103/RevModPhys.73.1031}, \href
  {http://esoads.eso.org/abs/2001RvMP...73.1031F} {73, 1031}

\bibitem[\protect\citeauthoryear{{Fletcher}, {Beck}, {Shukurov}, {Berkhuijsen}
  \& {Horellou}}{{Fletcher} et~al.}{2011}]{fletcher_11a}
{Fletcher} A.,  {Beck} R.,  {Shukurov} A.,  {Berkhuijsen} E.~M.,   {Horellou}
  C.,  2011, \mn@doi [\mnras] {10.1111/j.1365-2966.2010.18065.x}, \href
  {http://esoads.eso.org/abs/2011MNRAS.412.2396F} {412, 2396}

\bibitem[\protect\citeauthoryear{{Hanasz}, {W{\'o}lta{\'n}ski}  \&
  {Kowalik}}{{Hanasz} et~al.}{2009}]{hanasz_09a}
{Hanasz} M.,  {W{\'o}lta{\'n}ski} D.,   {Kowalik} K.,  2009, \mn@doi [\apjl]
  {10.1088/0004-637X/706/1/L155}, \href
  {http://esoads.eso.org/abs/2009ApJ...706L.155H} {706, L155}

\bibitem[\protect\citeauthoryear{{Heesen}, {Beck}, {Krause}  \&
  {Dettmar}}{{Heesen} et~al.}{2009a}]{heesen_09a}
{Heesen} V.,  {Beck} R.,  {Krause} M.,   {Dettmar} R.-J.,  2009a, \mn@doi
  [\aap] {10.1051/0004-6361:200810543}, \href
  {http://adsabs.harvard.edu/abs/2009A%26A...494..563H} {494, 563}

\bibitem[\protect\citeauthoryear{{Heesen}, {Krause}, {Beck}  \&
  {Dettmar}}{{Heesen} et~al.}{2009b}]{heesen_09b}
{Heesen} V.,  {Krause} M.,  {Beck} R.,   {Dettmar} R.-J.,  2009b, \mn@doi
  [\aap] {10.1051/0004-6361/200911698}, \href
  {http://adsabs.harvard.edu/abs/2009A%26A...506.1123H} {506, 1123}

\bibitem[\protect\citeauthoryear{{Heesen}, {Beck}, {Krause}  \&
  {Dettmar}}{{Heesen} et~al.}{2011}]{heesen_11a}
{Heesen} V.,  {Beck} R.,  {Krause} M.,   {Dettmar} R.-J.,  2011, \mn@doi [\aap]
  {10.1051/0004-6361/201117618}, \href
  {http://esoads.eso.org/abs/2011A%26A...535A..79H} {535, A79}

\bibitem[\protect\citeauthoryear{{Heesen}, {Brinks}, {Leroy}, {Heald}, {Braun},
  {Bigiel}  \& {Beck}}{{Heesen} et~al.}{2014}]{heesen_14a}
{Heesen} V.,  {Brinks} E.,  {Leroy} A.~K.,  {Heald} G.,  {Braun} R.,  {Bigiel}
  F.,   {Beck} R.,  2014, \mn@doi [\aj] {10.1088/0004-6256/147/5/103}, \href
  {http://esoads.eso.org/abs/2014AJ....147..103H} {147, 103}

\bibitem[\protect\citeauthoryear{{Heesen} et~al.,}{{Heesen}
  et~al.}{2015}]{heesen_15a}
{Heesen} V.,  et~al., 2015, \mn@doi [\mnras] {10.1093/mnrasl/slu168}, \href
  {http://esoads.eso.org/abs/2015MNRAS.447L...1H} {447, L1}

\bibitem[\protect\citeauthoryear{{Hodges-Kluck} \& {Bregman}}{{Hodges-Kluck} \&
  {Bregman}}{2013}]{hodges-kluck_13a}
{Hodges-Kluck} E.~J.,  {Bregman} J.~N.,  2013, \mn@doi [\apj]
  {10.1088/0004-637X/762/1/12}, \href
  {http://esoads.eso.org/abs/2013ApJ...762...12H} {762, 12}

\bibitem[\protect\citeauthoryear{{Hodges-Kluck} \& {Bregman}}{{Hodges-Kluck} \&
  {Bregman}}{2014}]{hodges-kluck_14a}
{Hodges-Kluck} E.~J.,  {Bregman} J.~N.,  2014, \mn@doi [\apj]
  {10.1088/0004-637X/789/2/131}, \href
  {http://esoads.eso.org/abs/2014ApJ...789..131H} {789, 131}

\bibitem[\protect\citeauthoryear{{Hughes}}{{Hughes}}{1991}]{hughes_91a}
{Hughes} P.~A.,  1991, {Beams and Jets in Astrophysics}.
Cambridge University Press, Cambridge, UK

\bibitem[\protect\citeauthoryear{{Hughes} et~al.,}{{Hughes}
  et~al.}{2014}]{hughes_14a}
{Hughes} T.~M.,  et~al., 2014, \mn@doi [\aap] {10.1051/0004-6361/201323245},
  \href {http://adsabs.harvard.edu/abs/2014A%26A...565A...4H} {565, A4}

\bibitem[\protect\citeauthoryear{{Hummel}, {Lesch}, {Wielebinski}  \&
  {Schlickeiser}}{{Hummel} et~al.}{1988}]{hummel_88a}
{Hummel} E.,  {Lesch} H.,  {Wielebinski} R.,   {Schlickeiser} R.,  1988, \aap,
  \href {http://esoads.eso.org/abs/1988A%26A...197L..29H} {197, L29}

\bibitem[\protect\citeauthoryear{{Hunter} et~al.,}{{Hunter}
  et~al.}{2012}]{hunter_12a}
{Hunter} D.~A.,  et~al., 2012, \mn@doi [\aj] {10.1088/0004-6256/144/5/134},
  \href {http://esoads.eso.org/abs/2012AJ....144..134H} {144, 134}

\bibitem[\protect\citeauthoryear{{Ipavich}}{{Ipavich}}{1975}]{ipavich_75a}
{Ipavich} F.~M.,  1975, \apj, \href
  {http://adsabs.harvard.edu/abs/1975ApJ...196..107I} {196, 107}

\bibitem[\protect\citeauthoryear{{Irwin}, {English}  \& {Sorathia}}{{Irwin}
  et~al.}{1999}]{irwin_99a}
{Irwin} J.~A.,  {English} J.,   {Sorathia} B.,  1999, \mn@doi [\aj]
  {10.1086/300827}, \href {http://esoads.eso.org/abs/1999AJ....117.2102I} {117,
  2102}

\bibitem[\protect\citeauthoryear{{Irwin} et~al.,}{{Irwin}
  et~al.}{2012}]{irwin_12a}
{Irwin} J.,  et~al., 2012, \mn@doi [\aj] {10.1088/0004-6256/144/2/43}, \href
  {http://esoads.eso.org/abs/2012AJ....144...43I} {144, 43}

\bibitem[\protect\citeauthoryear{{Jaffe} \& {Perola}}{{Jaffe} \&
  {Perola}}{1973}]{jaffe_73a}
{Jaffe} W.~J.,  {Perola} G.~C.,  1973, \aap, \href
  {http://esoads.eso.org/abs/1973A%26A....26..423J} {26, 423}

\bibitem[\protect\citeauthoryear{{Karachentsev}, {Makarov}  \&
  {Kaisina}}{{Karachentsev} et~al.}{2013}]{karachentsev_13a}
{Karachentsev} I.~D.,  {Makarov} D.~I.,   {Kaisina} E.~I.,  2013, \mn@doi [\aj]
  {10.1088/0004-6256/145/4/101}, \href
  {http://adsabs.harvard.edu/abs/2013AJ....145..101K} {145, 101}

\bibitem[\protect\citeauthoryear{{Kennicutt}, {Lee}, {Funes}, {Sakai}  \&
  {Akiyama}}{{Kennicutt} et~al.}{2008}]{kennicutt_08a}
{Kennicutt} Jr. R.~C.,  {Lee} J.~C.,  {Funes} Jos{\'e}~G. S.~J.,  {Sakai} S.,
  {Akiyama} S.,  2008, \mn@doi [\apjs] {10.1086/590058}, \href
  {http://adsabs.harvard.edu/abs/2008ApJS..178..247K} {178, 247}

\bibitem[\protect\citeauthoryear{{Kepley}, {M{\"u}hle}, {Everett}, {Zweibel},
  {Wilcots}  \& {Klein}}{{Kepley} et~al.}{2010}]{kepley_10a}
{Kepley} A.~A.,  {M{\"u}hle} S.,  {Everett} J.,  {Zweibel} E.~G.,  {Wilcots}
  E.~M.,   {Klein} U.,  2010, \mn@doi [\apj] {10.1088/0004-637X/712/1/536},
  \href {http://adsabs.harvard.edu/abs/2010ApJ...712..536K} {712, 536}

\bibitem[\protect\citeauthoryear{{Krause}}{{Krause}}{2009}]{krause_09a}
{Krause} M.,  2009, Rev. Mex. Astron. Astrofis., \href
  {http://adsabs.harvard.edu/abs/2009RMxAC..36...25K} {36, 25}

\bibitem[\protect\citeauthoryear{{Krause}}{{Krause}}{2011}]{krause_11a}
{Krause} M.,  2011, in Proceedings of `Magnetic Fields in the Universe: From
  Laboratory and Stars to Primordial Structures', eds. M. Soida, K.
  Otmianowska-Mazur, E.M. de Gouveia Dal Pino \& A. Lazarian.  (\mn@eprint
  {arXiv} {1111.7081})

\bibitem[\protect\citeauthoryear{{Krause}, {Wielebinski}  \& {Dumke}}{{Krause}
  et~al.}{2006}]{krause_06a}
{Krause} M.,  {Wielebinski} R.,   {Dumke} M.,  2006, \mn@doi [\aap]
  {10.1051/0004-6361:20053789}, \href
  {http://adsabs.harvard.edu/abs/2006A%26A...448..133K} {448, 133}

\bibitem[\protect\citeauthoryear{{Kulsrud} \& {Pearce}}{{Kulsrud} \&
  {Pearce}}{1969}]{kulsrud_69a}
{Kulsrud} R.,  {Pearce} W.~P.,  1969, \mn@doi [\apj] {10.1086/149981}, \href
  {http://adsabs.harvard.edu/abs/1969ApJ...156..445K} {156, 445}

\bibitem[\protect\citeauthoryear{{Lisenfeld} \& {V{\"o}lk}}{{Lisenfeld} \&
  {V{\"o}lk}}{2000}]{lisenfeld_00a}
{Lisenfeld} U.,  {V{\"o}lk} H.~J.,  2000, \aap, \href
  {http://esoads.eso.org/abs/2000A%26A...354..423L} {354, 423}

\bibitem[\protect\citeauthoryear{{Longair}}{{Longair}}{2011}]{longair_11a}
{Longair} M.~S.,  2011, {High Energy Astrophysics}.
Cambridge University Press, Cambridge, UK

\bibitem[\protect\citeauthoryear{{Mao}, {Zweibel}, {Fletcher}, {Ott}  \&
  {Tabatabaei}}{{Mao} et~al.}{2015}]{mao_15a}
{Mao} S.~A.,  {Zweibel} E.,  {Fletcher} A.,  {Ott} J.,   {Tabatabaei} F.,
  2015, \mn@doi [\apj] {10.1088/0004-637X/800/2/92}, \href
  {http://esoads.eso.org/abs/2015ApJ...800...92M} {800, 92}

\bibitem[\protect\citeauthoryear{{Marvil}, {Owen}  \& {Eilek}}{{Marvil}
  et~al.}{2015}]{marvil_15a}
{Marvil} J.,  {Owen} F.,   {Eilek} J.,  2015, \mn@doi [\aj]
  {10.1088/0004-6256/149/1/32}, \href
  {http://esoads.eso.org/abs/2015AJ....149...32M} {149, 32}

\bibitem[\protect\citeauthoryear{{McMullin}, {Waters}, {Schiebel}, {Young}  \&
  {Golap}}{{McMullin} et~al.}{2007}]{mcmullin_07a}
{McMullin} J.~P.,  {Waters} B.,  {Schiebel} D.,  {Young} W.,   {Golap} K.,
  2007, ASP Conf. Ser., \href {http://esoads.eso.org/abs/2007ASPC..376..127M}
  {376, 127}

\bibitem[\protect\citeauthoryear{{Mora} \& {Krause}}{{Mora} \&
  {Krause}}{2013}]{mora_13a}
{Mora} S.~C.,  {Krause} M.,  2013, \mn@doi [\aap]
  {10.1051/0004-6361/201321043}, \href
  {http://esoads.eso.org/abs/2013A%26A...560A..42M} {560, A42}

\bibitem[\protect\citeauthoryear{{Moss}, {Sokoloff}, {Beck}  \&
  {Krause}}{{Moss} et~al.}{2010}]{moss_10a}
{Moss} D.,  {Sokoloff} D.,  {Beck} R.,   {Krause} M.,  2010, \mn@doi [\aap]
  {10.1051/0004-6361/200913509}, \href
  {http://esoads.eso.org/abs/2010A%26A...512A..61M} {512, A61}

\bibitem[\protect\citeauthoryear{{Mulcahy} et~al.,}{{Mulcahy}
  et~al.}{2014}]{mulcahy_14a}
{Mulcahy} D.~D.,  et~al., 2014, \mn@doi [\aap] {10.1051/0004-6361/201424187},
  \href {http://esoads.eso.org/abs/2014A%26A...568A..74M} {568, A74}

\bibitem[\protect\citeauthoryear{{Murphy}, {Helou}, {Kenney}, {Armus}  \&
  {Braun}}{{Murphy} et~al.}{2008}]{murphy_08a}
{Murphy} E.~J.,  {Helou} G.,  {Kenney} J.~D.~P.,  {Armus} L.,   {Braun} R.,
  2008, \mn@doi [\apj] {10.1086/587123}, \href
  {http://esoads.eso.org/abs/2008ApJ...678..828M} {678, 828}

\bibitem[\protect\citeauthoryear{{Neininger} \& {Dumke}}{{Neininger} \&
  {Dumke}}{1999}]{neininger_99a}
{Neininger} N.,  {Dumke} M.,  1999, \mn@doi [Proc. Natl. Acad. Sci. USA]
  {10.1073/pnas.96.10.5360}, \href
  {http://esoads.eso.org/abs/1999PNAS...96.5360N} {96, 5360}

\bibitem[\protect\citeauthoryear{{Niklas} \& {Beck}}{{Niklas} \&
  {Beck}}{1997}]{niklas_97a}
{Niklas} S.,  {Beck} R.,  1997, \aap, \href
  {http://esoads.eso.org/abs/1997A%26A...320...54N} {320, 54}

\bibitem[\protect\citeauthoryear{{Niklas}, {Klein}  \& {Wielebinski}}{{Niklas}
  et~al.}{1997}]{niklas_97b}
{Niklas} S.,  {Klein} U.,   {Wielebinski} R.,  1997, \aap, \href
  {http://esoads.eso.org/abs/1997A%26A...322...19N} {322, 19}

\bibitem[\protect\citeauthoryear{{Oosterloo}, {Fraternali}  \&
  {Sancisi}}{{Oosterloo} et~al.}{2007}]{oosterloo_07a}
{Oosterloo} T.,  {Fraternali} F.,   {Sancisi} R.,  2007, \mn@doi [\aj]
  {10.1086/520332}, \href {http://esoads.eso.org/abs/2007AJ....134.1019O} {134,
  1019}

\bibitem[\protect\citeauthoryear{{Press}, {Teukolsky}, {Vetterling}  \&
  {Flannery}}{{Press} et~al.}{1992}]{press_92a}
{Press} W.~H.,  {Teukolsky} S.~A.,  {Vetterling} W.~T.,   {Flannery} B.~P.,
  1992, {Numerical recipes in FORTRAN. The Art of Scientific Computing}.
Cambridge University Press, Cambridge, UK

\bibitem[\protect\citeauthoryear{{Rau} \& {Cornwell}}{{Rau} \&
  {Cornwell}}{2011}]{rau_11a}
{Rau} U.,  {Cornwell} T.~J.,  2011, \mn@doi [\aap]
  {10.1051/0004-6361/201117104}, \href
  {http://esoads.eso.org/abs/2011A%26A...532A..71R} {532, A71}

\bibitem[\protect\citeauthoryear{{Rohlfs} \& {Wilson}}{{Rohlfs} \&
  {Wilson}}{2004}]{wilson_04a}
{Rohlfs} K.,  {Wilson} T.~L.,  2004, {Tools of Radio Astronomy}.
Springer, Berlin, Germany

\bibitem[\protect\citeauthoryear{{Rossa} \& {Dettmar}}{{Rossa} \&
  {Dettmar}}{2003a}]{rossa_03a}
{Rossa} J.,  {Dettmar} R.-J.,  2003a, \mn@doi [\aap]
  {10.1051/0004-6361:20030615}, \href
  {http://esoads.eso.org/abs/2003A%26A...406..493R} {406, 493}

\bibitem[\protect\citeauthoryear{{Rossa} \& {Dettmar}}{{Rossa} \&
  {Dettmar}}{2003b}]{rossa_03b}
{Rossa} J.,  {Dettmar} R.-J.,  2003b, \mn@doi [\aap]
  {10.1051/0004-6361:20030698}, \href
  {http://esoads.eso.org/abs/2003A%26A...406..505R} {406, 505}

\bibitem[\protect\citeauthoryear{{Rossa}, {Dettmar}, {Walterbos}  \&
  {Norman}}{{Rossa} et~al.}{2004}]{rossa_04a}
{Rossa} J.,  {Dettmar} R.-J.,  {Walterbos} R.~A.~M.,   {Norman} C.~A.,  2004,
  \mn@doi [\aj] {10.1086/422489}, \href
  {http://esoads.eso.org/abs/2004AJ....128..674R} {128, 674}

\bibitem[\protect\citeauthoryear{{Sault}, {Teuben}  \& {Wright}}{{Sault}
  et~al.}{1995}]{sault_95a}
{Sault} R.~J.,  {Teuben} P.~J.,   {Wright} M.~C.~H.,  1995, ASP Conf.\ Ser.,
  \href {http://adsabs.harvard.edu/abs/1995ASPC...77..433S} {77, 433}

\bibitem[\protect\citeauthoryear{{Schlafly} \& {Finkbeiner}}{{Schlafly} \&
  {Finkbeiner}}{2011}]{schlafly_11a}
{Schlafly} E.~F.,  {Finkbeiner} D.~P.,  2011, \mn@doi [\apj]
  {10.1088/0004-637X/737/2/103}, \href
  {http://esoads.eso.org/abs/2011ApJ...737..103S} {737, 103}

\bibitem[\protect\citeauthoryear{{Shapiro} \& {Field}}{{Shapiro} \&
  {Field}}{1976}]{shapiro_76a}
{Shapiro} P.~R.,  {Field} G.~B.,  1976, \mn@doi [\apj] {10.1086/154332}, \href
  {http://esoads.eso.org/abs/1976ApJ...205..762S} {205, 762}

\bibitem[\protect\citeauthoryear{{Soida}, {Krause}, {Dettmar}  \&
  {Urbanik}}{{Soida} et~al.}{2011}]{soida_11a}
{Soida} M.,  {Krause} M.,  {Dettmar} R.-J.,   {Urbanik} M.,  2011, \mn@doi
  [\aap] {10.1051/0004-6361/200810763}, \href
  {http://esoads.eso.org/abs/2011A%26A...531A.127S} {531, A127}

\bibitem[\protect\citeauthoryear{{Sokoloff}, {Bykov}, {Shukurov},
  {Berkhuijsen}, {Beck}  \& {Poezd}}{{Sokoloff} et~al.}{1998}]{sokoloff_98a}
{Sokoloff} D.~D.,  {Bykov} A.~A.,  {Shukurov} A.,  {Berkhuijsen} E.~M.,  {Beck}
  R.,   {Poezd} A.~D.,  1998, \mn@doi [\mnras]
  {10.1046/j.1365-8711.1998.01782.x}, \href
  {http://adsabs.harvard.edu/abs/1998MNRAS.299..189S} {299, 189}

\bibitem[\protect\citeauthoryear{{Strong}, {Moskalenko}  \& {Ptuskin}}{{Strong}
  et~al.}{2007}]{strong_07a}
{Strong} A.~W.,  {Moskalenko} I.~V.,   {Ptuskin} V.~S.,  2007, \mn@doi [Annu.\
  Rev.\ Nucl.\ Part.\ Sci.] {10.1146/annurev.nucl.57.090506.123011}, \href
  {http://esoads.eso.org/abs/2007ARNPS..57..285S} {57, 285}

\bibitem[\protect\citeauthoryear{{Tabatabaei} et~al.,}{{Tabatabaei}
  et~al.}{2013a}]{tabatabaei_13a}
{Tabatabaei} F.~S.,  et~al., 2013a, \mn@doi [\aap]
  {10.1051/0004-6361/201220249}, \href
  {http://esoads.eso.org/abs/2013A%26A...552A..19T} {552, A19}

\bibitem[\protect\citeauthoryear{{Tabatabaei}, {Berkhuijsen}, {Frick}, {Beck}
  \& {Schinnerer}}{{Tabatabaei} et~al.}{2013b}]{tabatabaei_13b}
{Tabatabaei} F.~S.,  {Berkhuijsen} E.~M.,  {Frick} P.,  {Beck} R.,
  {Schinnerer} E.,  2013b, \mn@doi [\aap] {10.1051/0004-6361/201218909}, \href
  {http://esoads.eso.org/abs/2013A%26A...557A.129T} {557, A129}

\bibitem[\protect\citeauthoryear{{T{\"u}llmann}, {Dettmar}, {Soida}, {Urbanik}
  \& {Rossa}}{{T{\"u}llmann} et~al.}{2000}]{tuellmann_00a}
{T{\"u}llmann} R.,  {Dettmar} R.-J.,  {Soida} M.,  {Urbanik} M.,   {Rossa} J.,
  2000, \aap, \href {http://adsabs.harvard.edu/abs/2000A%26A...364L..36T} {364,
  L36}

\bibitem[\protect\citeauthoryear{{T{\"u}llmann}, {Pietsch}, {Rossa},
  {Breitschwerdt}  \& {Dettmar}}{{T{\"u}llmann} et~al.}{2006}]{tuellmann_06a}
{T{\"u}llmann} R.,  {Pietsch} W.,  {Rossa} J.,  {Breitschwerdt} D.,   {Dettmar}
  R.-J.,  2006, \mn@doi [\aap] {10.1051/0004-6361:20052936}, \href
  {http://esoads.eso.org/abs/2006A%26A...448...43T} {448, 43}

\bibitem[\protect\citeauthoryear{{V\"olk}}{{V\"olk}}{1989}]{voelk_89a}
{V\"olk} H.~J.,  1989, \aap, \href
  {http://esoads.eso.org/abs/1989A%26A...218...67V} {218, 67}

\bibitem[\protect\citeauthoryear{{Wiegert} et~al.,}{{Wiegert}
  et~al.}{2015}]{wiegert_15a}
{Wiegert} T.,  et~al., 2015, \mn@doi [\aj] {10.1088/0004-6256/150/3/81}, \href
  {http://esoads.eso.org/abs/2015AJ....150...81W} {150, 81}

\bibitem[\protect\citeauthoryear{{van Haarlem} et~al.,}{{van Haarlem}
  et~al.}{2013}]{vanHaarlem_13a}
{van Haarlem} M.~P.,  et~al., 2013, \mn@doi [\aap]
  {10.1051/0004-6361/201220873}, \href
  {http://esoads.eso.org/abs/2013A%26A...556A...2V} {556, A2}

\makeatother
\end{thebibliography}
\appendix

\section{Numerical modelling of the cosmic ray transport equations}
In this appendix we describe how we model the cosmic ray transport equations
in order to create synthetic profiles of the non-thermal radio continuum
intensity.

\subsection*{Cosmic ray electron number density}
As explained in the main text of the paper, we aim to solve following two equations for the CRe number
density numerically:
\begin{equation}
  \frac{\upartial N(E,z)} {\upartial z} = \frac{1}{V}\left \lbrace
    \frac{\upartial}{\upartial E}\left [ b(E)
    N(E,z)\right ]\right \rbrace\qquad ({\rm Advection}),
\label{eq:conv2}
\end{equation}
\begin{equation}
  \frac{\upartial^2N(E,z)}{\upartial z^2} = \frac{1}{D}\left \lbrace\frac{\upartial}{\upartial
    E}\left [ b(E) N(E,z)\right ]\right \rbrace\qquad ({\rm Diffusion}),
\label{eq:diff2}
\end{equation}
where Equation~(\ref{eq:conv2}) describes pure cosmic ray advection and
Equation~(\ref{eq:diff2}) describes pure cosmic ray diffusion. The
advection speed $V$ is here assumed to be constant. We assume the
diffusion coefficient to be a function of the CRe energy as $D=D_0E_{\rm GeV}^\umu$,
where values of $\umu$ are thought to range between $0.3$ and $0.6$ \citep*{strong_07a}. The combined
synchrotron and IC radiation energy loss rate for a single CRe is given by \citep[e.g.][]{longair_11a}:
\begin{equation}
-\left (\frac{{\rm d}E}{{\rm d}t}\right )=b(E)=\frac{4}{3} \sigma_{\rm T} c \left (\frac{E}{m_{\rm
      e}c^2} \right )^2 (U_{\rm rad}+U_{\rm B}).
\label{eq:be2}
\end{equation}
Notably, the magnetic field strength is not constant, but falls of
exponentially with increasing distance from the disc plane. A magnetic field
strength of $B=B_0\exp(-z/h_{\rm B})$ leads
to a magnetic field energy density of $U_{\rm B}=B^2/(8\upi)=
B_0^2/(8\upi)\exp(-2z/h_{\rm B})$. The radiation energy density is $U_{\rm
  rad}=U_{\rm IRF} +U_{\rm CMB}$ (Sect.~\ref{urad}), where we assume a constant
ratio of $U_{\rm IRF}/U_{\rm B}$ everywhere. 

\begin{figure}
   \includegraphics[width=0.9\hsize]{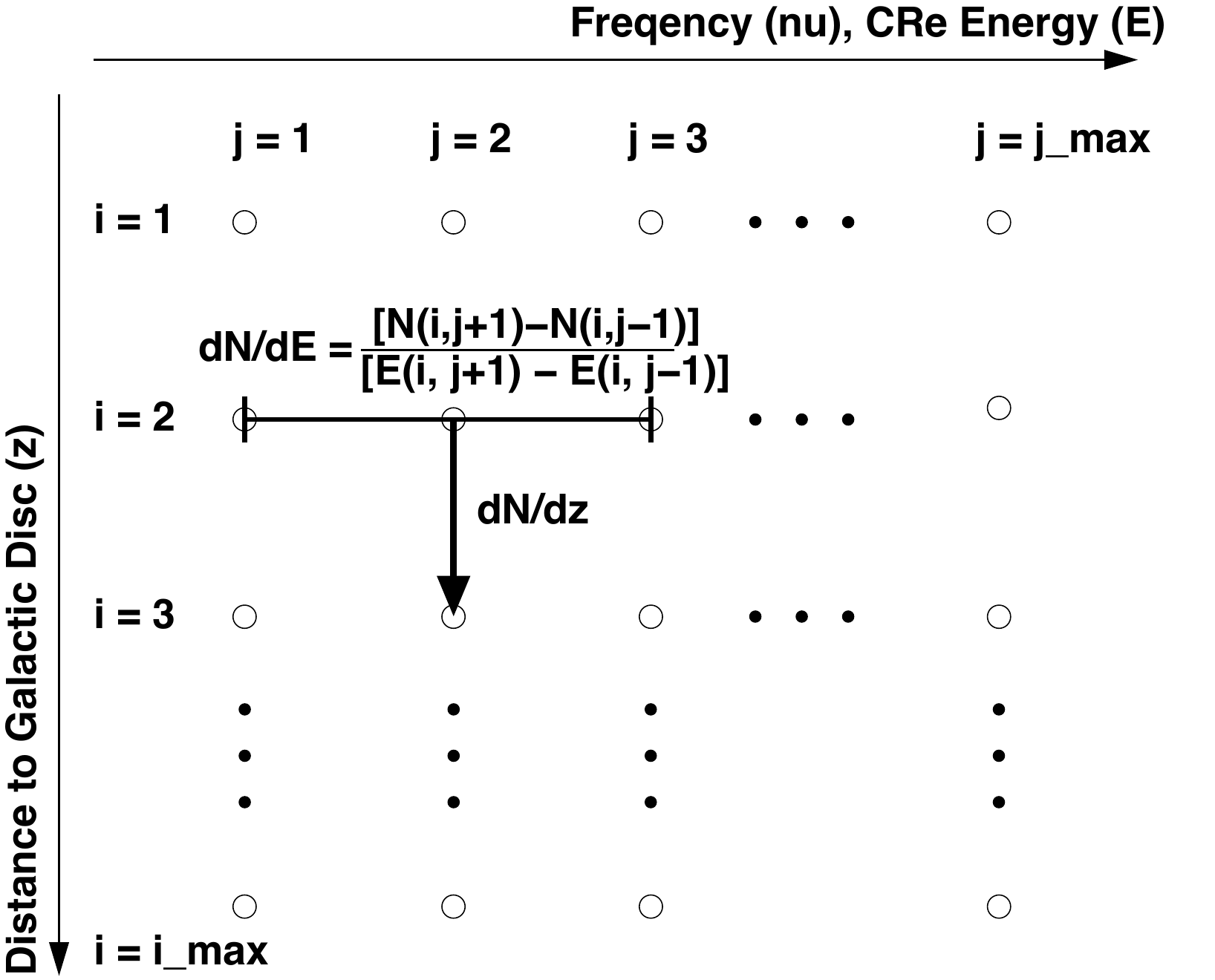}
    \caption{Numerical grid used for the discretization of the 1D cosmic ray transport equations.}
    \label{fig:grid}
\end{figure}

We solve Equations (\ref{eq:conv2}) and (\ref{eq:diff2}) numerically using the
method of
finite differences. To this end, we set up a numerical grid with $z_{\rm i}$ ($i=0,1,2,3\ldots i_{\rm
  max}$) and $\nu_{\rm j}$  ($j=0,1,2,3\ldots j_{\rm
  max}$), where $z_0=0$~kpc and $z_{\max}=4\ldots 7$~kpc with 200 grid points
so that $\Delta z=0.02\ldots 0.035$~kpc. The
frequency range is given by $\nu_0$=10~MHz and $\nu_{\rm max}$=1000~GHz, with
400 grid points logarithmically spaced so that $\nu_{j+1}/\nu_j=1.035$. We convert frequencies
into CRe energies, assuming that the CRe emit all their energy at the
critical frequency:
\begin{equation}
  \nu_{\rm c} = 0.01608 \left ( \frac{E}{\rm GeV} \right )^2
  \left ( \frac{B}{\rm\umu G} \right ) ~{\rm GHz}.
\label{eq:crit}
\end{equation}
This results in grid points of $E_j$  ($j=0,1,2,3\ldots j_{\rm
  max}$), where $E_0 = 0.24$~GeV and $E_{\rm max}= 79$~GeV for a magnetic
field strength of $B_0=10~\umu\rm G$. Thus, we have a two-dimensional grid for the CRe electron number density
$N(E_j,z_i)$.  A sketch of the numerical
grid used is shown in Fig.~\ref{fig:grid}. 
As inputs to the right hand side of Equations~(\ref{eq:conv2})
and (\ref{eq:diff2}) we have:
\begin{equation}
  \frac{\upartial}{\upartial E}\left [ b(E) N(E,z)\right ] =
 \frac{\upartial b}{\upartial E}  N(E,z)  + b(E) \frac{\upartial N}{\upartial E}.
\end{equation}
In case of advection, using the discretization of values on the numerical grid, we obtain:
\begin{equation}
  N_{i+1,j} = N_{i,j} + \frac{1}{V} \left [2
    b(E_{i,j})E_{i,j}^{-1}  N_{i,j} + b(E_{i,j}) \frac{N_{i,j+1} - N_{i,j-1}}
    {E_{i,j+1} - E_{i,j-1}}\right ] \Delta z.
\label{eq:conv3}
\end{equation}
In case of diffusion, we introduce $y=\upartial N/\upartial z$, so that:
\begin{eqnarray}
  \nonumber
  y_{i+1,j} & = & y_{i,j} + \frac{1}{D} \left [2
    b(E_{i,j})E_{i,j}^{-1}  N_{i,j} + b(E_{i,j}) \frac{N_{i,j+1} - N_{i,j-1}}
    {E_{i,j+1} - E_{i,j-1}}\right ] \Delta z \\
  N_{i+1,j} & = & N_{i,j} + y_{i,j} \Delta z.
\label{eq:diff3}
\end{eqnarray}
The inner boundary condition at $z=0$~kpc is $N(E,z)=N_0E^{-\gamma}$, setting
$N_0=1$, and obtaining $\gamma$ from the relation $\alpha_{\rm nt}=(1-\gamma)/2$,
where the non-thermal spectral index can be measured in the galactic disc
plane. Equations~(\ref{eq:conv3}) and (\ref{eq:diff3}) can be integrated from the
inner boundary using a Runge--Kutta scheme \citep[e.g.][]{press_92a}. 

\subsection*{Synchrotron intensity}
Usually, the synchrotron intensity is calculated with the assumption that the
CRe can be described as
power-law so that $N(E,z) = N_o(z)E^{-\gamma(z)}$. In such a case the
synchrotron intensity of an ensemble of CRe is:
\begin{equation}
  I_{\rm nt}(\nu) = ({\rm constants}) N_0(z) B^{(\gamma+1)/2.0} \nu^{-(\gamma-1)/2}.
\end{equation}
Thus, one obtains the relation $\alpha_{\rm nt}=(1-\gamma)/2$, already used
above. However, the assumption that the CRe number density is a power-law is
not true if spectral ageing is important. We expect an exponential cut-off at high frequencies. A proper
calculation of synchrotron intensities has to convolve the synchrotron
emission function of a single CRe with the CRe number density:
\begin{equation}
  I_{\rm nt}(\nu) = \int_0^\infty j(\nu) N(E,z) {\rm d}E.
\end{equation}
We introduce the ratio $x=\nu/\nu_{\rm c}$, where $\nu_{\rm c}$ is again the
critical frequency as defined in Equation~(\ref{eq:crit}). The synchrotron
emissivity of a single ultra-relativistic CRe is:
\begin{equation}
  j(x) = ({\rm constants}) B F(x).
\end{equation}
We have used the tabulation of the function $F(x)$ from
\citet{wilson_04a}. In order to calculate the synchrotron intensity, we have
to integrate over the ratio $x$ of observing frequency to critical
frequency. It is also important to keep in mind that the critical frequency is
a function of the magnetic field strength and thus of the distance to the
disc. First, we convert from the integral over the energy to an integral over $x$:
\begin{equation}
  {\rm d}x = \frac{\nu}{\nu_{\rm c}^2} {\rm
    d}\nu_{\rm c} \quad x = \frac{A}{E^2} \quad {\rm d}E = \frac{1}{2} A^{1/2}
  x^{-3/2} {\rm d}x.
\end{equation}
We can now integrate over $x$:
\begin{equation}
  I_{\rm nt}(\nu) = \frac{1}{2A^{1/2}}\int_0^\infty j(x) N(x) x^{-3/2}{\rm d}x.
\end{equation}
Because our numerical grid is set up in frequency space, we write:
\begin{equation}
  x^{-3/2} {\rm d}x = \frac{{\rm d}\nu_{\rm c}}{\nu_{\rm c}^{1/2} \nu^{1/2}}.
\end{equation}
We can now integrate over the frequency space:
\begin{equation}
  I_{\rm nt}(\nu) = \frac{1}{2A^{1/2}}\int_0^\infty j(x) N(x) \frac{{\rm d}\nu_{\rm c}}{\nu_{\rm c}^{1/2} \nu^{1/2}}.
\end{equation}
\begin{figure}
 \includegraphics[width=0.99\hsize]{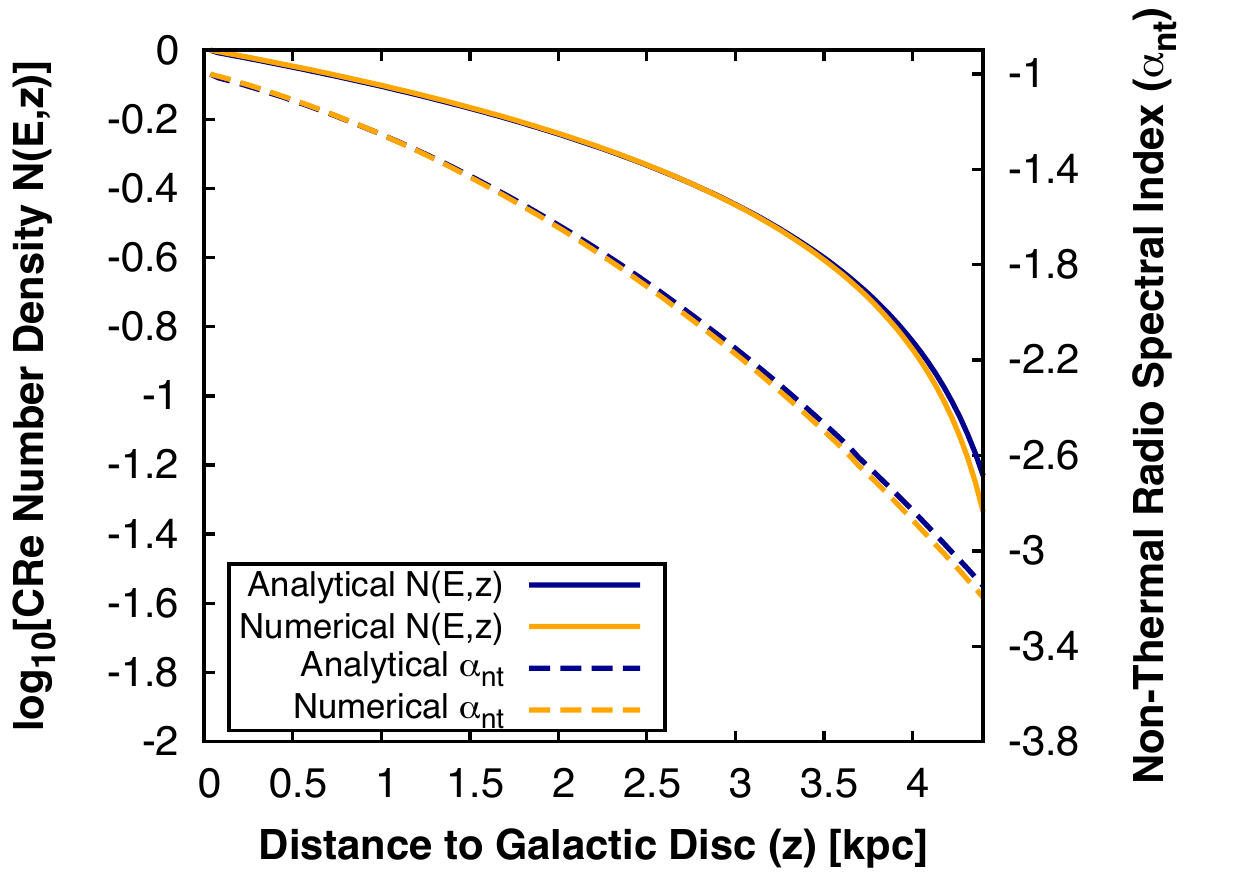}
    \caption{Profiles of the CRe number density at $\lambda$22~cm (equivalent
      to a CRe energy of $2.9$~GeV) for a
      constant magnetic field and advective cosmic ray transport ($B=10~\umu\rm G$, $U_{\rm
        rad}/U_{\rm B}=0.31$, $V=150~\rm km\,s^{-1}$, $\gamma_{\rm
        inj}=3.0$). Compared are the analytical and 
      numerical solutions, which are plotted on the left-hand side
      $y$-axis. The corresponding
       non-thermal radio spectral indices between $\lambda\lambda$ 22 and 6~cm are plotted on the right-hand side $y$-axis.}
    \label{fig:ute}
\end{figure}
\begin{figure}
 \includegraphics[width=0.99\hsize]{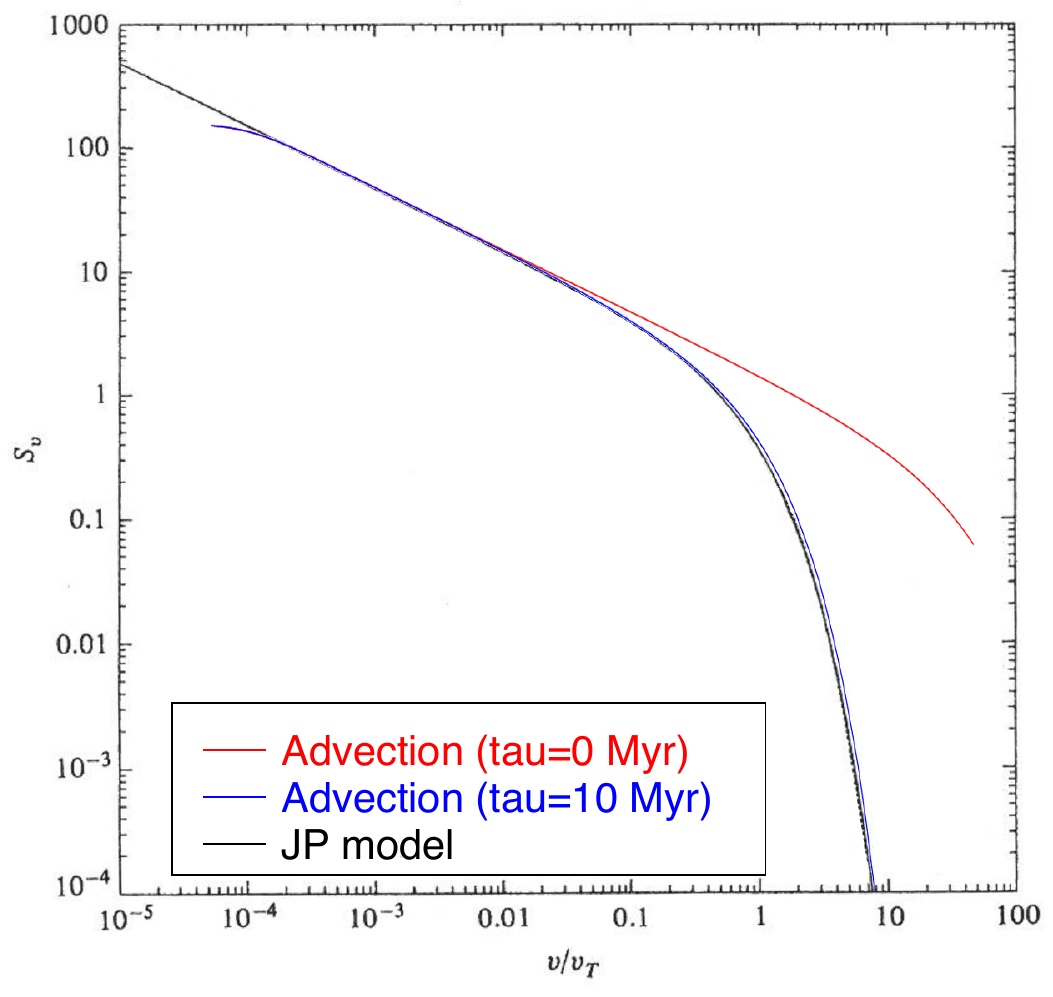}
    \caption{Comparison of our advection model with a JP spectrum taken from
      \citet{hughes_91a}. The upper, red line shows the injection spectrum
      ($\gamma_{\rm inj}=2.0$, $\tau=0$~Myr) and the lower, blue line shows the spectrally aged
      spectrum ($\tau=10$~Myr). The lower, black line shows the JP model to which our
      spectrum should be compared. The flux densities $S_\nu$ are on
      an arbitrary scale. The frequencies $\nu$ are normalized to the break
      frequency $\nu_{\rm T}=20.7$~GHz ($B=10~\umu\rm G$,
      $U_{\rm IRF}/U_{\rm B}=0$). IC losses in the CMB at redshift zero are
      taken into account.}
    \label{fig:syn}
\end{figure}
\subsection*{Comparison with reference models}
We can test the numerical solution of the advection
equation for a constant magnetic field strength, for which
there is an analytical solution for the CRe number distribution \citep[e.g.][]{longair_11a}:
\begin{equation}
  N(E,z) = N_0 \left ( 1 - \frac{bz}{E} \right )^{\gamma-1}.
\label{eq:analytical}
\end{equation}
In Fig.~\ref{fig:ute}, we compare the analytical with the numerical
solution. We find that the electron number densities agree to within 1~per cent
at distances from the disc $z<3.0$~kpc, where the non-thermal spectral index
is $\alpha_{\rm nt}>-2.2$.  Even at larger distances
where the spectral index steepens to a very steep $-3.0$, the agreement is
better than 6 per cent. Since this is the regime we are mostly interested
in, our numerical solution is of sufficient accuracy.

A second test is to compare our model with the widely used Jaffe--Perola \citep[JP;][]{jaffe_73a}
model, which predicts a spectrum of the non-thermal radio continuum intensity
as function of frequency. This model used the CRe number distribution of
Equation~(\ref{eq:analytical}) and convolves it with the synchrotron emission
spectrum of a single CRe. In Fig.~\ref{fig:syn} we compare the synchrotron
intensities from our modelling with the JP spectrum as presented in
\citet{hughes_91a}. The upper, red line shows the injection spectrum and the
lower, blue
line the spectrally aged spectrum at a spectral age of 10~Myr. The
theoretical predication by the JP model is shown as the lower, black line. The frequency
is normalized with respect to the break frequency $\nu_{\rm brk}$ for which we find \citep[e.g.][]{hughes_91a}:
\begin{equation}
  \nu_{\rm T} = \frac{2.52 \times 10^3 [ B /10~\umu{\rm G}]} {([B / 10~\umu {\rm G}]^2
    +  [B_{\rm CMB} /10~\umu{\rm G}]^2)^2[\tau / \rm Myr]^2}~{\rm GHz} \mbox{.}
\label{eq:nu_brk}
\end{equation}
Here, the equivalent CMB magnetic field strength at redshift zero is
$B_{\rm CMB} = 3.2~\rm \umu G$. For our case ($B=10~\umu\rm G$, $U_{\rm
  IRF}=0$, $\tau=10$~Myr), the break frequency is $\nu_{\rm T}=20.7~\rm
GHz$. At the break frequency the spectrally aged spectrum is deficient by a
factor of four compared with the injection spectrum. We find a very good
agreement between our spectrum and the reference spectrum from
\citet{hughes_91a}.

\section{Data tables}
\label{data_tables}
In this appendix we present tables of the data
used in the paper for the analysis. Tables \ref{tab:sh_n7090} and
\ref{tab:sh_n7462} contain the non-thermal scale heights and CRe radiation lifetimes in
NGC~7090 and 7462, respectively. Tables \ref{tab:si_n7090} and
\ref{tab:si_n7462} contain the vertical profiles of the non-thermal radio
spectral index in NGC~7090 and 7462, respectively. Quantities for the
northern/southern haloes are designated as `N/S'. Finally, in
Table~\ref{tab:b_n7090} we present vertical profiles in NGC~7090 of the linearly polarized
emission together, degrees of polarization together with the total ($B$), ordered ($B_{\rm ord}$) and turbulent ($B_{\rm turb}$) magnetic
field strengths.

\begin{table*}
\caption{Non-thermal radio continuum scale heights measured from
  two-component exponential fits and CRe radiation
  lifetimes in the northern (N) and southern (S) halo of NGC~7090.\label{tab:sh_n7090}}
\begin{tabular}{lcccc cccc ccc}
\hline\hline
Offset & \multicolumn{4}{c}{Thin disc (kpc)} & \multicolumn{4}{c}{Thick disc (kpc)} & \multicolumn{2}{c}{Rad.\ lif.\ (Myr)}\\
(kpc) & $\lambda$22~cm (N) & $\lambda$6~cm (N) & $\lambda$22~cm (S) & $\lambda$6~cm (S) & $\lambda$22~cm (N) & $\lambda$6~cm (N) & $\lambda$22~cm (S) & $\lambda$6~cm (S) & $t_{\rm rad22}$ & $t_{\rm rad6}$ \\
\hline
$-3.4$ & $0.34\pm 0.08$ & $0.31\pm 0.08$ & $0.44\pm 0.05$ & $0.23\pm 0.13$ & $1.54\pm 0.35$ & $1.18\pm 1.00$ & $2.65\pm 1.85$ & $0.92\pm 1.85$ & $28.0\pm 2.0$ & $15.3\pm 1.0$ \\
$-1.7$ & $0.41\pm 0.05$ & $0.47\pm 0.09$ & $0.39\pm 0.05$ & $0.21\pm 0.08$ & $1.57\pm 0.12$ & $1.31\pm 0.89$ & $2.17\pm 0.47$ & $1.06\pm 0.47$ & $20.5\pm 2.0$ & $11.2\pm 1.0$ \\
$0.0$  & $0.31\pm 0.03$ & $0.34\pm 0.05$ & $0.38\pm 0.04$ & $0.18\pm 0.12$ & $1.92\pm 0.12$ & $2.04\pm 0.48$ & $1.52\pm 0.29$ & $0.97\pm 0.29$ & $17.2\pm 2.0$ & $9.4 \pm 1.0$ \\
$1.7$  & $0.34\pm 0.18$ & $0.47\pm 0.15$ & $0.50\pm 0.10$ & $0.40\pm 0.07$ & $1.67\pm 0.18$ & $2.15\pm 0.55$ & $1.40\pm 1.00$ & $1.77\pm 1.00$ & $21.0\pm 3.0$ & $11.5\pm 1.0$ \\
$3.4$  & $0.63\pm 0.13$ & $0.21\pm 0.99$ & $0.04\pm 0.46$ & $0.46\pm 0.25$ & $2.33\pm 0.55$ & $2.01\pm 1.19$ & $1.35\pm 0.25$ & $1.39\pm 0.25$ & $26.0\pm 2.0$ & $13.9\pm 1.0$ \\
\hline
\end{tabular}
\end{table*}

\begin{table*}
\caption{Non-thermal radio continuum scale heights measured from one-component
  Gaussian fits and CRe radiation
  lifetimes in the northern (N) and southern (S) halo of NGC~7462.\label{tab:sh_n7462}}
\begin{tabular}{lcccc cc}
\hline\hline
& \multicolumn{4}{c}{Thick disc (kpc)}  & \multicolumn{2}{c}{Rad.\ lif.\ (Myr)}\\
Offset (kpc) & $\lambda$22~cm (N) & $\lambda$6~cm (N) & $\lambda$22~cm (S) & $\lambda$6~cm (S) & $t_{\rm rad22}$ & $t_{\rm rad6}$\\
\hline
$-4.8$ & $2.17\pm 0.15$ & $1.58\pm 0.08$ & $1.96\pm 0.11$ & $2.01\pm 0.13$ & $29.5\pm 2.0$ & $16.1\pm 1.0$ \\
$-2.4$ & $1.11\pm 0.15$ & $1.11\pm 0.04$ & $1.42\pm 0.05$ & $1.34\pm 0.03$ & $21.2\pm 2.0$ & $11.6\pm 1.0$ \\
$ 0.0$ & $1.36\pm 0.07$ & $1.23\pm 0.08$ & $1.35\pm 0.04$ & $1.32\pm 0.03$ & $19.0\pm 2.0$ & $10.4\pm 1.0$ \\
$ 2.4$ & $1.89\pm 0.09$ & $2.27\pm 0.11$ & $1.79\pm 0.05$ & $1.61\pm 0.05$ & $25.6\pm 2.0$ & $14.0\pm 1.0$ \\
$ 4.8$ & $1.56\pm 0.09$ & $1.94\pm 0.24$ & $1.09\pm 0.06$ & $1.41\pm 0.12$ & $32.6\pm 2.0$ & $17.8\pm 1.0$ \\
\hline                     
\end{tabular}
\end{table*}

\begin{table*}
\caption{Vertical profiles of the non-thermal radio continuum
  intensities (at $\lambda\lambda$ 22 and 6~cm) and of the radio spectral
  index (between $\lambda\lambda$ 22 and 6~cm) in the northern (N) and southern (S) halo of NGC~7090.\label{tab:si_n7090}}
\begin{tabular}{lcccccccc}
\hline\hline
$|z|$  & $I_{\rm nt22}$ (N) & $I_{\rm nt6}$ (N) & $\alpha_{\rm nt}$ (N) & $I_{\rm nt22}$(S) & $I_{\rm nt6}$ (S) & $\alpha_{\rm nt}$ (S)\\
(kpc) & \multicolumn{2}{c}{($\umu\rm Jy\,beam^{-1}$)} & & \multicolumn{2}{c}{($\umu\rm Jy\,beam^{-1}$)}\\
\hline
$0.1750$ & $1266\pm 64$ & $380\pm 19$ & $-0.99\pm 0.07$ & $1214\pm 61$ & $336\pm 17$ & $-1.06\pm 0.07$\\
$0.5250$ & $922 \pm 47$ & $260\pm 13$ & $-1.05\pm 0.07$ & $836 \pm 43$ & $201\pm 11$ & $-1.18\pm 0.07$\\
$0.8750$ & $585 \pm 30$ & $150\pm 8 $ & $-1.12\pm 0.08$ & $485 \pm 26$ & $102\pm 6 $ & $-1.29\pm 0.09$\\
$1.2250$ & $378 \pm 21$ & $85 \pm 6 $ & $-1.23\pm 0.10$ & $277 \pm 16$ & $58 \pm 5 $ & $-1.29\pm 0.11$\\
$1.5750$ & $287 \pm 16$ & $54 \pm 5 $ & $-1.37\pm 0.12$ & $187 \pm 12$ & $39 \pm 4 $ & $-1.30\pm 0.14$\\
$1.9250$ & $249 \pm 15$ & $48 \pm 4 $ & $-1.37\pm 0.13$ & $148 \pm 11$ & $33 \pm 4 $ & $-1.23\pm 0.16$\\
$2.2750$ & $230 \pm 14$ & $49 \pm 4 $ & $-1.28\pm 0.13$ & $118 \pm 10$ & $32 \pm 4 $ & $-1.07\pm 0.17$\\
$2.6250$ & $189 \pm 12$ & $41 \pm 4 $ & $-1.25\pm 0.14$ & $78  \pm 9 $ & $19 \pm 4 $ & $-1.18\pm 0.24$\\
$2.9750$ & $147 \pm 11$ & $32 \pm 4 $ & $-1.25\pm 0.16$ & $34  \pm 8 $ & $1  \pm 4 $ & $-3.17\pm 4.19$\\
$3.3250$ & $122 \pm 10$ & $21 \pm 4 $ & $-1.47\pm 0.21$ & $32  \pm 8 $ & $3  \pm 4 $ & $-2.08\pm 1.24$\\
$3.6750$ & $101 \pm 9 $ & $15 \pm 4 $ & $-1.60\pm 0.27$ & $52  \pm 8 $ & $10 \pm 4 $ & $-1.36\pm 0.39$\\
$4.0250$ & $85  \pm 9 $ & $12 \pm 4 $ & $-1.60\pm 0.31$ & $59  \pm 9 $ & $11 \pm 4 $ & $-1.40\pm 0.36$\\
\hline
\end{tabular}
\end{table*}

\begin{table*}
\caption{Vertical profiles of the non-thermal radio continuum
  intensities (at $\lambda\lambda$ 22 and 6~cm) and of the radio spectral
  index (between $\lambda\lambda$ 22 and 6~cm) in the northern (N) and southern (S) halo of NGC~7462.\label{tab:si_n7462}}
\begin{tabular}{lcccccccc}
\hline\hline
$|z|$  & $I_{\rm nt22}$ (N) & $I_{\rm nt6}$ (N) & $\alpha_{\rm nt}$ (N) & $I_{\rm nt22}$(S) & $I_{\rm nt6}$ (S) & $\alpha_{\rm nt}$ (S)\\
(kpc) & \multicolumn{2}{c}{($\umu\rm Jy\,beam^{-1}$)} & & \multicolumn{2}{c}{($\umu\rm Jy\,beam^{-1}$)}\\
\hline
$0.225$ & $638\pm 33$ & $142\pm 8$ & $-1.24\pm 0.06$ & $716\pm 36$ & $157\pm 8$ & $-1.25\pm 0.06$\\
$0.675$ & $468\pm 24$ & $109\pm 6$ & $-1.20\pm 0.06$ & $581\pm 30$ & $132\pm 7$ & $-1.22\pm 0.06$\\
$1.125$ & $302\pm 17$ & $76 \pm 5$ & $-1.14\pm 0.07$ & $400\pm 21$ & $98 \pm 5$ & $-1.16\pm 0.06$\\
$1.575$ & $182\pm 11$ & $45 \pm 3$ & $-1.16\pm 0.08$ & $243\pm 14$ & $57 \pm 4$ & $-1.20\pm 0.07$\\
$2.025$ & $107\pm 9 $ & $21 \pm 3$ & $-1.33\pm 0.13$ & $144\pm 10$ & $30 \pm 3$ & $-1.30\pm 0.10$\\
$2.475$ & $79 \pm 8 $ & $11 \pm 3$ & $-1.66\pm 0.22$ & $67 \pm 8 $ & $14 \pm 3$ & $-1.28\pm 0.18$\\
$2.925$ & $66 \pm 8 $ & $5  \pm 3$ & $-2.19\pm 0.46$ & $25 \pm 7 $ & $4  \pm 3$ & $-1.49\pm 0.54$\\
$3.375$ & $39 \pm 7 $ & $3  \pm 3$ & $-2.25\pm 0.82$ & $19 \pm 7 $ & $2  \pm 3$ & $-1.92\pm 1.12$\\
\hline
\end{tabular}
\end{table*}

\begin{table*}
\caption{Vertical profiles of the linearly polarized intensity, degree
  of polarization (at $\lambda\lambda$ 22 and 6~cm, respectively) and magnetic field strengths in NGC~7090. Positive values of $z$ denote the northern and negative ones the southern halo.\label{tab:b_n7090}}
\begin{tabular}{lcccccccc}
\hline\hline
$z$  & $PI_{22}$ & $PI_{6}$ & $p_{22}$ & $p_{6}$ & $\alpha_{\rm nt}$ &  $B$ & $B_{\rm ord}$ & $B_{\rm turb}$\\
(kpc) & ($\umu\rm Jy\,beam^{-1}$) & ($\umu\rm Jy\,beam^{-1}$) & & & & ($\umu\rm G$) & ($\umu\rm G$) & ($\umu\rm G$)\\
\hline
$-6.65$ & $13\pm  11$ & $4 \pm 5$ & $0.12\pm 0.13$ & $-$            & $-$              & $0.70\pm 0.24$ & $0.26\pm 0.25$ & $0.65\pm 0.28$\\
$-5.95$ & $30\pm  11$ & $9 \pm 5$ & $0.27\pm 0.18$ & $-$            & $-3.55\pm 15.55$ & $0.83\pm 0.26$ & $0.46\pm 0.24$ & $0.69\pm 0.36$\\
$-5.25$ & $31\pm  11$ & $10\pm 5$ & $0.21\pm 0.12$ & $0.81\pm 1.90$ & $-2.01\pm 1.92 $ & $0.99\pm 0.29$ & $0.51\pm 0.23$ & $0.85\pm 0.36$\\
$-4.55$ & $34\pm  11$ & $2 \pm 5$ & $0.17\pm 0.08$ & $0.06\pm 0.19$ & $-1.67\pm 0.92 $ & $1.18\pm 0.32$ & $0.54\pm 0.20$ & $1.05\pm 0.37$\\
$-3.85$ & $55\pm  12$ & $12\pm 5$ & $0.24\pm 0.08$ & $0.34\pm 0.33$ & $-1.59\pm 0.75 $ & $1.41\pm 0.34$ & $0.76\pm 0.24$ & $1.18\pm 0.44$\\
$-3.15$ & $73\pm  12$ & $20\pm 5$ & $0.26\pm 0.07$ & $0.41\pm 0.26$ & $-1.43\pm 0.53 $ & $1.67\pm 0.37$ & $1.21\pm 0.33$ & $1.16\pm 0.64$\\
$-2.45$ & $80\pm  12$ & $30\pm 5$ & $0.16\pm 0.03$ & $0.27\pm 0.09$ & $-1.25\pm 0.24 $ & $1.99\pm 0.40$ & $1.18\pm 0.32$ & $1.61\pm 0.55$\\
$-1.75$ & $73\pm  12$ & $39\pm 5$ & $0.07\pm 0.01$ & $0.16\pm 0.03$ & $-1.24\pm 0.12 $ & $2.38\pm 0.42$ & $1.08\pm 0.23$ & $2.12\pm 0.49$\\
$-1.05$ & $47\pm  12$ & $41\pm 5$ & $0.02\pm 0.00$ & $0.07\pm 0.01$ & $-1.17\pm 0.07 $ & $2.83\pm 0.45$ & $0.86\pm 0.16$ & $2.70\pm 0.47$\\
$-0.35$ & $33\pm  11$ & $25\pm 5$ & $0.01\pm 0.00$ & $0.02\pm 0.00$ & $-1.09\pm 0.06 $ & $6.46\pm 0.36$ & $1.13\pm 0.16$ & $6.36\pm 0.36$\\
$0.35$  & $28\pm  11$ & $14\pm 5$ & $0.01\pm 0.00$ & $0.01\pm 0.00$ & $-1.06\pm 0.06 $ & $6.46\pm 0.36$ & $0.81\pm 0.18$ & $6.41\pm 0.36$\\
$1.05$  & $23\pm  11$ & $32\pm 5$ & $0.01\pm 0.00$ & $0.04\pm 0.01$ & $-1.11\pm 0.07 $ & $2.84\pm 0.45$ & $0.68\pm 0.13$ & $2.76\pm 0.46$\\
$1.75$  & $30\pm  11$ & $37\pm 5$ & $0.02\pm 0.01$ & $0.10\pm 0.02$ & $-1.24\pm 0.10 $ & $2.50\pm 0.47$ & $0.92\pm 0.20$ & $2.32\pm 0.51$\\
$2.45$  & $52\pm  12$ & $31\pm 5$ & $0.05\pm 0.01$ & $0.15\pm 0.03$ & $-1.30\pm 0.14 $ & $2.20\pm 0.47$ & $0.96\pm 0.24$ & $1.98\pm 0.54$\\
$3.15$  & $96\pm  12$ & $22\pm 5$ & $0.14\pm 0.02$ & $0.17\pm 0.05$ & $-1.34\pm 0.20 $ & $1.94\pm 0.47$ & $0.90\pm 0.23$ & $1.72\pm 0.54$\\
$3.85$  & $108\pm 13$ & $13\pm 5$ & $0.25\pm 0.05$ & $0.15\pm 0.08$ & $-1.35\pm 0.31 $ & $1.71\pm 0.46$ & $0.96\pm 0.28$ & $1.41\pm 0.59$\\
$4.55$  & $92\pm  12$ & $4 \pm 5$ & $0.37\pm 0.11$ & $0.06\pm 0.07$ & $-1.01\pm 0.39 $ & $1.50\pm 0.44$ & $1.06\pm 0.36$ & $1.07\pm 0.72$\\
$5.25$  & $74\pm  12$ & $7 \pm 5$ & $0.54\pm 0.26$ & $0.12\pm 0.11$ & $-0.72\pm 0.56 $ & $1.32\pm 0.43$ & $1.14\pm 0.50$ & $0.66\pm 1.21$\\
$5.95$  & $66\pm  12$ & $13\pm 5$ & $0.55\pm 0.30$ & $1.00\pm 2.32$ & $-1.85\pm 1.93 $ & $1.16\pm 0.41$ & $0.96\pm 0.46$ & $0.66\pm 0.99$\\
$6.65$  & $31\pm  11$ & $9 \pm 5$ & $0.26\pm 0.17$ & $-$            & $-$              & $1.03\pm 0.39$ & $0.58\pm 0.32$ & $0.85\pm 0.52$\\
\hline       
\end{tabular}
\end{table*}

\bsp	
\label{lastpage}
\end{document}